%% file: main.tex
\pdfoutput=1

\documentclass[sigconf,screen,nonacm]{acmart} 

\usepackage[many]{tcolorbox}
\usepackage{algorithmic}
\usepackage{graphicx}
\usepackage{textcomp}
\usepackage{xspace}

\def\BibTeX{{\rm B\kern-.05em{\sc i\kern-.025em b}\kern-.08em
    T\kern-.1667em\lower.7ex\hbox{E}\kern-.125emX}}
    
 \usepackage[normalem]{ulem}
 \usepackage{fancyhdr}
 \usepackage[labelfont=bf, textfont=bf, skip=2pt]{caption}
 \usepackage[english]{babel}
 \usepackage{booktabs}
 \usepackage{multirow}
 \usepackage{url}
 \usepackage{pifont}
 \usepackage{enumitem}
 \usepackage{color, colortbl}
 \usepackage{xargs}                    
 \usepackage{marginnote}
 \usepackage{clipboard}
 \usepackage{subcaption}
 \usepackage{setspace}
 \usepackage{blindtext}
 \usepackage{imakeidx}
 \usepackage{soul}
 \usepackage[binary-units=true]{siunitx}
 \usepackage[italic]{mathastext}
 \usepackage{flushend}
 \usepackage{dblfloatfix}
 \usepackage{hyperref}
 \usepackage[acronym,nonumberlist,nowarn]{glossaries}
     \glsdisablehyper
    \loadglsentries{acronyms}

\hypersetup{
  colorlinks   = true, 
  urlcolor     = blue, 
  linkcolor    = black, 
  citecolor   = black 
}

\setcopyright{none}

\begin{document}

\include{macros}

\title[Extending Memory Capacity in Consumer Devices with Emerging Non-Volatile Memory: An Experimental Study]{\vspace{-15pt}Extending Memory Capacity in Consumer Devices with \\Emerging Non-Volatile Memory: An Experimental Study \vspace{-10pt}}

\newcommand{\tsc}[1]{\textsuperscript{#1}} 
\newcommand{\affilETH}{\tsc{1}}
\newcommand{\affilUIUC}{\tsc{2}}
\newcommand{\affilGoogle}{\tsc{3}}
\newcommand{\affilRivos}{\tsc{4}}
\settopmatter{authorsperrow=1} 

\author{
 {
  Geraldo F. Oliveira\affilETH \qquad
  Saugata Ghose\affilUIUC \qquad
  Juan Gómez-Luna\affilETH \qquad
  Amirali Boroumand\affilGoogle \qquad
 }
}
\author{
 {
  Alexis Savery\affilGoogle \qquad
  Sonny Rao\affilRivos \qquad
  Salman Qazi\affilGoogle \qquad
  Gwendal Grignou\affilGoogle \qquad
 }
}
\author{
 {
  Rahul Thakur\affilGoogle \qquad
  Eric Shiu\affilRivos \qquad
  Onur Mutlu\affilETH 
 }
}
\affiliation{
\institution{
      \vspace{8pt}
      \affilETH ETH Zürich \qquad
      \affilUIUC University of Illinois Urbana-Champaign \qquad
      \affilGoogle Google \qquad 
      \affilRivos Rivos
  }
  \country{}
}
 
\renewcommand{\authors}{Geraldo F. Oliveira, Saugata Ghose, Juan Gómez-Luna, Amirali Boroumand, Alexis Savery, Sonny Rao, Salman Qazi, Gwendal Grignou, Rahul Thakur, Eric Shiu, and Onur Mutlu}

\begin{abstract}
DRAM scalability is becoming a limiting factor to the available memory capacity in consumer devices. As a potential solution, manufacturers have introduced emerging non-volatile memories (NVMs) into the market, which can be used to increase the memory capacity of consumer devices by augmenting or replacing DRAM. In this work, we provide the first analysis of the impact of extending the main memory space of consumer devices using off-the-shelf NVMs. We equip real web-based Chromebook computers with the Intel Optane solid-state drive (SSD), which contains state-of-the-art low-latency NVM, and use the NVM as swap space. We analyze the performance and energy consumption of the Optane-equipped Chromebooks, and compare this with (i) a baseline system with double the amount of DRAM than the system with the NVM-based swap space; and (ii) a system where the Intel Optane SSD is naively replaced with a state-of-the-art NAND-flash-based SSD. Our experimental analysis reveals that while Optane-based swap space provides a cost-effective way to alleviate the DRAM capacity bottleneck in consumer devices, naive integration of the Optane SSD leads to several system-level overheads, mostly related to (1) the Linux block I/O layer, which can negatively impact overall performance; and (2) the off-chip traffic to the swap space, which can negatively impact energy consumption. To reduce the Linux block I/O layer overheads, we tailor several system-level mechanisms (i.e., the I/O scheduler and the I/O request completion mechanism) to the currently-running application’s access pattern. To reduce the off-chip traffic overhead, we leverage an operating system feature (called Zswap) that allocates some DRAM space to be used as a compressed in-DRAM cache for data swapped between DRAM and the Intel Optane SSD, significantly reducing energy consumption caused by the off-chip traffic to the swap space. We conclude that emerging NVMs are a cost-effective solution to alleviate the DRAM capacity bottleneck in consumer devices, which can be further enhanced by tailoring system-level mechanisms to better leverage the characteristics of our workloads and the NVM.
\end{abstract}


\renewcommand{\shortauthors}{G. F. Oliveira et al.}
\renewcommand{\shorttitle}{Extending Memory Capacity in Consumer Devices with Emerging Non-Volatile Memory: An Experimental Study}

\maketitle
\input{sections/1-introduction.tex}
\input{sections/2-background.tex}
\input{sections/3-performance-evaluation.tex}
\input{sections/4-system-optimization.tex}

\input{sections/6-relatedwork.tex}

\input{sections/7-conclusion.tex}

\section*{Acknowledgments}
    We thank the SAFARI Research Group members for valuable feedback and the stimulating intellectual environment they provide. We acknowledge support from the SAFARI \sgiii{Research} Group's industrial partners, especially ASML, Facebook, Google, Huawei, Intel, Microsoft, and VMware. We acknowledge support from the Semiconductor Research Corporation and the ETH Future Computing Laboratory. \sgiii{This research started at Google, during Geraldo F. Oliveira's internship, and continued as a successful collaboration between Google and SAFARI \gfiii{since then.}}

\bibliographystyle{IEEEtran}
\bibliography{references}

\end{document}

%% file: acronyms.tex
\newacronym{ADI}{ADI}{Alternating Direction Implicit}
\newacronym{AGU}{AGU}{Address Generation Unit}
\newacronym[longplural={Access History Tables}]{AHT}{AHT}{Access History Table}
\newacronym{AHT-R}{AHT-R}{Access History Table - Read}
\newacronym{AHT-W}{AHT-W}{Access History Table - Write-Back}
\newacronym{AIP}{AIP}{Access Interval Predictor}
\newacronym{AL}{AL}{Added Latency for column accesses}
\newacronym{ASIC}{ASIC}{Application Specific Integrated Circuit}
\newacronym{AVX}{AVX}{Advanced Vector Extensions}
\newacronym[longplural={Basic Block Vectors}]{BBV}{BBV}{Basic Block Vector}
\newacronym{BHT}{BHT}{Branch History Table}
\newacronym{HBM}{HBM}{Hybrid Bandwidth Memory}
\newacronym{DDR4}{DDR4}{Double-Data Rate 4}

\newacronym{NN}{NN}{Neural Network}
\newacronym{BTB}{BTB}{Branch Target Buffer}
\newacronym{CAS}{CAS}{Column Address Strobe}
\newacronym{AMAT}{AMAT}{Average Memory Access Time}
\newacronym{CCD}{CCD}{Column to Column Delay}
\newacronym{AI}{AI}{Arithmetic Intensity}
\newacronym[longplural={Columnar Database Systems}]{CDBS}{CDBS}{Columnar Database System}
\newacronym[longplural={Chip Multiprocessors}]{CMP}{CMP}{Chip Multiprocessor}
\newacronym{CMT}{CMT}{Chip Multithreading}
\newacronym[longplural={Coarse-Grain Reconfigurable Arrays}]{CGRA}{CGRA}{Coarse-Grain Reconfigurable Array}
\newacronym{C-RAM}{C-RAM}{Computational-RAM}
\newacronym{CWD}{CWD}{Column Write Delay}
\newacronym{DBI}{DBI}{Dynamic Binary Instrumentation}
\newacronym[longplural={Database Management Systems}]{DBMS}{DBMS}{Database Management System}
\newacronym{DDR}{DDR}{Double Data Rate}
\newacronym{DEWP}{DEWP}{Dead Line and Early Write-Back Predictor}
\newacronym{DRAM}{DRAM}{Dynamic Random Access Memory}
\newacronym{DSBP}{DSBP}{Dead Sub-Block Predictor}
\newacronym{DSM}{DSM}{Decomposed Storage Manage}
\newacronym{FAW}{FAW}{Four row Activation Window}
\newacronym{FP}{FP}{Floating-Point}
\newacronym{EDP}{EDP}{Energy-Delay Product}
\newacronym{FCFS}{FCFS}{First-Come First-Serve}
\newacronym{FIFO}{FIFO}{Fist In, First Out}
\newacronym{FSM}{FSM}{Finite State Machine}
\newacronym{FPGA}{FPGA}{Field-Programmable Gate Array}
\newacronym[longplural={Functional Units}]{FU}{FU}{Functional Unit}
\newacronym{GAg}{GAg}{Global Adaptive branch prediction using one Global PHT}
\newacronym{GAs}{GAs}{Global Adaptive branch prediction using per-Set PHT}
\newacronym{GCC}{GCC}{GNU Compiler Collection}
\newacronym{GEMS}{GEMS}{General Execution-driven Multiprocessor Simulator}
\newacronym[longplural={General Purpose Processors}]{GPP}{GPP}{General Purpose Processor}
\newacronym{HMC}{HMC}{Hybrid Memory Cube}
\newacronym{HIVE}{HIVE}{HMC Instruction Vector Extensions}
\newacronym{IATAC}{IATAC}{Inter-Access Time per Access Count}
\newacronym{ILP}{ILP}{Instruction Level Parallelism}
\newacronym{IPC}{IPC}{Instructions per Cycle}
\newacronym{ISA}{ISA}{Instruction Set Architecture}
\newacronym{IRAM}{IRAM}{Intelligent RAM}
\newacronym{KIPS}{KIPS}{Kilo Instructions per Second}
\newacronym{LDS}{LDS}{Linked Data Structure}
\newacronym{LFSR}{LFSR}{Linear Feedback Shift Register}
\newacronym{LFMR}{LFMR}{Last-to-First Miss-Ratio}
\newacronym{MLP}{MLP}{Memory-Level Parallelism}

\newacronym{LLC}{LLC}{last-level cache}
\newacronym{LRU}{LRU}{Least Recently Used}
\newacronym{LSU}{LSU}{Load Store Unit}
\newacronym{LTP}{LTP}{Last-Touch Predictor}
\newacronym{LvP}{LvP}{Live-time Predictor}
\newacronym{LWP}{LWP}{Last Write Predictor}
\newacronym{MARSS}{MARSS}{Micro Architectural and System Simulator}
\newacronym{McPAT}{McPAT}{Multi-core Power, Area, and Timing}
\newacronym{MIPS}{MIPS}{Microprocessor without Interlocked Pipeline Stages}
\newacronym{MOB}{MOB}{Memory Order Buffer}
\newacronym{MMU}{MMU}{Memory Management Unit}
\newacronym{MPKI}{MPKI}{Misses per Kilo-Instruction}
\newacronym{MSHR}{MSHR}{Miss-Status Handling Registers}
\newacronym{NAS}{NAS}{Numerical Aerodynamic Simulation}
\newacronym{NDP}{NDP}{Near-Data Processing}
\newacronym{NMP}{NMP}{Near Memory Processor}
\newacronym{NoC}{NoC}{Network-on-Chip}
\newacronym{NPB}{NPB}{NAS Parallel Benchmark}
\newacronym{NUCA}{NUCA}{Non-Uniform Cache Architecture}
\newacronym{NUMA}{NUMA}{Non-Uniform Memory Access}
\newacronym{OoO}{OoO}{Out-of-Order}
\newacronym{OpenMP}{OpenMP}{Open Multi-Processing}
\newacronym{OS}{OS}{Operating System}
\newacronym{PAg}{PAg}{Per-address Adaptive branch prediction using one Global PHT}
\newacronym{PAs}{PAs}{Per-address Adaptive branch prediction using per-Set PHT}
\newacronym{PC}{PC}{Program Counter}
\newacronym[longplural={Pointer-Chasing Engines}]{PCE}{PCE}{Pointer-Chasing Engine}
\newacronym{PCM}{PCM}{Performance Counter Monitor}
\newacronym{PIM}{PIM}{Processing-in-Memory}
\newacronym[longplural={Pattern History Tables}]{PHT}{PHT}{Pattern History Table}
\newacronym{RAPL}{RAPL}{Running Average Power Limit}
\newacronym{RAS}{RAS}{Row Address Strobe}
\newacronym{RAT}{RAT}{Registers Alias Table}
\newacronym{RC}{RC}{Row Cycle}
\newacronym{RCD}{RCD}{RAS to CAS Delay}
\newacronym[longplural={Row-based Database Systems}]{RDBS}{RDBS}{Row-based Database System}
\newacronym{ROB}{ROB}{Reorder Buffer}
\newacronym{RRD}{RRD}{Row to Row activation Delay}
\newacronym{RP}{RP}{Row Precharge}
\newacronym{RTL}{RTL}{Register Transfer Level}
\newacronym{RTP}{RTP}{Read To Precharge}
\newacronym{RVU}{RVU}{Reconfigurable Vector Unit}
\newacronym{SDP}{SDP}{Skewed Dead-Block Predictor}
\newacronym{SESC}{SESC}{Superescalar Simulator}
\newacronym{SSE}{SSE}{Streaming SIMD Extensions}
\newacronym{SFP}{SFP}{Spatial Footprint Predictor}
\newacronym{SiNUCA}{SiNUCA}{Simulator of Non-Uniform Cache Architectures}
\newacronym{SMT}{SMT}{Simultaneous Multi-Threading}
\newacronym{SIMD}{SIMD}{Single Instruction Multiple Data}
\newacronym{SPEC}{SPEC}{Standard Performance Evaluation Corporation}
\newacronym{SoC}{SoC}{System-on-Chip}
\newacronym{SPP}{SPP}{Spatial Pattern Predictor}
\newacronym{SRAM}{SRAM}{Static Random Access Memory}
\newacronym{SSOR}{SSOR}{Symmetric Successive Over-Relaxation}
\newacronym{SSV}{SSV}{Search Set Vector}
\newacronym{TLB}{TLB}{Translation Look-aside Buffer}
\newacronym{TLP}{TLP}{Thread Level Parallelism}
\newacronym[longplural={Through-Silicon Vias}]{TSV}{TSV}{Through-Silicon Via}
\newacronym{VWQ}{VWQ}{Virtual Write Queue}
\newacronym{WTR}{WTR}{Write Recovery time}
\newacronym{WR}{WR}{Write To Read delay time}

\newacronym[longplural={non-volatile memories}]{NVM}{NVM}{non-volatile memory}

%% file: macros.tex
\newcommand{\circled}[1]{\tikz[baseline=(char.base)]{\node[shape=circle,draw,inner sep=0pt,fill=black, text=white] (char) {#1};}}

\definecolor{lightblue}{rgb}{0.980, 0.956, 0.623}

\newcommand{\tempcommand}[1]{\renewcommand{\arraystretch}{#1}}

\definecolor{Gray}{gray}{0.9}

\newcommand{\ltzswapoff}{4.5\xspace}
\newcommand{\ltzswapon}{8.3\xspace}
\newcommand{\ltzswapoffhf}{2.4\xspace}
\newcommand{\ltzswaponhf}{4.4\xspace}

\newcommand{\gfi}[1]{\textcolor{black}{#1}}
\newcommand{\gfii}[1]{\textcolor{black}{#1}} 
\newcommand{\gfiii}[1]{\textcolor{black}{#1}} 
\newcommand{\gfiv}[1]{\textcolor{black}{#1}} 
\newcommand{\gfv}[1]{\textcolor{black}{#1}} 

\definecolor{caribbeangreen}{rgb}{0.0, 0.8, 0.6}
\newcommand{\juan}[1]{\textcolor{black}{#1}} 
\newcommand{\jgl}[1]{\textcolor{blue}{JGL: #1}} 

\newcommand{\sgi}[1]{\textcolor{black}{#1}} 
\newcommand{\sgii}[1]{\textcolor{black}{#1}} 
\newcommand{\sgiii}[1]{\textcolor{black}{#1}} 

\newcommand{\ieeea}[1]{\textcolor{black}{#1}} 
\newcommand{\ieeearev}[1]{\textcolor{black}{#1}} 
\newcommand{\ieeearevi}[1]{\textcolor{black}{#1}} 

\newcommand{\sgfb}[1]{\textcolor{black}{#1}} 

\newcommand{\revdel}[1]{}

\newif\ifrevision
\revisiontrue

\newcommand\revision[1][0]{}
\newcommand\revi[1][0]{}
\ifrevision
\newcommand\revii[1][0]{}
\else
\newcommand{\revii}[1]{\textcolor{black}{#1}}
\fi

\newif\ifrevisionvii
\revisionviitrue

\ifrevisionvii
    \newcommand{\ieeearevii}[1]{\textcolor{black}{#1}} 
    \sethlcolor{white}

\else
    \newcommand{\ieeearevii}[1]{\textcolor{blue}{#1}} 

    \sethlcolor{lightblue}
\fi

\newif\ifcamerareadysubmit
\camerareadysubmittrue

\ifcamerareadysubmit
    \newcommand{\gfcr}[1]{\textcolor{black}{#1}} 
\else
    \newcommand{\gfcr}[1]{\textcolor{blue}{#1}} 
\fi

\definecolor{dollarbill}{rgb}{0.52, 0.73, 0.4}
\newcommand{\prtag}[1]{\lfbox[padding=1pt, border-color=red, background-color=red!40]{\textbf{\small #1}}}

\newcommandx{\unsure}[2][1=]{\todo[linecolor=red,backgroundcolor=red!25,bordercolor=red,#1, size=\tiny]{#2}}
\newcommandx{\change}[2][1=]{\todo[linecolor=blue,backgroundcolor=blue!25,bordercolor=blue,#1,size=\tiny]{#2}}
\newcommandx{\feedback}[2][1=]{\todo[linecolor=yellow,backgroundcolor=yellow!25,bordercolor=yellow,#1]{#2}}
\newcommandx{\improvement}[2][1=]{\todo[linecolor=Plum,backgroundcolor=Plum!25,bordercolor=Plum,#1]{#2}}
\newcommandx{\thiswillnotshow}[2][1=]{\todo[disable,#1]{#2}}
\newcommandx{\completedRevision}[2][1=]{\todo[disable,backgroundcolor=red,#1]{#2}}
\newcommandx{\dataSource}[2][1=]{\todo[disable,backgroundcolor=red,#1]{#2}}
\newcommandx{\info}[2][1=]{\todo[linecolor=dollarbill,backgroundcolor=dollarbill!25,bordercolor=dollarbill,#1, size=\tiny]{#2}}

\newcommand{\boxbegin} {
	\begin{tcolorbox}[enhanced, frame hidden, colback=gray!50, breakable]
}

\newcommand{\boxend} {
	\end{tcolorbox}
}

\newcommand{\yboxbegin} {
	\begin{tcolorbox}[breakable, enhanced, frame hidden,
	enlarge top by=-0.25cm,
   enlarge bottom by=-0.1cm,
	colback=yellow!50]
}

\newcommand{\yboxend} {
	\end{tcolorbox}
}

\newcommand{\ar}{\noindent \textbf{Author response:}}
\newcommand{\aact}{\noindent \ding{122} \textbf{Author action:}}

%% file: sections/1-introduction.tex
\section{Introduction} 
\label{sec_introduction}

\label{r1.1}\Copy{R1.1}{\ieeearev{The number and diversity of consumer devices (e.g., smartphones, tablets, \revdel{laptops,} Chromebooks~\mbox{\cite{chromebook}}, and wearable devices) are growing rapidly~\mbox{\gfii{\cite{emarketer2016slowing,reddi2018two,armqualcomm2014,halpern2016mobile,CanalysN72}}}.} The number of consumer devices has surpassed the number of desktop computers~\cite{reddi2018two}. For example, web-based computers, such as  Chromebooks, account for 58\% of all computer shipments to schools in the United States~\cite{heater2017chromebook}. These devices have different design constraints than traditional computers due to \gfi{their} limited area, power dissipation restrictions, and target market. Therefore, it is essential to guarantee low cost per device while maintaining \gfi{good} performance \gfi{and high-quality user experience}.} 

One critical component of consumer devices is the main memory (typically consisting of DRAM~\gfi{\cite{dennard1968dram,mutlu2014memorybook}}), which is used not only as the working memory space but also as storage for \gfii{least-recently-used \gfii{(i.e., cold)} memory blocks} 
(i.e., as swap space~\gfi{\cite{bovet2005understanding, tanenbaum1997operating,lecturevirtualmemory}}). As consumer devices grow in sophistication, many target applications handle increasing amounts of data and require larger main memory capacity to avoid significant performance issues~\gfii{\cite{halpern2016mobile,badr2020mocktails,boroumand2018google,mohan2017storage,amiraliphd,nelsonsize,lebeck2020end}}. Unfortunately, it is becoming increasingly challenging to increase DRAM capacity inside consumer devices due to
the worsening reliability, cost, and performance issues as manufacturers 
scale DRAM technology \gfi{to higher storage capacity levels}~\gfi{\cite{mutlu2013memory,mutlu2015research,kim2014flipping,mutlu2019rowhammer,kim2020revisiting,mutlu2015main,  kang2014co, hong2010memory, kanev_isca2015, mutlu2017rowhammer, ghose2018your,liu2013experimental,frigo2020trrespass,liu2012raidr,patel2017reach,qureshi2015avatar,mandelman2002challenges, khan2014efficacy,khan2016parbor,khan2017detecting,lee2015adaptive,lee2017design,chang2017understandingphd,chang2017understandingsigmetrics,chang2016understanding,chang2014improving,meza2015revisiting,david2011memory,deng2011memscale,yauglikcci2022understanding,orosa2021deeper,hassan2021uncovering}}. 

As a potential solution to the DRAM scalability challenge, manufacturers have introduced emerging non-volatile memories (NVMs) into the market, which can be used to expand the memory capacity of consumer devices by augmenting or replacing DRAM~\gfi{\cite{lee2009architecting,
qureshi2009scalable,
lee2010phase,
lee2010phasecacm,
kultursay2013evaluating,
zhou2009durable,
wong2010phase, 
meza2012case, 
meza2013case, 
song2020improving,
song2021aging, 
song2019enabling,
atwood2018pcm,
bock2011analyzing,
burr2008overview,
du2013bit,
ferreira2010increasing,
jiang2012fpb,
jiang2013hardware,
kannan2016energy,
qureshi2011pay,
qureshi2010improving,
qureshi2010morphable,
sebastian2017temporal,
wang2015exploit,
yue2013accelerating,
zhou2012writeback,
zhou2013writeback,
yoon2013techniques,
DAC-2009-DhimanAR,
wang2013low,
chen2010advances,
diao2007spin,
hosomi2005novel,
raychowdhury2009design,akinaga2010resistive,wong2012metal,yang2013memristive,kund2005conductive,bondurant1990ferroelectronic}}. However, these NVM-based devices are still slower than DRAM~\gfi{\cite{harris2020ultra,wu2021storage,lee2010phasecacm,lee2010phase,lee2009architecting,wang2019panthera,salkhordeh2019analytical,yoon2012row,meza2012enabling}}. For example, the state-of-the-art Intel Optane SSD (\sgi{solid-state drive})~\cite{h10}, which is a low-latency NVM-based SSD device \gfi{(i.e., a\gfiii{n} SSD device that uses NVM as its primary persistent storage media)}, \sgi{has an access latency that is two orders of magnitude slower than that}
\gfi{of} DRAM~\gfi{\cite{izraelevitz2019basic,psaropoulos2019bridging}} (but \gfi{it} is still \gfi{one order of magnitude} 
faster than traditional NAND-flash-based SSDs~\gfi{\cite{harris2020ultra,lee2019asynchronous,zhang2018performance,chien2018characterizing,yang2020exploring,hady2017platform,wu2019exploiting,imamura2018reducing,wu2021storage}}), while providing a better cost-per-byte (\$1.50 per GB~\cite{optanePrice} vs.\ \$5 per GB for \gfi{DRAM~\cite{dramcost}}). Previous works propose two different ways to integrate an NVM device into state-of-the-art computers \sgi{in order to alleviate DRAM scalability issues}. The first method uses the device as byte-addressable \gfi{main} memory, \gfi{which} directly replaces DRAM. In this method, the system can access the NVM device directly using load and store instructions~\gfi{\cite{peng2019system, metzler2015prototyping, hassan2015energy, chauhan2016nvmove,kultursay2013evaluating, lee2009architecting, lee2010phase,qureshi2009scalable,zhou2009durable,DAC-2009-DhimanAR,yoon2014efficient,yoon2012row,li2017utility,meza2012enabling,zhang2021chameleondb,bae20182b,kim2018optimized,wu2021storage}}. \gfi{T}he second method \gfi{uses the} NVM as a block \gfi{storage} device that \gfi{replaces} a NAND-flash-based SSD~\gfi{\cite{cai2017error,luo2018improving,luo2018heatwatch,cai2017vulnerabilities,luo2016enabling,cai2015read,lecturenand}} or a magnetic hard drive, using the same access protocol and interface \gfi{\cite{harris2020ultra,lee2019asynchronous,zhang2018performance,chien2018characterizing,yang2020exploring,hady2017platform,wu2019exploiting,ke2018lirs,liu2020nvm,imamura2018reducing,han2020splitkv,papagiannis2020optimizing,jia2020flash,bae20182b,wu2019towards}}. 
\sgi{Both methods alleviate DRAM scalability issues by taking advantage of the higher density and lower cost-per-byte that NVMs offer over DRAM.} However, \gfi{\emph{entirely replacing DRAM}} with NVM in consumer devices imposes large system integration and design challenges (e.g., due to the high write access latency and limited endurance of NVM~\gfiii{\cite{lee2009architecting,lee2010phase,lee2010phasecacm}}).
 
\sgi{Integrating} emerging NVM-based SSDs \gfi{in consumer devices} can open up new opportunities for memory management. Traditional desktop and enterprise computers employ a swap space~\gfi{\cite{bovet2005understanding, tanenbaum1997operating,lecturevirtualmemory,lebeck2020end}} to increase the total main memory space available in the system \gfi{beyond the capacity of available DRAM}. In \gfi{such systems}, the system moves cold \gfii{memory blocks}  present in DRAM to a swap space, which usually \gfi{exists in} a high-latency and high-capacity storage device (e.g., \gfi{NAND-flash-based SSD,} magnetic hard drive). However, consumer devices \gfi{usually} do \gfi{\emph{not}} employ a swap space~\gfii{\cite{zhong2014building, kim2015cause, liu2017non, zhong2017building, kim2019analysis, zhu2017smartswap, kim2018comparison,zhong2014dr,kim2017application,kim2019ezswap,kim2020maintaining,guo2015mars,liang2020acclaim}}, since accessing \gfi{a} storage device directly impacts system performance and user experience due to the device's high access latenc\gfi{y}. 
Since \gfi{emerging} NVM-based SSDs have \sgi{an order of magnitude lower access latency} than \gfi{the commonly-used fast storage devices, i.e., } NAND-flash-based SSDs, they have the potential to enable the use of a swap space in consumer devices. Recent works~\gfi{\mbox{\cite{zhong2014building, kim2015cause, liu2017non, zhong2017building, kim2019analysis, zhu2017smartswap, kim2018comparison,zhong2014dr}}} propose extending the total main memory space available to applications by using NVM as swap space for DRAM in mobile systems. However, no prior work analyzes the implications of enabling a \emph{real} NVM-based swap space in \emph{real} consumer devices.

In this work, we provide the \emph{first} analysis of the impact of extending the main memory space of consumer devices using off-the-shelf NVMs. We \emph{extensively} examine system performance and energy consumption when the NVM device is used as swap space for DRAM main memory to effectively extend the main memory capacity. Our empirical analyses lead us to several observations and insights that can be useful for the design of future systems and NVMs.

For our experimental evaluation, we equip real web-based Chromebook computers~\gfi{\cite{chromebook}} with the Intel Optane SSD~\gfi{\cite{h10}}. Our target workloads are interactive applications, \sgi{with a focus on} the Google Chrome~\mbox{\cite{chrome}} web browser. We choose such workloads for two reasons. First, in interactive applications, the system needs to respond to user inputs at a target output latency to provide a satisfactory user experience. Second, in Chromebooks, the Chrome browser serves as the main interface to execute services for the user. We compare the performance and energy consumption of interactive workloads running on our Chromebook with NVM-based swap space, where the Intel Optane SSD capacity is used as swap space to extend main memory capacity, against two state-of-the-art systems: (i) a baseline system with double the amount of DRAM than the system with the NVM-based swap space, which resembles current consumer devices but has high manufacturing cost due to the \gfi{large DRAM capacity and} relatively high cost-per-bit of DRAM; and (ii) a system where the Intel Optane SSD is naively replaced with a state-of-the-art (yet slower) off-the-shelf NAND-flash-based SSD, which we use as a swap space of equivalent size as the NVM-based swap space. The NAND-flash-based SSD provides a cheap alternative to extend the main memory space, but it can penalize system performance due to its high access latency. We use a memory \gfii{capacity} pressure test~\cite{memorypressure} to measure the impact of the new \gfi{NVM} swap space \gfi{on user tasks that consist of} loading, scrolling, and switching between Chrome \gfi{browser} tabs. We measure how the NVM device increases the 99th-percentile latency (i.e., tail latency) \gfi{of} each \gfi{task} and the total number of \gfi{Chrome} tabs \gfi{that the user} can open \gfi{without} discarding old tabs. We divide our \gfi{system evaluation, analysis, and optimization} into two \gfi{major parts}: (1) \emph{Evaluating NVM for Consumer Devices}, and (2) \emph{System Optimization}. 

\textbf{Evaluating NVM for Consumer Devices.} In the first \gfi{part of our work}, we compare the baseline system \gfii{with} \gfi{the system where we extend the main memory space with the Intel Optane SSD \gfii{and} the system where we extend the main memory space with the NAND-flash-based SSD. \label{r1.2}\Copy{R1.2}{We make four major observations. \gfi{First, w}e observe that extending the main memory space with the Intel Optane SSD improves the average performance of interactive workloads (measured as the latency of switching across Chrome browser tabs) compared to the baseline system with twice the amount of DRAM. The NVM-based swap space enables the system to leverage a larger aggregate main memory space than the baseline system while also reducing system cost. However, extending the main memory space with the Intel Optane SSD increases the number of violations of the application's target output latency by 2.6$\times$ (on average) once the memory traffic between DRAM and the Intel Optane SSD exceeds a threshold, which happens under high system load (e.g., a large number of opened Chrome browser tabs). \ieeearev{Since the Intel Optane SSD is integrated to the system via the high-latency off-chip bus, constantly moving data between DRAM and the Intel Optane SSD directly impacts browser performance.}}} \gfi{Second, w}e observe that accessing the Intel Optane SSD through the power-hungry off-chip bus \emph{significantly} increases energy consumption. We mitigate this issue by allocating some DRAM space to be used as a compressed in-DRAM cache for data swapped between DRAM and the Intel Optane SSD, which reduces the number of accesses to the Intel Optane SSD by \gfii{up to} 2.11$\times$ and, as a result, improves energy efficiency. \gfi{Third,} extending the main memory space even with the slow NAND-flash-based SSD provides performance benefits compared to the baseline system. However, due to the high access latency of the NAND-flash-based SSD, the number of violations of the application's target output latency increases compared to the baseline system and the system with the Intel Optane SSD's NVM-based swap space. \gfi{Fourth, we \sgi{observe} that the Linux block I/O layer becomes a major source of performance overhead when \sgi{the} main memory space is extended using NVM, primarily due to (i)~I/O scheduling bottlenecks created by the mismatch between our workloads' I/O access patterns and the default I/O scheduling policy; and (ii)~overheads related to the asynchronous operation of the I/O request completion mechanism. We mitigate some of these overheads by proposing two system optimizations that can better leverage the characteristics of our workloads and the NVM.}

\textbf{System Optimization.} \gfi{In the second part of our work, we mitigate some of the system-level overheads we identify in the first part of our work by proposing two system optimizations that can better leverage the characteristics of our workloads and the NVM.} First, we employ different I/O schedulers that better match our workloads' I/O access patterns, which improves performance. Second, we change the default asynchronous I/O request completion model to a hybrid I/O request completion model that adaptively switches from synchronous to asynchronous operation. The baseline asynchronous I/O request completion model entails non-trivial overheads (e.g., latency overheads of interrupt and context switch). The hybrid I/O request completion model partially avoids these overheads by allowing the process to synchronously wait for the completion of I/O requests for a determined period of time. As a result, in the best case, the process \sgi{needs to wait for only the low access latency of NVM, instead of incurring the large latency} overheads of the asynchronous I/O request completion model.

We make the following key contributions in this work:
\begin{itemize}[itemsep=0pt, topsep=0pt, leftmargin=*]
    \item \sgi{We perform the first experimental analysis of the impact of using off-the-shelf non-volatile memory (NVM) for swap space in a real consumer device. Our studies highlight how a \gfiii{state-of-the-art NVM-based} Intel Optane SSD can be used effectively to extend the total main memory space available to interactive applications such as the Google Chrome web browser.}
    
    \item \sgi{We demonstrate that using a state-of-the-art off-the-shelf NVM-based SSD as swap space can improve the performance of interactive applications compared to increasing DRAM capacity, but that naively integrating the NVM-based SSD leads to system-level overheads. These overheads primarily arise from the Linux block I/O layer and the off-chip traffic to the SSD.}
    
    \item \sgi{We identify two system optimizations that can mitigate some of the system-level overheads that occur when using an NVM-based SSD as swap space. Both of these optimizations adapt system-level mechanisms to application and runtime behavior, and improve the overall performance of the system.}
\end{itemize}

%% file: sections/2-background.tex
\section{Background and Motivation}
\label{sec_background}

\gfi{We} provide the background and motivation required to understand the main components of our experimental setup \gfi{and evaluated workloads}. First, we discuss how the memory system impacts the performance of the Google Chrome web browser~\cite{chrome} (Section~\mbox{\ref{sec_background_chrome_profile}}).  Second, we investigate how users interact with consumer devices and how this interaction contributes to memory \sgii{capacity} pressure (Section~\mbox{\ref{sec_background_memory_pressure}}). Third, we explain the main characteristics of the Intel Optane SSD device~\cite{intel2018900p} (Section~\ref{sec_background_03}). 

\subsection{Google Chrome Web Browser in Consumer Devices}
\label{sec_background_chrome_profile}

The web browser is one of the main applications of consumer devices. Due to its significance, several web browser applications are present in many mobile benchmark \gfi{suites}~\gfi{\cite{gutierrez2011full,huang2014moby,pandiyan2013performance, boroumand2018google}}. \gfi{The} Google Chrome \gfi{web browser}~\cite{chrome}, which has over a billion active users and the largest share of the mobile browsing market~\gfii{\cite{popper2017google,googleshare}}, is one of the most relevant web browsers available. Therefore, we investigate Google Chrome performance in this work. 

Chrome performance can be defined based on three key metrics: (i)~the time it takes to load a \gfii{web} page; (ii)~the smoothness of scrolling a \gfii{web} page \gfii{(i.e., whether or not the user can perceive discontinuous movements or jumps when moving up/down inside a web page, measured in \sgii{frames per second})}\gfi{;} and (iii)~how quickly \gfi{the browser} can switch between \gfii{web} pages \gfi{in different browser tabs} (i.e., its \emph{tab switch latency}). Loading a new \gfi{web} page and scrolling through a \gfii{web} page are highly compute-intensive operations~\cite{boroumand2018google}, since the main \sgi{operations that} Chrome performs during their execution are rendering~\gfi{\cite{blink}} and rasterization~\gfi{\cite{skia}}, respectively. In this work, we are primarily interested in evaluating the impact of the swap space on Chrome performance. Therefore, the most relevant metric for our analysis is the tab switch \sgi{latency,} since switching between a recently\gfi{-}opened \gfi{web} page and a \gfi{previously-opened} \gfi{web} page \gfi{(i.e., a \gfi{web} page that \gfii{is open in an inactive tab})} will likely result in a page fault when the system runs out of memory. 

In Chrome, each \gfi{tab} represents a single process \gfi{associated with the web page displayed in the
tab}, which \gfiii{improves} reliability and security~\gfii{\cite{reis2009isolating,barth2008security}}. When the user switches between \gfi{tabs}, the browser executes two main tasks. First, it executes a context-switch between the \sgi{currently-opened} \gfi{tab} and the requested \gfi{tab}. Second, it executes a load operation of the requested \gfii{web page}, which involves loading data frames related to the requested \gfii{web page} from memory and rendering the requested \gfi{web page}. The latency from the time a user clicks a \gfii{web page} to the time the \gfi{web} page is rendered on the screen is crucial since it impacts user satisfaction. This time is mostly dominated by how fast the system can load the data frames related to the requested \gfii{web page} from memory or disk. However, storing a large number of \sgii{web pages} has become a challenge for consumer devices for two main reasons. First, the total size of a single \gfi{web} page has been growing in recent years due to the increased use of images, JavaScript, and video in web pages~\cite{httparchieve}. Second, users tend to \gfi{keep} many \gfi{web pages} \gfi{open} \gfi{concurrently} during web browsing, \gfi{leading to many open tabs.} \sgi{This} \gfi{results in} a demand for larger memory space \gfi{required to keep the \gfii{web page} in physical memory} in \gfi{modern} systems. 

\begin{figure}[ht]
  \centering
  \includegraphics[width=0.95\linewidth]{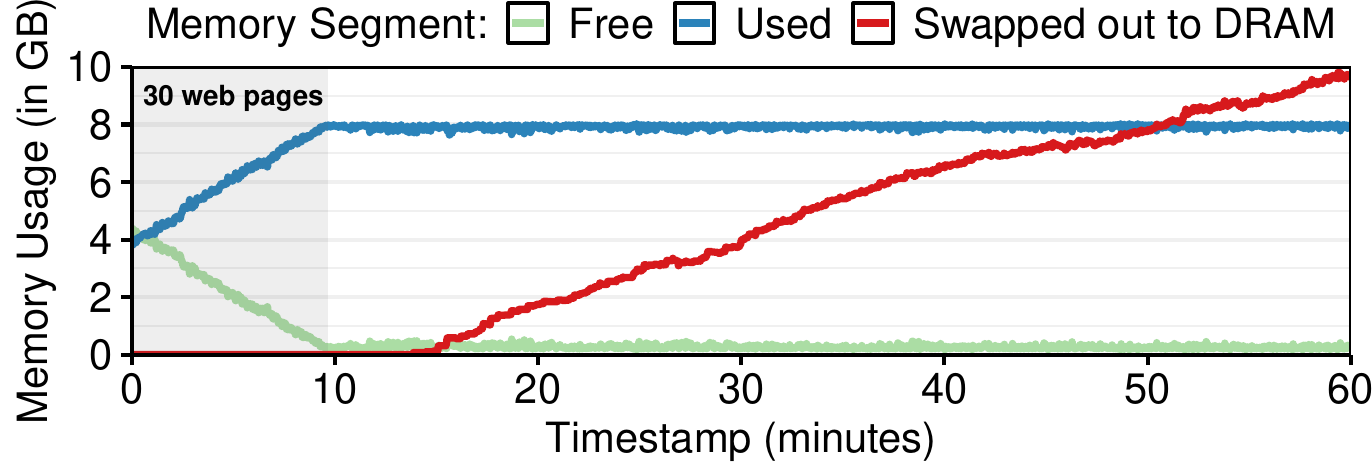}
  \caption{System memory \gfi{usage} while loading \gfi{30} Chrome \gfii{web} pages.}
  \label{fig_memory_consumption}
  \vspace{-15pt}
\end{figure}

To understand the impact of the number of open \gfi{web} pages and memory consumption, we evaluate how many \gfii{web} pages it is possible to open in a system with \SI{8}{\giga\byte} of DRAM before \gfi{the system} runs out of \gfi{physical} memory. Figure~\ref{fig_memory_consumption} shows the memory profile of a test that continually opens new Chrome \gfii{web} pages during one hour of execution. We observe that the system runs out of \gfi{physical} memory \gfi{(i.e., \sgi{free} memory)} by opening only 30~\gfii{web} pages \gfi{(within the \sgi{course} of almost 10~minutes)}. When this happens, the system enters a \emph{memory \gfii{capacity} pressure} state, leading to increased swap activity \gfi{(as shown \sgi{by} the increasing \gfii{red} line in Figure~\ref{fig_memory_consumption})} and, consequently, performance degradation \gfi{(not shown in Figure~\ref{fig_memory_consumption})}. \gfi{We conclude that there is a clear need to provide more memory capacity to support more \sgi{concurrently-open \gfii{web} pages and, hence,} better user experience in consumer (i.e., mobile) devices.}

A traditional way of expanding the memory space in modern systems is to enable page swapping~\gfi{\cite{bovet2005understanding, tanenbaum1997operating,lecturevirtualmemory}}. However, mobile devices usually disable page swapping due to \sgi{the} \gfi{large} performance \sgi{penalty and user experience degradation imposed on the system by high-latency storage devices}~\gfi{\cite{zhong2014building, kim2015cause, liu2017non, zhong2017building, kim2019analysis, zhu2017smartswap, kim2018comparison,zhong2014dr,kim2017application,kim2019ezswap,kim2020maintaining,guo2015mars,liang2020acclaim}}. Instead, a kernel module contiguously verifies the memory space and terminates processes to make room for incoming memory requests. This approach is called ``\emph{low memory killer}''~\gfi{\cite{rientjes2010oom, collins2011android,lebeck2020end}}. Recently, Google enabled a swap alternative for its mobile devices (i.e., Google Pixel smartphones~\cite{pixel} and Google Chromebooks~\cite{chromebook}). In these devices, the \sgi{operating system} (OS) enables an in-DRAM compressed swap space, called \sgi{\emph{ZRAM}}~\cite{jennings2013transparent}. When enabling ZRAM in the system, the OS reserves a \gfi{fraction} of \gfi{the} DRAM space to be used as a swap device. Pages are compressed before being \gfi{moved from the working region in main memory to ZRAM (i.e., \emph{swapped \ieeea{out}})}, and decompressed before being \gfi{moved from ZRAM to the working region in main memory (i.e., \emph{swapped \ieeea{in}})}. \gfi{By using compression in ZRAM}, the system can increase the capacity of the swap space by the compression ratio (\ieeea{e.g.,} by 3:1~\gfi{\cite{shiu2015system,shiu2017driving,pekhimenko2013linearly}}).

\subsection{The Impact of Main Memory Capacity Pressure on Consumer Devices}
 \label{sec_background_memory_pressure}

\Copy{R2/1A}{We design an experiment to understand the impact of \gfi{main} memory \gfi{capacity} pressure on consumer devices. Our experiment aims to characterize: (1)~how users utilize a web browser; (2)~how often users suffer from high response latencies from interactive workloads; and (3)~how often users push the system into a state of memory \gfi{capacity} pressure. For this purpose, we distributed Chromebook devices}\footnote{\Copy{R2/1B}{\ieeearev{The Chromebook devices we use for our experiment consist of off-the-shelf Chromebook devices with \mbox{\SI{8}{\giga\byte}} DRAM capacity, }\sgfb{of which} \mbox{\SI{4}{\giga\byte}} are reserved to enable an in-DRAM compressed swap space.}}\label{ft:r2.1}\Copy{R2/1C}{ to 114 different users \gfi{at Google}, whom we asked to perform their daily activities using the Chromebook devices.} We picked the users randomly from a major division at the company that employs thousands of people. We monitored their activity by periodically collecting the following system information over a period of three months:
\label{r2.2}\Copy{R2/2}{
\begin{itemize}
    \item \emph{Number of Chrome tabs opened:} We recorded  the number of \gfii{open} tabs across all Chrome windows. Data samples were reported every 5 minutes. In total, we collected 19,487 data samples. 
    \item \emph{Tab switch latency:} We collected the tab switch latencies for each  tab switch the user performed during their activity. Data samples were reported at each individual tab switch. In total, we collected 62,243 data samples.
    \item \emph{Memory \gfi{capacity} pressure level:} 
    \sgi{We periodically (i.e., every five seconds) sampled the current state of the memory to determine which of the following three memory \gfii{capacity} pressure levels the memory was currently experiencing:}
    \emph{no memory \gfii{capacity}  pressure}, \emph{moderate memory \gfiii{capacity} pressure}, and \emph{critical memory \gfiii{capacity} pressure}. The memory \gfii{capacity} pressure level is defined as follows. First, Chrome calculates the amount of $fill$ memory, as \sgfb{$fill =1 -  \frac{(mem\_free + \frac{swap\_free}{RAM\_vs\_swap\_weight})}{(mem\_total + \frac{swap\_total}{RAM\_vs\_swap\_weight})}$;} \ieeearev{ where \textit{mem\_free} is the amount of main memory space currently free, \textit{mem\_total} is the total amount of main memory space (free and occupied), \textit{swap\_free} is the total amount of memory space in the swap \sgfb{device that is currently} free, \textit{swap\_total} is the total amount of swap space, and $RAM\_vs\_swap\_weight$ accounts for the relative ease (considering memory access latency) of allocating RAM directly versus having to swap its contents out first. This parameter has a default value of ``4'', which the operating system empirically has defined.} If \sgi{\emph{fill}} ranges between 60\% and 95\%, the system is under moderate memory \gfii{capacity}  pressure. If \gfi{\emph{fill}} is greater than or equal to 95\%, it is under critical memory \gfii{capacity} pressure~\cite{MemoryCo67}. In total, we collected 1,571,701 data samples.
\end{itemize}
}

\begin{figure}[ht]
\vspace{-5pt}
\begin{subfigure}{\linewidth}
  \centering
  \includegraphics[width=0.95\linewidth]{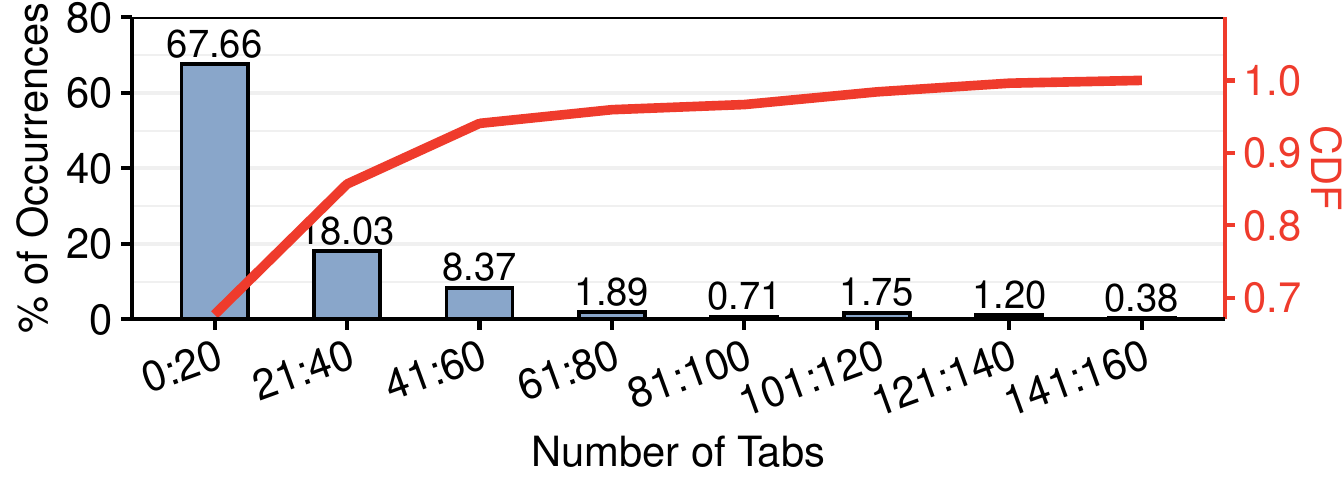}  
  \vspace{-6pt}
  \caption{Distribution of the number of open tabs.}
  \label{fig_step_1_tabs_per_user_and_tab_switch_latencya}
\end{subfigure}
\par\bigskip 
\begin{subfigure}{\linewidth}
  \centering
  \includegraphics[width=0.95\linewidth]{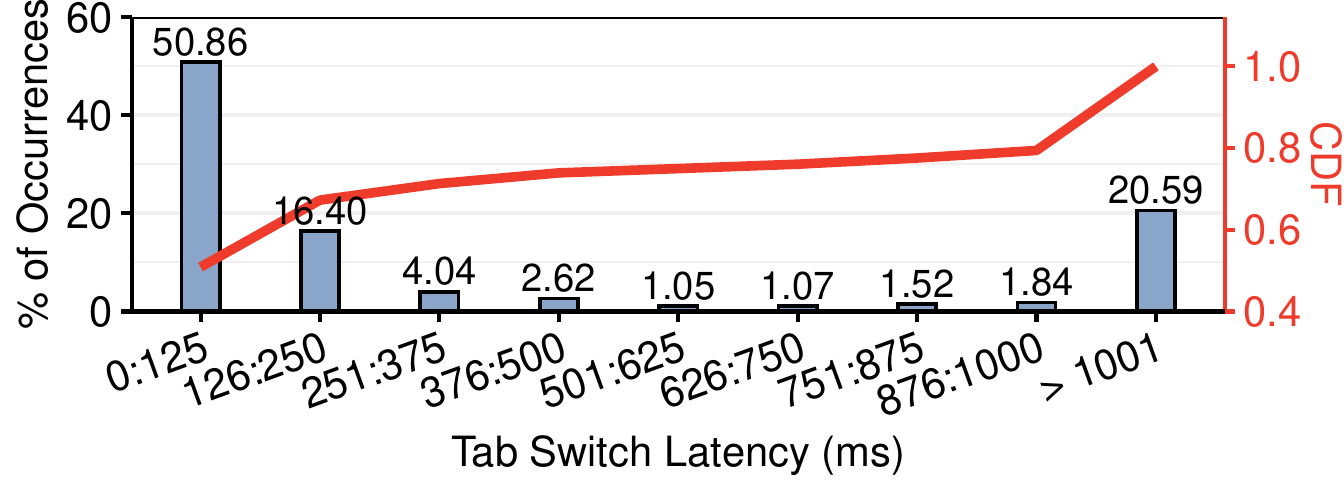}  
  \vspace{-5pt}
  \caption{Distribution of the number of tabs switch latencies.}
  \label{fig_step_1_tabs_per_user_and_tab_switch_latencyb}
\end{subfigure}
 \caption{Distribution of the number of \gfi{open} tabs and tab switch latencies.}
 \vspace{-10pt}

\end{figure}

Figure~\ref{fig_step_1_tabs_per_user_and_tab_switch_latencya} shows how many Chrome tabs users opened during our experiments. We observe that users had up to 20 Chrome tabs open in \gfcr{68}\% of the samples; 21 to 40 Chrome tabs open in 18\% of the samples; 41 to 80 Chrome tabs open in 10\% of the samples; and 81 to 160 Chrome tabs open in 4\% of the samples (no user had more than 160 tabs open at any time). Even though \gfcr{4}\% is a relatively low number of occurrences, it represents 710 sample points where the users kept a large number of Chrome tabs open.


Figure~\ref{fig_step_1_tabs_per_user_and_tab_switch_latencyb} shows the \gfi{distribution of the} tab switch latenc\gfi{y} the users experienced. We observe that users experienced a tolerable latency (i.e., a tab switch latency less than \SI{250}{\milli\second}\footnote{\gfv{We use \SI{250}{\milli\second} for the tolerable tab switch latency, since a rule of thumb in the web performance community is to provide visual feedback in under \SI{250}{\milli\second} to keep the user engaged~\cite{grigorikPerf,lohrImpatient}}.}) in \gfi{67.3}\% of our samples. However, the users experienced unacceptable latency from the system (i.e., a tab switch latency greater than or equal to \SI{250}{\milli\second}) in \gfi{32.7}\% of the samples. The tab switch latency was larger than 1 second for \gfi{20.59}\% of the samples. We \gfi{periodically} collected the \gfi{memory capacity pressure level of the system} to verify that the high tab switch latency was due to high memory \gfiii{capacity} pressure. Our experiment shows that 35.8\% \gfi{of our samples}, the system experiences moderate to critical memory \gfii{capacity} pressure (563,143 data samples in moderate memory \gfii{capacity} pressure and 42 data samples in critical memory \gfiii{capacity} pressure  from a total 1,571,701 data samples).

With this experiment, we conclude that real users, for a considerable fraction of \gfi{their usage time of the Chrome web browser}, often push the system to points that induce moderate to critical memory \gfii{capacity} pressure, leading to large \gfi{and often unacceptable} response \gfi{times} for interactive workloads.

\subsection{Intel Optane SSD}
\label{sec_background_03}

In the past \gfiii{several} decades, \gfiii{various} works~\gfi{\cite{lee2009architecting,
qureshi2009scalable,
lee2010phase,
lee2010phasecacm,
kultursay2013evaluating,
zhou2009durable,
wong2010phase, 
meza2012case, 
meza2013case, 
song2020improving,
song2021aging, 
song2019enabling,
atwood2018pcm,
bock2011analyzing,
burr2008overview,
du2013bit,
ferreira2010increasing,
jiang2012fpb,
jiang2013hardware,
kannan2016energy,
qureshi2011pay,
qureshi2010improving,
qureshi2010morphable,
sebastian2017temporal,
wang2015exploit,
yue2013accelerating,
zhou2012writeback,
zhou2013writeback,
yoon2013techniques,
DAC-2009-DhimanAR,
wang2013low,
chen2010advances,
diao2007spin,
hosomi2005novel,
raychowdhury2009design,akinaga2010resistive,wong2012metal,yang2013memristive,kund2005conductive,bondurant1990ferroelectronic}} have investigated how to employ novel \gfiii{data} \gfi{storage} technologies (e.g., \gfi{phase-change memory, PCM~\cite{wong2010phase, lee2010phase,lee2009architecting,qureshi2009scalable,lee2010phasecacm,zhou2009durable,atwood2018pcm,bock2011analyzing,burr2008overview,du2013bit,ferreira2010increasing,jiang2012fpb,jiang2013hardware,kannan2016energy,qureshi2011pay,qureshi2010improving,qureshi2010morphable,sebastian2017temporal,wang2015exploit,yue2013accelerating,zhou2012writeback,zhou2013writeback,song2020improving,song2019enabling,yoon2013techniques,DAC-2009-DhimanAR,meza2012case,song2021aging}}; \gfi{spin-transfer torque magnetic RAM, STT-MRAM}~\gfi{\cite{wang2013low, kultursay2013evaluating,chen2010advances,diao2007spin,hosomi2005novel,raychowdhury2009design,meza2012case}}; \gfi{metal-oxide resistive RAM, ReRAM}~\cite{akinaga2010resistive,wong2012metal,yang2013memristive,Chi2016,song2018graphr,song2017pipelayer,yao2017face,hu2016dot}; \gfi{conductive bridging RAM, CBRAM~\cite{kund2005conductive,gopalan2011demonstration,jana2015conductive,cha2020conductive}; ferroelectric RAM, FeRAM~\cite{bondurant1990ferroelectronic,scott1989ferroelectric,scott2007applications,mikolajick2001feram}}) to build fast non-volatile memories. Intel and Micron recently announced the first \gfiii{widely-available} commercial NVM device based on the 3D \sgi{XPoint} non-volatile \gfi{memory} technology~\cite{webb20163d}, called Intel Optane~\cite{intel2018900p}. Intel \gfiii{provides} two different memory devices based on Optane: (1)~the Intel Optane SSD~\cite{intel2018900p}, and (2)~the Intel Optane DC Persistent DIMM~\cite{cutress2016intel}. The key difference between these two devices is their system interface. For the Intel Optane SSD, the device has a system interface similar to current NAND-based flash \gfi{memory} devices, where the system communicates to the device via the PCIe bus~\gfi{\cite{specification2002pci}}. This configuration provides one order of magnitude \gfi{lower} latency than traditional NAND\gfi{-flash-based} SSD\gfi{s}~\gfi{\cite{harris2020ultra,lee2019asynchronous,zhang2018performance,chien2018characterizing,yang2020exploring,hady2017platform,wu2019exploiting,imamura2018reducing,wu2021storage}}. For the Intel Optane DC Persistent DIMM, the device is integrated into the system with a DIMM-based interface, similar to DRAM devices. The system directly accesses the device using load/store requests at the byte granularity~\gfi{\cite{izraelevitz2019basic,patil2019performance,gill2019single,wu2020lessons,weiland2019early,shanbhag2020large,mironov2019performance,peng2019system,yang2020empirical,benson_perma_2022,xiang2022characterizing}}. This configuration provides a much \gfi{lower} access latency, on the order of \gfii{hundreds of} nanoseconds \gfii{(around \SI{169}{\nano\second} for sequential reads~\cite{izraelevitz2019basic,yang2020empirical,benson_perma_2022,xiang2022characterizing})}\gfi{, but comes at a high cost, 5$\times$ the cost of \gfii{the Intel Optane SSD} in dollars-per-bit~\cite{optanePrice,IntelOpt84}}. 

Even though the Intel Optane DC Persistent DIMM can provide significant benefits for future systems due to its performance characteristics, there are several challenges to solve before leveraging such devices in future systems, \gfi{including: (1) the need for system mechanisms for} proper data placement between DRAM and the Optane \gfiii{DIMM} device~\gfi{\cite{bittman2020twizzler,wang2019panthera,salkhordeh2019analytical,li2017utility,yoon2012row,meza2012enabling}}, \gfi{(2) difficulties in} fabricating \gfi{printed circuit boards (PCBs)} for mobile platforms that can accommodate the Optane DIMM\gfi{, and (3) accommodating the high cost-per-bit of the Intel Optane DIMM in cost-sensitive mobile systems}. \gfii{Even though prior} work\gfii{s}~\gfi{\cite{zhong2014building, kim2015cause, liu2017non, zhong2017building, zhu2017smartswap, kim2018comparison}} propose\gfii{, using simulation models,} to integrate byte-addressable NVM devices as swap space for mobile systems\gfii{, we choose the Intel Optane SSD for our studies for two main reasons.} \gfi{First, \gfii{due to the} manufacturing difficulties, including high manufacturing cost, and open \gfii{system-level} challenges \gfii{that} need to be solved \gfii{before} integrating the Optane DIMM as swap space in consumer devices. Second, \gfii{since} the goal of \gfi{our} work is to evaluate the performance implications of emerging NVM devices in \emph{real} consumer devices.}

\section{Experimental Setup and Methodology}
\label{sec_experimental_setup}
In this work, we characterize the performance of interactive workloads running on consumer devices. Our target device is the Google Chromebook web-based computer.  We use the Asus Chromebox \gfi{3}~\cite{chromebox} for our experiments, \sgi{as} it is not physically possible to integrate the Intel Optane SSD module in \gfi{regular} Chromebook due to its limited PCIe lanes. The Asus Chromebox runs the same operating system as the Chromebook device (ChromeOS~\cite{wright2009ready}), and has a similar hardware configuration. The device is equipped with a \sgi{7th-generation} Intel Core i3-7100U processor~\gfi{\cite{i3}}, \SI{8}{\giga\byte} DDR4 memory~\gfi{\cite{skhynixddr4}}, and a \SI{32}{\giga\byte} \sgi{NAND-flash-based} SSD~\gfi{\cite{transcendssd}}. \gfi{ChromeOS uses up to 50\% of the DRAM capacity (i.e., \SI{4}{\giga\byte}) to enable \sgiii{an in-DRAM compressed swap space called ZRAM, capable of holding up to \SI{12}{\giga\byte} of compressed data (assuming a 3:1 compression ratio~\cite{shiu2015system,shiu2017driving,pekhimenko2013linearly})}}.\footnote{\ieeea{We use a 3:1 ZRAM compression rate as an empirically-evaluated upper bound observed by prior works~\cite{shiu2015system,shiu2017driving,pekhimenko2013linearly}. In practice, ZRAM compression rate varies with the pages getting swapped out. We observe an average ZRAM compression rate of 1.14:1, with a maximum of 3:1, and a minimum 0.001:1 from our analysis.}}
\Copy{R1/3A}{We modify the system \sgi{by} \gfi{\gfiii{(1)~}removing the in-DRAM compressed swap space} and \gfi{including} an Intel Optane SSD module, \gfi{which the system uses} as \gfi{the} swap device \gfi{for DRAM}\gfiii{; and (2)~}reducing the DRAM size to \SI{4}{\giga\byte}, \sgi{to hold the non-swap-space DRAM capacity constant}. We use the Intel Optane H10~\mbox{\cite{h10}} module for our experiments. It  contains a \SI{16}{\giga\byte} Intel Optane SSD device and a \SI{256}{\giga\byte} Intel QLC \sgi{3D-NAND-flash-based} SSD. We modify the Intel Optane H10 firmware to avoid using the \sgi{NAND-flash-based} SSD during our experiments.}\footnote{\label{r1.3}\Copy{R1/3B}{\ieeearev{We selected the Optane H10 module for our experiments because it was the only Optane device in stock at the time \ieeearevi{ we performed the studies}\ieeearev{. Based on the technical specifications~\mbox{\cite{h10,m10}}, 
the H10 module combines the Intel Optane M10 module~\mbox{\cite{m10}} (i.e., an M.2 module containing only the Intel Optane SSD) with a QLC 3D-NAND-flash-based SSD. In our initial tests, we did \emph{not} observe any performance impact on the raw performance of the H10 module with our modified firmware. Our modified firmware only disables the  QLC 3D-NAND-flash-based SSD in the H10 module.}}}\label{ft:R1/3B}}

\gfi{We compare the performance and energy consumption of interactive workloads running on our Chromebook using three system configurations, as Table~\ref{table_parameters} describes:}
\begin{itemize}[itemsep=0pt, topsep=0pt, leftmargin=*]
    \item \gfi{\textit{Baseline:} a baseline system with \juan{8 GB of DRAM.} \gfiv{\SI{4}{\giga\byte} \juan{are used as main memory, which is uncompressed, and the other} \SI{4}{\giga\byte} 
    \juan{are} used as an in-DRAM compressed swap space (ZRAM), which can house up to \SI{12}{\giga\byte} of actual data, assuming a 3:1 compression ratio~\cite{shiu2015system,shiu2017driving,pekhimenko2013linearly};}} 
    
    \item \gfi{\textit{Optane:} a system with \SI{4}{\giga\byte} of main memory, and \SI{16}{\giga\byte} of \sgi{Intel Optane SSD} swap space;}
    
    \item \gfi{\textit{NANDFlash:} a system with \SI{4}{\giga\byte} of main memory, and \SI{16}{\giga\byte} of \sgi{NAND-flash-based SSD} swap space.}
\end{itemize} 

\begin{table*}[h]
\caption{\gfi{Evaluated system configurations.}}
\label{table_parameters}
\tempcommand{1.2}
\centering
\resizebox{0.95\linewidth}{!}{
\begin{tabular}{|c||c|l|c|c|}
\hline
\rowcolor{Gray}
\multicolumn{5}{|c|}{\textbf{Swap Space Configurations}}                    \\ \hline \hline
\textbf{Configuration}         & \textbf{DRAM Capacity}         & \multicolumn{1}{c|}{\textbf{Swap Space Device}} & \textbf{Swap Space Size} & \textbf{Effective Memory Capacity} \\ \hline
\textbf{Baseline}  & \SI{8}{\giga\byte} & \multicolumn{1}{c|}{In-DRAM Compressed Swap Space (ZRAM~\cite{jennings2013transparent})}      & \SI{12}{\giga\byte} & \SI{16}{\giga\byte}    \\ \hline
\textbf{Optane}    & \SI{4}{\giga\byte} & \multicolumn{1}{c|}{Intel Optane SSD (H10 Module)~\cite{h10}}        & \SI{16}{\giga\byte}  & \SI{20}{\giga\byte}   \\ \hline
\textbf{NANDFlash} & \SI{4}{\giga\byte} & \multicolumn{1}{c|}{NVMe NAND-flash-based SSD}                                     & \SI{16}{\giga\byte}  & \SI{20}{\giga\byte}   \\ \hline

\hline \hline
\rowcolor{Gray}
\multicolumn{5}{|c|}{\textbf{Common System Parameters}}              \\ \hline \hline
\multicolumn{2}{|c|}{}                                          & \multicolumn{3}{l|}{Asus Chromebox~\cite{chromebox}; 7th-generation Intel Core i3-7100U processor~\cite{i3};} \\
\multicolumn{2}{|c|}{\multirow{-2}{*}{\textbf{Hardware Setup}}} & \multicolumn{3}{l|}{DDR4 main memory~\cite{skhynixddr4}; \SI{32}{\giga\byte} NAND-flash-based SSD~\cite{skhynixddr4} for storage}               \\ \hline
\multicolumn{2}{|c|}{}    & \multicolumn{3}{l|}{Operating System:  ChromeOS~\cite{wright2009ready}; kernel version 4.14} \\
\multicolumn{2}{|c|}{\multirow{-2}{*}{\textbf{Software Setup}}} & \multicolumn{3}{l|}{Test Automation Tool: Chromium Project's memory capacity pressure test~\cite{memorypressure}} \\ \hline
\end{tabular}
}
\end{table*}

One of the main obstacles we faced during our analysis was creating the correct experimental setup. This is challenging for three main reasons, mostly related to the lack of a standard benchmark suite for consumer devices~\cite{badr2020mocktails} and the lack of automation tools for real-world experiments \gfi{on mobile devices like Chromebooks:}

\emph{Challenge 1: Executing real-world workloads.} Popular interactive workloads for mobile devices are proprietary \gfi{(e.g., \sgi{social networks such as Facebook~\cite{facebook} and Instagram~\cite{instagram}, messengers such as WhatsApp~\cite{whatsapp} and Telegram~\cite{telegram}, document readers such as Adobe Acrobat Reader~\cite{adobe}, games such as Minecraft~\cite{minecraft}})}, and their source code is not \gfi{openly} available. This limits the scope of our analysis, since we can only analyze \sgi{such} applications as a black box, often making it unclear which specific system resources an application uses and why. Prior works~\gfi{\cite{gutierrez2011full,huang2014moby,pandiyan2013performance,guthaus2001mibench,lee1997mediabench}} put effort into creating benchmark suites for mobile applications. However, they are outdated and often include only a small number of  kernels from \gfi{a small set of} application\gfi{s}.

\emph{Challenge 2: \sgii{Automating execution and enabling reproducibility}.}  There is a lack of tools for automating the execution of mobile workloads~\cite{badr2020mocktails}. This is critical when evaluating an entire system, since experiments need to be executed multiple times to reduce system-level noise (e.g., due to OS tasks, uncontrollable network response times). Without automation, it is difficult to launch \gfi{and execute} applications in an automatic \gfi{and easily reproducible} manner, deal with network traffic, and mimic user interactions with the system, for example.

\emph{Challenge 3: Stressing the \gfi{main} memory \gfi{capacity}.} As prior works show~\cite{lebeck2020end} and as we observe in our analysis, running a single application is \gfi{usually} not  enough to stress the \gfi{main} memory \gfi{capacity} and create swap activity, which we aim to study in this work. This happens because many popular interactive applications have a memory footprint of \sgi{only a few} hundreds of megabytes~\cite{lebeck2020end}, which \sgi{fit} within the main memory of the device. Even though we could mix \gfi{an increasingly large number of different} applications until we \gfi{place the system under} memory \gfi{capacity} pressure, this procedure would be hard to automate, since each application requires different user interactions, and generating random combinations of workloads could dramatically change our analysis.

To overcome these challenges, we rely on the infrastructure that the open-source Chromium project~\mbox{\cite{chromium}} provides to automate the execution of web-based processes. Specifically, we use the \sgi{Chromium project's} open-source memory \gfiii{capacity} pressure test~\mbox{\cite{memorypressure}} to evaluate the system. The test has three phases. In the first phase (memory pressure), the test opens multiple Chrome tabs \sgi{in} Chrome until the first tab discard occurs \gfi{(i.e., when Chrome terminates the process associated with an open tab)}. In the second phase (cold switch), the test opens the least-recently-used tabs (called cold tabs) to induce page faults. In the third phase (heavy load), the test executes tab switches to measure system performance under heavy memory \gfi{capacity pressure}.  A tab discard happens when Chrome observes that the system is running out of memory. To calculate the amount of available memory, Chrome computes $available\_mem = available\_RAM + num\_swap\_pages / RAM\_vs\_swap\_weight$, \gfi{where $num\_swap\_pages$ defines the number of available (free and non-defective) page slots in all active swap areas, and} $RAM\_vs\_swap\_weight$ accounts for the relative ease of allocating RAM directly versus having to swap its contents out first.

Using the memory \gfii{capacity} pressure test in our analysis allows us to overcome all three main challenges \gfi{we discuss above}. We mitigate the first challenge (i.e., \gfi{executing real-world workloads}) by \gfi{using} the \gfi{open-source and commonly-used} Chrome web browser as our primary workload driver. \revision{Doing so} brings two main advantages for our experimental setup. First, we can fully understand Chrome's internal structure since it is an open-source tool, and can understand the system resources it demands using different profiling tools (e.g., \emph{perf} profiler~\cite{perf}). Second, we create an environment where the system executes distinct tasks concurrently, since the user utilizes Chrome as the primary interface to execute different services, and Chrome creates a new process for each new Chrome tab. We mimic a multiprocess system that runs different workloads and stresses different segments of the system, by opening different Chrome tabs that execute different services.  These services include Google web services (e.g., YouTube~\cite{youtube}, Google Maps~\cite{maps}, Google Sheets~\cite{sheets}, Google Docs~\cite{docs}), Facebook~\cite{facebook}, \sgi{and} Twitter~\cite{twitter}, each \revision{of which} demands different computational sources. For example, when loading YouTube as one of our Chrome tabs, the newly-created process executes tasks related to web browsing (e.g.,  texture tiling, color blitting) and YouTube-related tasks, such as video decoding, locally~\cite{chromiummedia}. We profile our system using the \emph{perf} profiling tool while executing our memory \gfiii{capacity} pressure test\gfi{, to characterize} the workloads and tasks executed by our setup that are \emph{unrelated} to web browsing tasks. A non-exhaustive list of workloads executed in our evaluation setup is: 

\begin{itemize}[noitemsep, leftmargin=*, topsep=0pt]
    \item Video decoding using the FFmpeg~\cite{ffmpegDo29} and libvpx~\cite{TheWebMP86} libraries, which are used to execute the VP8/VP9~\cite{grange2016vp9} video decoder;
    \item Web Media Player~\cite{chromiummedia} to support HTML5 video playback~\cite{HTMLStan14}, including audio/video decoders supported by the Mojo System API~\cite{Mojodocs82} and the Video Acceleration API~\cite{VaAPI61}; 
    \item Audio rendering utilizing Chromium's audio rendering API~\cite{chromiummedia};
    \item GIF decoding/encoding using the SkGifCodec~\cite{srccodec76};
    \item V8 JavaScript engine~\cite{V8JavaSc93}.
\end{itemize}

\noindent
We observe that our experimental setup includes two of the workloads also evaluated by prior work on consumer devices~\cite{boroumand2018google} (Chrome web browser, VP9 video decoding). In addition, our setup includes several workloads not covered by prior work~\cite{boroumand2018google} (e.g., Web Media Player, audio rendering, GIF decoding/encoding, V8 JavaScript engine) that are commonly employed in consumer devices. \gfi{O}ur system setup is sufficient for our study, since our goal  is not to provide optimizations for a particular workload, but rather to understand the impact of new memory technology in a real system while running real applications.

We mitigate the second challenge (i.e., the lack of \gfi{tools for automation and reproducibility}) by leveraging the memory \gfiii{capacity} pressure test's capability to load a new Chrome tab, scroll through a tab, and perform tab switches across \gfi{open} tabs without user's intervention. This is possible since the test utilizes ChromeOS' Tast integration-testing framework~\cite{TastTast15}. The framework provides APIs that allow the test code to interact with elements of the user interface through the chrome.automation library~\cite{chromeau55}.

We mitigate the third challenge (i.e., stressing the memory system to induce enough swap activity) by launching enough Chrome tabs until the system experiences moderate to critical memory \gfi{capacity} pressure \gfi{(see Section~\ref{sec_background_memory_pressure})}. We can easily control the amount of memory \gfi{capacity} pressure in the system since we can easily predict the memory footprint of opening a new Chrome tab. \gfi{A}s we discuss in Section~\ref{sec_background_memory_pressure}, we quantitatively analyze a range of memory \gfi{capacity} pressure conditions that are experienced during real user activity \gfi{in a user study across 114 users}.

\gfi{\bf{Metrics.}} We use two key metrics throughout our analysis in this paper: (i)~\emph{tab count}\gfi{, which is} the number of Chrome tabs our memory \gfii{capacity} pressure test can open before a tab discard happens; and (ii)~\emph{tab switch latency}\gfi{, which is} the latency of switching across different tabs that are already open. The tab count metric \gfi{provides} an indication of the memory \gfi{capacity} pressure the system can support. The tab switch latency metric is relevant \gfi{and important} for two main reasons. First, \gfi{it} directly \sgi{impacts} the user experience \gfi{by affecting \sgi{the} response time to the user}. Second, \gfi{the tab switch latency metric provides a clear indication of the performance impact on our interactive workloads of moving memory \gfii{blocks} \sgi{back and forth between} main memory and the swap space. B}y switching to a \sgi{previously-opened} tab, whose memory pages have been \gfi{moved from main memory to the swap \gfiii{space}}, \gfi{we} force the system to load data from the swap space. \gfi{Then, the latency associated with the data movement from the swap space to main memory is accounted \sgi{for} in the final tab switch latency.} 

%% file: sections/3-performance-evaluation.tex
\section{Evaluating Intel Optane SSD for Consumer Devices}
\label{sec_step1}

In this section, we evaluate the performance implications of employing \sgi{an Intel Optane SSD in} consumer devices. First, we evaluate the impact \sgi{on Chrome \gfi{web browser} performance} of reducing DRAM size while \gfi{using the} Intel Optane SSD \gfi{as swap space for DRAM} (Section~\ref{sec_step1_1}). Second, we leverage a compressed in-DRAM cache to reduce tail latency and energy consumption in \sgi{a} system equipped with the Intel Optane SSD (Section~\ref{sec_step1_2}). Third, we evaluate the impact of replacing \sgi{the} Intel Optane SSD with a cheaper \gfi{state-of-the-art} NAND-\sgi{flash-}based \gfi{SSD} (Section~\ref{sec_step3}).  

Throughout \gfi{the evaluations we conduct in this} section, we push the system to a critical memory \gfii{capacity} pressure state by opening as many Chrome tabs as possible (i.e., until the first tab discard happens). We evaluate such an extreme state since our goal in this section is to fully understand the benefits and drawbacks of employing the Intel Optane SSD device in our system. In addition, we report and analyze the impact of the Intel Optane SSD device \gfi{as swap space} at moderate memory \gfii{capacity} pressure states, where the system has fewer Chrome tabs open.

\subsection{Effect of NVM as a Swap Space}
\label{sec_step1_1}

\gfi{W}e evaluate (i)~the number of \gfii{Chrome} tabs the memory \gfii{capacity} pressure test can open before a tab discard happens, and (ii) the tab switch latency for the baseline (i.e., the system with \SI{8}{\giga\byte} of DRAM and ZRAM as the swap device) and the Optane (i.e., the system with \SI{4}{\giga\byte} of DRAM and the Intel Optane SSD as the swap device) configurations. During our analysis, we observe that the number of \gfii{open} tabs by the Optane configuration is 24\% larger than the number of \gfii{open} tabs \gfii{by} the baseline configuration our memory \gfii{capacity} pressure test (164 vs.\ 132 \gfii{open} tabs for \gfii{the} Optane \gfii{configuration} and \gfii{the} baseline \gfii{configuration}, respectively). We observe that this increase in the number of open tabs in the Optane configuration is due to the increase in total memory space \gfi{provided by} the Optane configuration. In the Optane configuration, the total memory space available is \SI{4}{\giga\byte} of DRAM plus \SI{16}{\giga\byte} of swap (thus, \SI{20}{\giga\byte} of \sgi{effective main} memory space). In the baseline configuration, this value is 25\% lower, since even though the system has \SI{8}{\giga\byte} of DRAM, \gfii{up \gfcr{to}} 50\% of DRAM space is reserved for the \gfi{in-DRAM} compressed swap space \gfi{(i.e., ZRAM)}. With a\gfii{n up to} 3:1 compression ratio~\gfii{\cite{shiu2015system,shiu2017driving,pekhimenko2013linearly}}, the total memory space in the baseline configuration becomes \SI{4}{\giga\byte} of DRAM \sgi{plus} \SI{12}{\giga\byte} of swap space (thus, \SI{16}{\giga\byte} of \sgi{effective main} memory space).

\begin{figure}[ht]
  \centering
  \includegraphics[width=\linewidth]{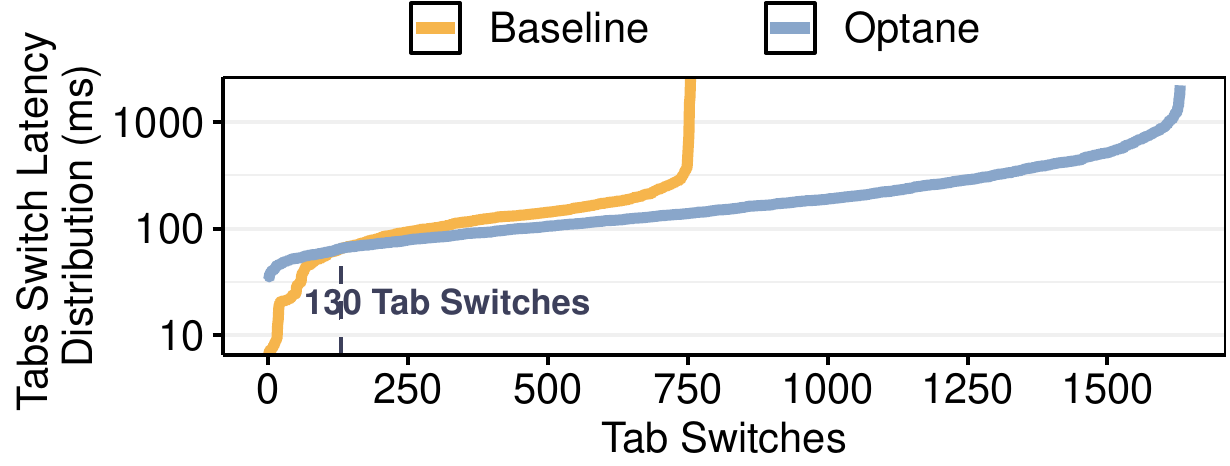}  
  \caption{Tab switch latency distribution: Baseline (ZRAM)  vs.\ Optane.}
  \label{fig_step_1_zram_optane_tab_latency}
  
\end{figure}

\label{r2.3}\Copy{R2/3}{Figure~\ref{fig_step_1_zram_optane_tab_latency} shows the tab switch latency distribution for the baseline and Optane configurations. The figure depicts the tab switch ID\ieeearev{, i.e., an identifier for a given sorted tab switch latency} (x-axis) and the sorted tab switch latency (y-axis, in logarithmic scale).  We draw two observations.}

First, we observe that the tab switch latency of the baseline configuration is lower than that of the Optane configuration when the number of tab switches is lower than 130.  This is because the Optane configuration has half of the DRAM space as the baseline  configuration. As a result, the Optane configuration experiences more page faults than the baseline configuration. In our experiments, the baseline configuration experiences 451 page faults until 130 tab switches, which are serviced by the ZRAM space. In contrast, the Optane configuration experiences 10$\times$ the page faults of the baseline until it hits 130 tab switches, which significantly reduces Optane performance \gfi{at low tab switch counts}. It is important to highlight that there are fewer page faults for the baseline configuration than for the Optane configuration when the number of tab switches is low, because the physical memory space the OS reserves for ZRAM is \gfi{\emph{not}} statically allocated. Initially, during execution, the baseline configuration enjoys a larger memory space than the Optane configuration since the baseline configuration has \SI{8}{\giga\byte} of physical memory available as working memory. This leads to a lower number of page faults experienced by the baseline configuration than by the Optane configuration when the tab count (and therefore the number of tab switches) is low. However, as swap activity increases with larger number of \gfii{open tabs}, the OS allocates physical memory for the compressed swap space utilized by the baseline. The OS allocates physical memory for the ZRAM swap space until the total \gfi{swap space} size reaches a predefined threshold (\SI{4}{\giga\byte} \sgi{of physical DRAM space} in our setup). \gfi{After the maximum swap space threshold is reached}, the baseline configuration can leverage \emph{only} the remaining \SI{4}{\giga\byte} of physical memory as working memory, which leads to a \gfi{much} larger number of page faults.

Second, after 130 tab switches, the Optane configuration provides \gfi{a lower tab switch latency} than the baseline configuration. That is due to how ZRAM operates once the system runs out of memory\gfi{:} Chrome pages need to be constantly swapped in\gfi{to}/out \gfi{of} the ZRAM space. This incurs high CPU overhead and data movement between DRAM regions since the processor need\gfi{s} to frequently \gfi{(1)~find cold pages to move from main memory to ZRAM (i.e., swap pages \gfii{into} the ZRAM space), which requires the processor to execute data compression operations; and (2)~find and move requested cold pages from ZRAM to main memory in case of a page fault (i.e., swap pages \gfii{out} \gfiii{of} the ZRAM space), which requires the processor to execute data decompression operations.} In contrast, the Optane configuration \gfi{greatly} alleviates CPU usage caused by compression\gfi{/decompression} activities. \gfi{When memory pages are swapped \sgi{into/out of} the Optane-based swap space, the processor needs to \emph{only} issue asynchronous read/write requests to the \gfii{Intel} Optane \gfii{SSD} device. We observe that by eliminating the CPU time the system would spend on compression\gfi{/decompression} activities, the page fault latency in the Optane configuration is 35\% lower than in the baseline \gfiii{ZRAM} configuration, which leads to lower tab switch latencies in the Optane configuration. \gfii{We conclude that the Optane configuration provides better performance than the ZRAM configuration for high-enough tab counts.}}

\begin{figure}[ht]
  \vspace{-5pt}
  \centering
  \includegraphics[width=\linewidth]{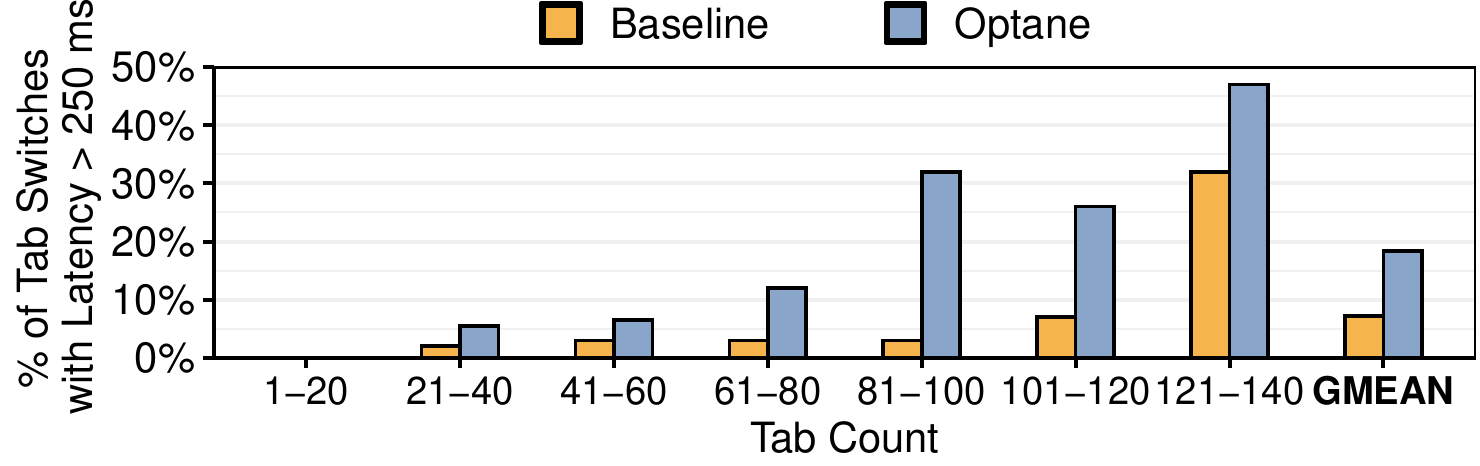}  
  \caption{High-latency tab switch distribution: Baseline (ZRAM)  vs.\ Optane.}
  \label{fig_step_1_zram_optane_latency_dist}
  \vspace{-5pt}
\end{figure}

Even though \gfi{using the} Intel Optane SSD \gfi{as a swap space} provides benefits for average tab switch latency \gfi{(especially at high tab counts)}, it can also harm \gfcr{tail latency} performance. Figure~\ref{fig_step_1_zram_optane_latency_dist} shows the distribution of tab switches with latency larger than \SI{250}{\milli\second} for both \gfi{the baseline and Optane} configurations. We make two observations. First, the \gfi{fraction} of tab switches with a latency larger than \SI{250}{\milli\second} for the baseline configuration is 7.1\%, on average during the execution of the memory \gfii{capacity} pressure test. \gfi{In contrast, i}n the Optane configuration, the \gfi{fraction} of high-latency tab switches is 18.4\%, on average. Thus, in the Optane configuration, the \gfi{fraction} of high-latency tab switches is 2.6$\times$ those in the baseline configuration. Second, for low tab counts (until 60~tabs), \ieeea{the percentage of high-latency tab switches for the Optane configuration is 
(i) the \emph{same} as in the baseline configuration for 1--20 tab counts (0\% vs.\ 0\% for the Optane and baseline configurations, respectively); and 
(ii) \emph{slightly} larger than the baseline configuration for 21--40 and 40--61 tab counts (high-latency tab switches
make up 5.5\% vs.\ 2\% for 21--40 tab counts, and 6.5\% vs.\ 3\% for 41--60 tab counts, for the Optane and baseline configurations, respectively).}
However, after 61~tabs, the number of tab \gfi{switches} with high latency increases significantly for the Optane configuration. \gfi{To understand the reason of this increase \gfiii{in} tab switches with high latency in the Optane configuration, we analyze the page fault latency distribution during the execution of our memory \gfii{capacity} pressure next.}

\begin{figure}[ht]
  \vspace{-5pt}
  \centering
  \includegraphics[width=\linewidth]{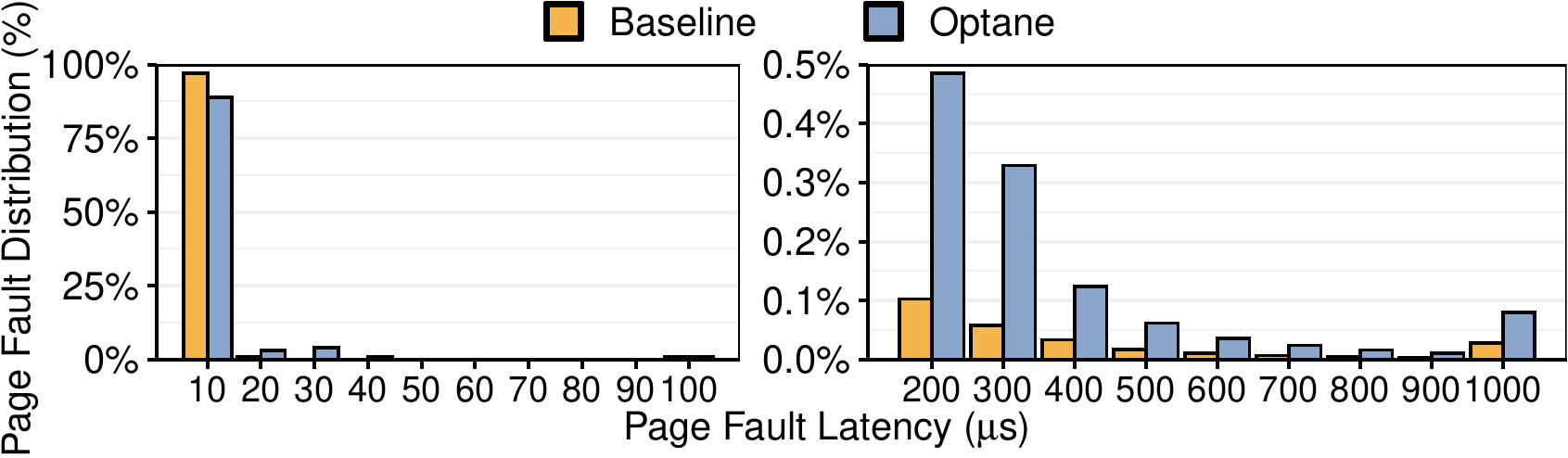}  
  \caption{\gfii{Page fault latency distribution.}}
  \label{fig_step_1_do_page_fault}
\vspace{-5pt}
\end{figure}

\label{r1.4}\Copy{R1/4}{\gfi{Figure~\ref{fig_step_1_do_page_fault} shows the page fault latency distribution for all page faults in the baseline and Optane configurations during the execution of our \gfii{memory capacity pressure} test. We make two observations. First,} when analyzing the page fault latency, we observe that \ieeearev{at least 98\% of the page faults}\revdel{most of the page faults (at least 98\%)} have a response latency of less than \SI{10}{\micro\second} for both configurations (\gfi{Figure~\ref{fig_step_1_do_page_fault}}, left). \gfi{Second,} when \gfi{examining} the tail of the latency distribution (\gfi{Figure~\ref{fig_step_1_do_page_fault}}, right), \gfi{the} Optane \gfi{configuration} has 4.43$\times$ \gfi{the number of page faults of the baseline} \gfii{(i.e., the number of page faults with a latency larger than \SI{10}{\micro\second} in the Optane configuration is 4.43$\times$ that of the baseline configuration)}. \ieeearev{We conclude that the increase in high-latency tab switches for the Optane configuration (in Figure~\mbox{\ref{fig_step_1_zram_optane_latency_dist}}) is due to} \sgfb{less than 2\% of the page faults, with these page faults having a high response latency} (more than \mbox{\SI{10}{\micro\second}}).}

\begin{figure}[ht]
  \centering
  \includegraphics[width=\linewidth]{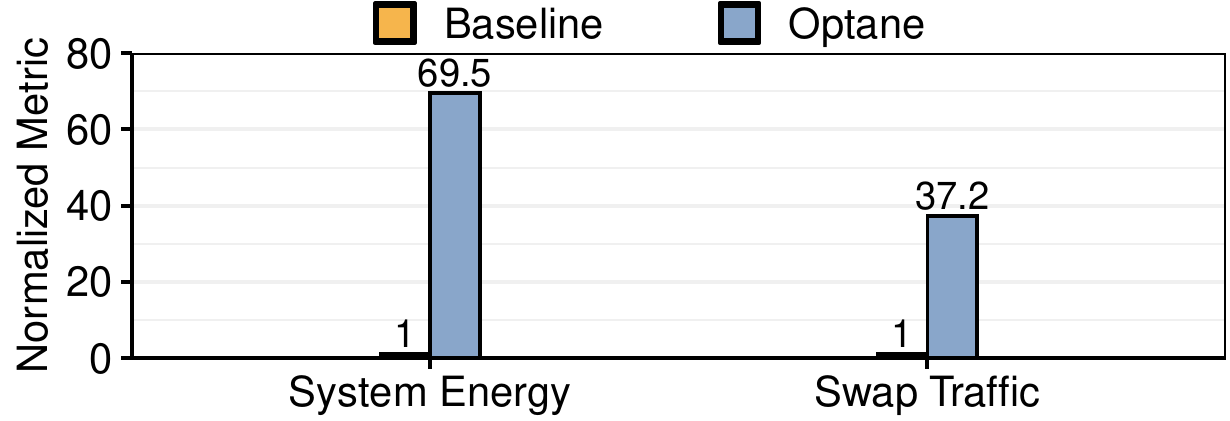}  
  \caption{\gfii{Energy and swap traffic. Y-axis values are normalized to the baseline configuration.}}
  \label{fig_step_1_zram_optane_energy}
\end{figure}

\gfi{Figure~\ref{fig_step_1_zram_optane_energy}} \sgi{compares} the impact of \gfi{the baseline} and Optane \gfi{configurations} on \gfii{average \ieeea{memory subsystem}} energy consumption and swap traffic \gfi{(i.e., the total number of bytes swapped \sgi{into/out of} the swap space)}. \gfi{To model Intel Optane energy consumption, we assume that the device uses PCM\gfi{-}based memory cells, as previous \sgi{work suggests}~\cite{choe2017intel}. Then, we gather \sgi{the} read/write energy consumption of PCM-based memory devices \gfi{and DRAM,} as reported by previous \sgi{work}~\cite{chen2012energy}\gfi{,} to model the energy consumption of our system\ieeea{, i.e., \SI{4.4}{\pico\joule}/\SI{2.47}{\pico\joule} read energy-per-bit and  \SI{5.5}{\pico\joule}/\SI{14.03}{\pico\joule} (set)--\SI{19.73}{\pico\joule} (reset) write energy-per-bit for DRAM/PCM}.}\footnote{\ieeea{We evaluate \emph{only} the energy of the memory subsystem in our analyses. This does \emph{not} include the energy consumed by the processor.}} In the figure, we normalize both metrics to the baseline values (y = 1.0 in the plot). We observe that the Optane configuration increases energy consumption and swap traffic by 69.5$\times$ and 37.2$\times$\gfiii{, respectively,} \gfii{compared to the baseline configuration}. This is due to two main reasons. First, writing one bit to Optane consumes up to \gfii{3.6$\times$ the energy of writing one bit to DRAM~\cite{chen2012energy}}. Even though reading one bit from Optane consumes 56\% less energy than reading one bit from DRAM~\cite{chen2012energy}, in our analysis, \gfi{we find that} the majority \sgi{(i.e., 54\%)} of the accesses to \gfi{the Intel} Optane \gfi{SSD} \gfi{are} write accesses.  Second, the Optane configuration generates \gfi{much} more swap activity \gfi{than the baseline}, since \gfi{the Optane configuration} has half the DRAM size of the baseline configuration. \gfii{We conclude that using the Intel Optane SSD as a swap space to DRAM \gfiii{severely} penalizes system energy consumption.}

\gfi{Summarizing our findings, the Optane configuration provides benefits compared to the baseline system, since it enables a large number of tab switches due to an increase in the main memory space. However, it has significant drawbacks in terms of tail latency, system energy and swap traffic, compared to the baseline with double the amount of DRAM. Most of these downsides come from the large number of accesses, especially write accesses, to the Intel Optane SSD. To solve these problems, we next examine multiple techniques that can (i)~reduce the swap traffic in the Optane configuration (Section~\ref{sec_step1_2}) and (ii)~improve overall system performance for the Optane configuration (Section~\ref{sec_step2}).}

\subsection{Reducing Tail Latency by Enabling a Compressed RAM Cache}
\label{sec_step1_2}

\sgi{The Intel Optane SSD} can improve overall performance for consumer devices since it enables \gfi{an extended} memory space. However, \gfi{as we \sgi{show} in Section~\ref{sec_step1_1},} it can also \gfi{negatively} impact tail latency and \gfi{system} energy consumption due to the need to issue high-latency and power-hungry I/O requests to access the device. \gfi{In addition}, NVM devices \gfi{\sgi{such as} the Intel Optane SSD} suffer from \gfi{limited} endurance~\gfiii{\cite{lee2009architecting,lee2010phase,lee2010phasecacm}}. \gfi{As a result, \sgi{the} large number of write operations caused by \sgi{page swapping} can degrade system reliability.}

To overcome \gfi{these} issues \gfi{introduced by the Intel Optane SSD, we aim to reduce the number of accesses to it. To this end}, we augment the Optane configuration with Zswap~\cite{zolnierkiewicz2013efficient}, an in-DRAM \gfi{swap} cache used to store compressed cold-pages. Zswap takes \gfi{memory} pages that are in the process of being swapped out \gfi{from main memory to the swap device (i.e., the Intel Optane SSD)} and attempts to compress them into a dynamically\gfiii{-}allocated \gfi{D}RAM-based memory pool. \gfii{When a page fault happens, the OS checks if the \sgii{requested} memory page is stored in the Zswap cache (i.e., Zswap load hit); \sgii{if the page is not in the Zswap cache, the OS} loads the memory page from the swap space (i.e., Zswap load miss). In case of a Zswap load hit, the OS (1)~loads and \gfii{decompresses} the memory page stored \gfiii{in} the Zswap cache, and (2)~\gfiii{services the page fault request by writing the decompressed memory page into DRAM.}} The motivation behind Zswap is to trade CPU cycles for a potential reduction \sgi{in} I/O requests. \gfi{Zswap \sgi{can improve performance} if the read requests are serviced faster from the in-DRAM compressed \gfi{swap} cache than from the swap device.}\footnote{\gfi{Even though both ZRAM and Zswap use an in-DRAM compressed memory space to operate, they are fundamentally different mechanisms. While ZRAM is an in-DRAM compressed \emph{swap space}, Zswap is an in-DRAM \emph{swap cache}. The system sees ZRAM as a swap device and Zswap as a cache for memory pages swapped \sgi{into/out of} the swap device. We enable Zswap \sgi{only} in the Optane and NANDFlash configurations, \gfiii{since the goal of enabling Zswap is to reduce I/O traffic to \emph{off-chip swap devices} (i.e., the Intel Optane SSD and the NAND-flash-based SSD in our experiments). We} \sgi{indicate Zswap-enabled configurations explicitly throughout the paper (e.g., the Optane configuration with Zswap is labeled \emph{Optane+Zswap})}.}}

To fully leverage Zswap in the Optane configuration, we first need to tailor two important parameters related to Zswap execution. First, as explained in Section~\ref{sec_background}, a tab discard happens when Chrome observes that the system is running out of memory, which is computed based on the \texttt{RAM\_vs\_swap\_weight} parameter \gfi{(i.e., the relative ease of allocating DRAM pages directly versus having to swap its contents out first\gfiv{; see Section~\ref{sec_experimental_setup}})}. This parameter is empirically defined for ZRAM as 4. Similarly, we empirically evaluate which value \texttt{RAM\_vs\_swap\_weight} enables the \gfii{memory capacity pressure} test to open more tabs before \gfi{the first} tab discard \gfi{happens}. We run tests varying the \texttt{RAM\_vs\_swap\_weight} value from 1 to 8, and we observe that \texttt{RAM\_vs\_swap\_weight} equals to 1 provides the largest number of \gfii{open} tabs for the Zswap configuration. \gfi{Second}, we need to understand the impact of the maximum Zswap \gfi{cache} size (\gfi{i.e.,} \texttt{max\_pool\_size}) on the tab switch latency. \gfi{To this end}, we run the memory \gfii{capacity} pressure test while varying the \texttt{max\_pool\_size} from 0\% to 50\% of the total DRAM size. As expected, we observe that the tab switch latency increase\gfi{s} with the  \texttt{max\_pool\_size}. \gfi{W}e \sgi{find} that setting \texttt{max\_pool\_size} to 20\% of the total DRAM \gfi{capacity} provide\gfi{s} a good balance between reduction \gfi{in} I/O traffic and tab switch latency.

\begin{figure}[h]
   \centering
   \includegraphics[width=\linewidth]{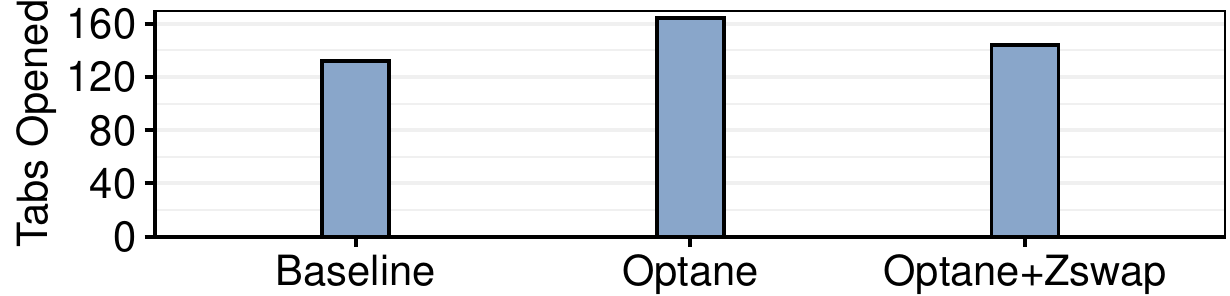}  
   \caption{Number of \gfi{open} \sgi{tabs}: \gfi{Baseline} vs.\ Optane vs.\ \gfi{Optane with} Zswap enabled (Optane+Zswap).}
   \label{fig_step_1_zswap_optane_open_tabs}
  \end{figure}

Figure~\ref{fig_step_1_zswap_optane_open_tabs} compares the total number of \gfii{open} tabs \gfi{before a tab discard happens in the baseline \sgi{and Optane configurations, and}} when \gfi{we} \gfi{enable} Zswap \gfi{in our Optane configuration \sgi{(labeled Optane+Zswap)}. W}e observe that by enabling Zswap,  the number of \gfi{open} tabs reduces by 12\% compared to the Optane configuration \sgi{without Zswap enabled}. \gfi{B}y enabling Zswap, we effectively reduce the total memory space available, thus causing a discard to happen sooner.

\begin{figure}[ht]
\begin{subfigure}{\linewidth}
  \centering
  \includegraphics[width=\linewidth]{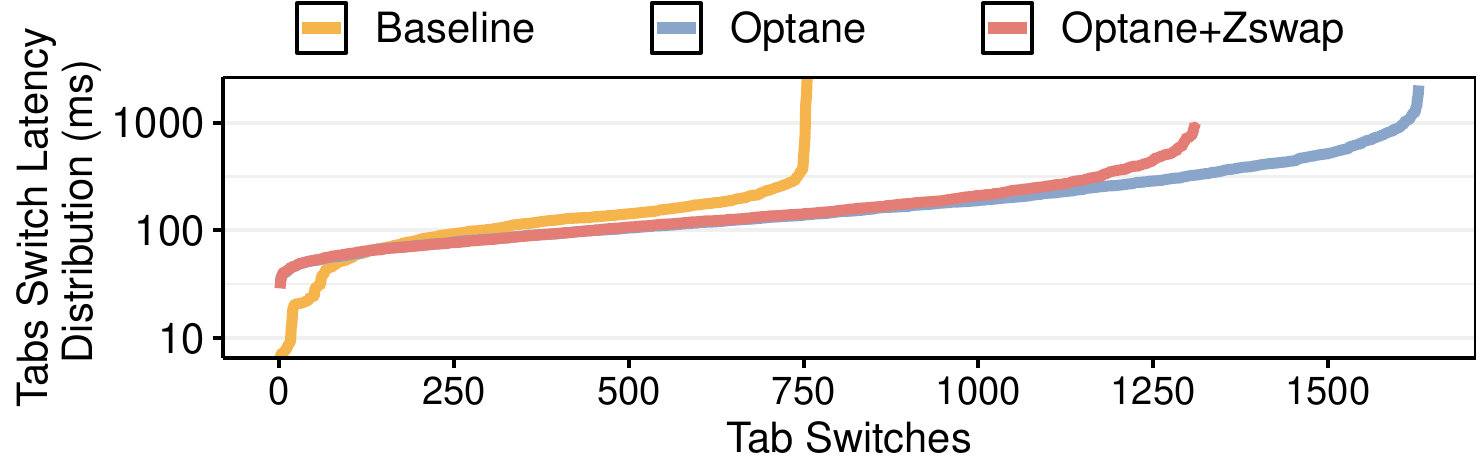}  
  \vspace{-15pt}
  \caption{Tab switch latency distribution.}
  \label{fig_step_1_zswap_optane_tab_latency}
\end{subfigure}
\par\bigskip
\begin{subfigure}{\linewidth}
  \centering
  \includegraphics[width=\linewidth]{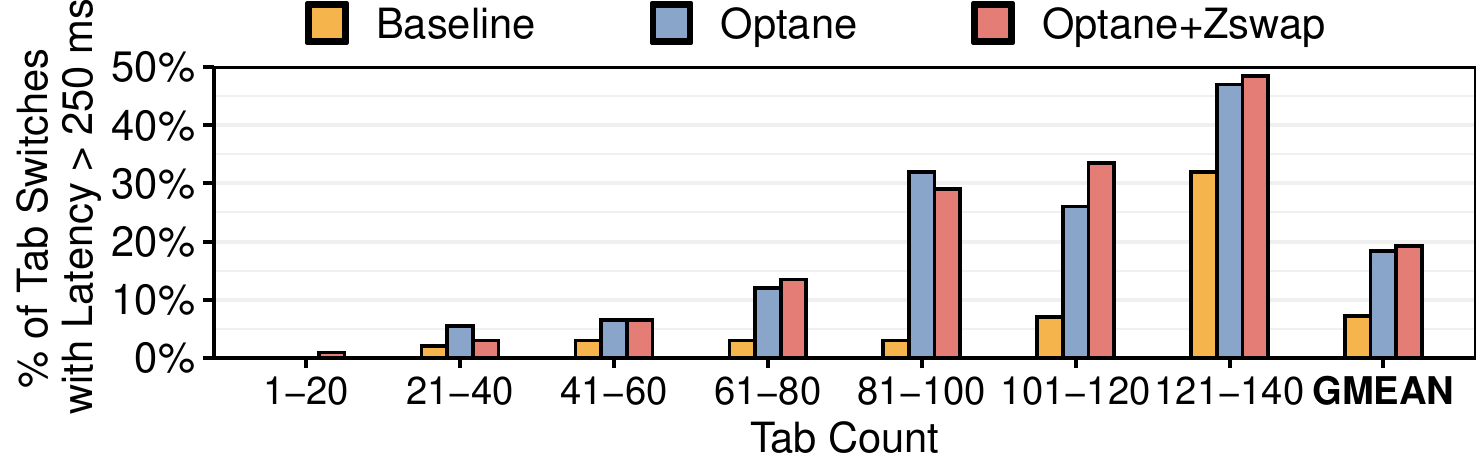} 
  \vspace{-10pt}
  \caption{High-latency tab switch distribution.}
  \label{fig_step_1_zswap_optane_latency_dist}
\end{subfigure}
\caption{Tab switch latency: Baseline vs.\ Optane vs.\ Optane+Zswap.}
\end{figure}

Figure~\ref{fig_step_1_zswap_optane_tab_latency} shows the impact of enabling Zswap on the tab switch latency. \gfi{Enabling} Zswap maintains a similar overall tab switch latency as the Optane configuration \sgi{without Zswap}. Figure~\ref{fig_step_1_zswap_optane_latency_dist} shows the impact that Zswap has on high-latency tab switches. \gfi{We make three observations. First, o}n average \gfi{across all tab counts}, enabling Zswap leads to a modest increase of 4\% on the number of tab \gfi{switches} with latency larger than \SI{250}{\milli\second} (up to 29\% for \gfi{tab counts of} 101--120). \gfi{Second}, the large difference in the \gfi{fraction} of \sgi{tabs} with high latency \gfi{tab switches} between Optane and Optane+Zswap configurations happens when swap activity increases, and the swap cache gets full. \gfi{W}hen the swap cache gets full, the system needs to \gfi{(}1)~free an entry in the swap cache, \gfi{(}2)~decompress the selected entry and evict it back to the swap device, \sgi{and} \gfi{(}3)~compress the new page and insert it in the swap cache. These operations represent the worst-case latency for a Zswap operation. \gfi{Third}, \gfi{tab counts \sgi{of}} 21--40 and 41--60 have a similar fraction of tab switches whose latencies are unacceptable \gfi{in \sgi{both the} Optane and Optane+Zswap configurations}. Enabling Zswap reduces the number of high-latency tab switches for \gfi{tab counts of} 21--40, \sgi{resulting in} a similar \gfi{fraction \sgi{of high-latency tabs}} as the baseline. For 21--40 tabs, the number of high latency tab switches is 2\%, 5.5\%, and 3\%, for the baseline, Optane, and Optane+Zswap configurations, respectively. \gfi{W}e conclude that when a page request hits in the Zswap cache, the system can provide \sgi{a lower} tab switch latency.

\begin{figure}[ht]
    \centering
    \includegraphics[width=\linewidth]{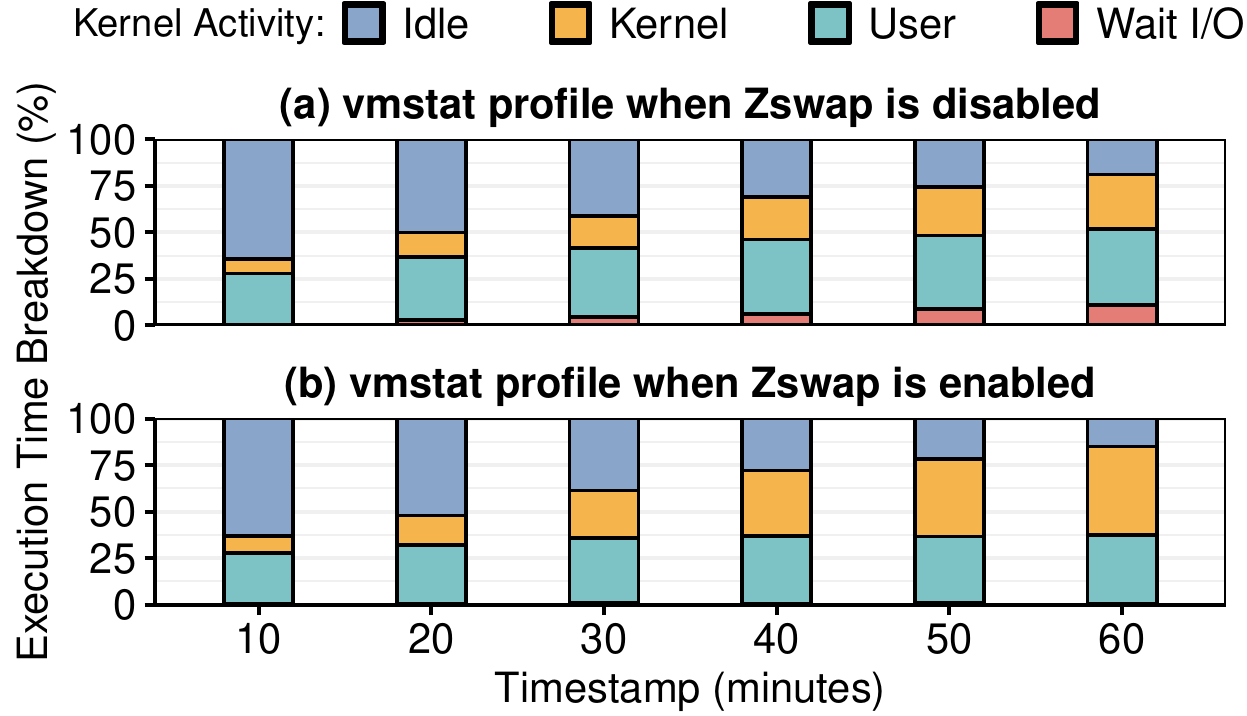}    \caption{System \gfi{execution time breakdown} during a memory \gfii{capacity} pressure test.}
    \label{fig_step_1_perf_zswap}
\end{figure}

To fully understand the impact of enabling Zswap in the \gfi{Optane} \gfi{configuration}, we monitor the CPU utilization of the system for both \gfi{Optane and Optane+Zswap} configurations. For this, we execute the memory \gfii{capacity} pressure test for both configuration\gfi{s} and monitor the CPU utilization with the \texttt{vmstat} tool~\gfi{\cite{tanaka2005monitoring}} for one hour. Figure~\ref{fig_step_1_perf_zswap} depicts the CPU utilization for both configurations. \gfi{We make two observations. First,} we observe that, when we disable Zswap (Figure~\ref{fig_step_1_perf_zswap}a), the CPU spends part of the execution \gfi{time} waiting for I/O operations to complete. Towards the end of the execution of the memory \gfii{capacity} pressure test, the I/O \gfi{waiting} time increases dramatically, since the swap activity also \gfi{greatly} increases. \gfi{Second}, \gfi{we observe that by enabling Zswap,} the system \gfi{spends \gfiii{a} smaller fraction of its time waiting for} I/O operations, as Figure~\ref{fig_step_1_perf_zswap}b shows. On the other hand, it also spends \gfi{a larger fraction of its} time on kernel activity due to Zswap compression/decompression execution, which reduces \gfi{the fraction of time spen\gfiii{t} on u}ser activity \gfi{and  penalize\gfiii{s} Chrome \gfiii{browser} performance. We conclude that the increase in kernel activity caused by Zswap cache compression/decompression operations is the \emph{primary} cause of \gfiii{the} increase \gfiii{in} the fraction of high-latency tab switches in the Optane+Zswap configuration.} 

\begin{figure}[ht]
\begin{subfigure}{\linewidth}
  \centering
  \includegraphics[width=\linewidth]{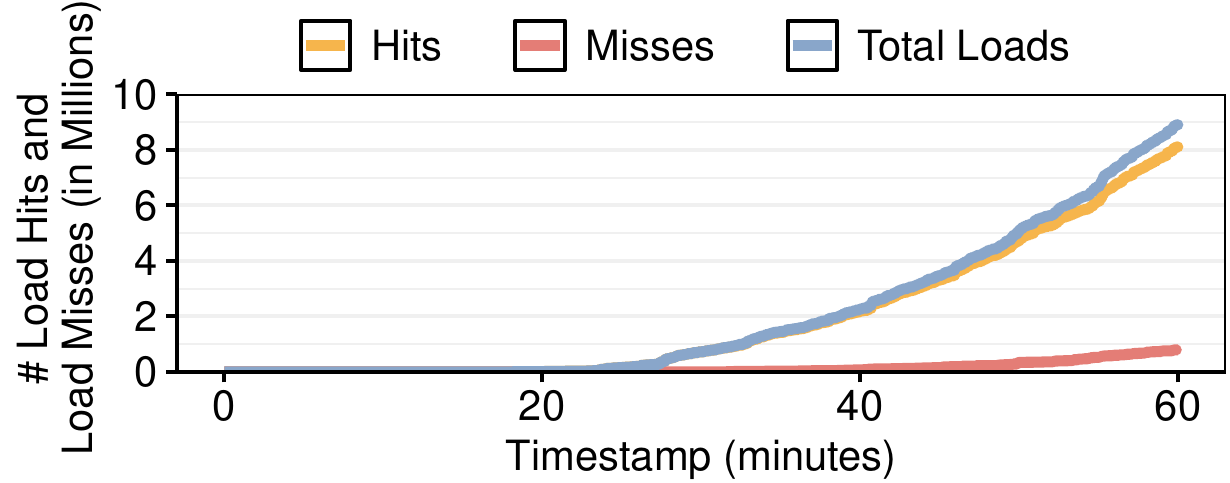}  
  \vspace{-15pt}
  \caption{Zswap cache load hits and load misses.}
  \label{fig_step_1_zswap_optane_hita}
\end{subfigure}
\par\bigskip
\begin{subfigure}{\linewidth}
  \centering
  \includegraphics[width=\linewidth]{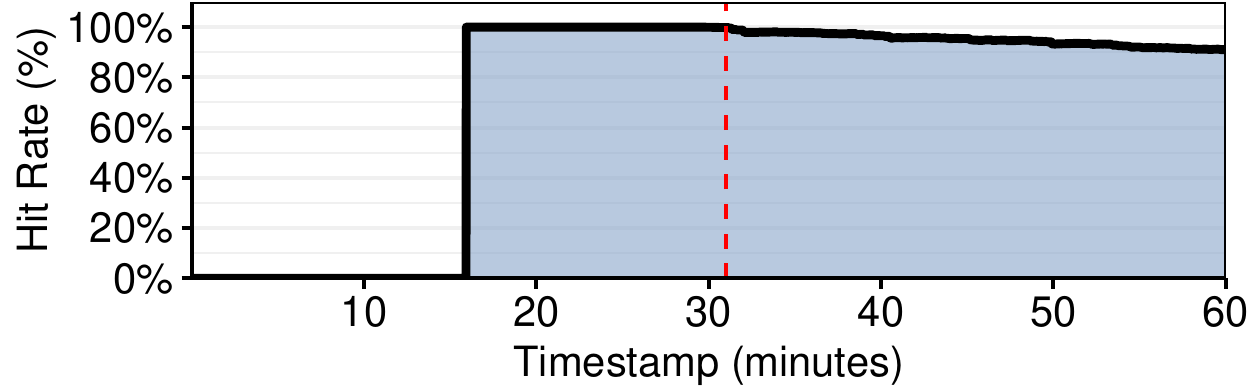}  
  \caption{Zswap cache hit rate.}
  \label{fig_step_1_zswap_optane_hitb}
\end{subfigure}
 \caption{Zswap cache performance.}
 \label{fig_step_1_zswap_optane_hit}
\end{figure}

To evaluate the \gfi{effectiveness} of the Zswap cache, we \gfi{analyze} the \gfi{Zswap} cache behavior during the execution of the memory \gfii{capacity} pressure \sgi{test.} Figure~\ref{fig_step_1_zswap_optane_hita} shows the number of load hits, load misses, and total loads \sgi{that} the Zswap cache services. Figure~\ref{fig_step_1_zswap_optane_hitb} shows the hit \gfi{rate} of the Zswap cache over the same time. We make the \gfi{two} observations based \gfi{on Figure~\ref{fig_step_1_zswap_optane_hit}}. First, \sgi{the} Zswap cache provides a high hit \gfi{rate} of \gfii{97\%}, on average \gfii{during the execution of the memory capacity pressure test}. In fact, \sgi{we see in} Figure~\ref{fig_step_1_zswap_optane_hitb} that the hit \gfi{rate} is close to \gfi{100\%} from 16 minutes to 30 minutes. \gfi{Second, we observe that} \gfi{the hit rate} drops \gfi{from 100\% at 30 minutes to \gfii{91}\% at 60 minutes}. We observe that at \gfi{30 minutes}, \sgi{the} Zswap cache gets full, which requires the system to evict old pages frequently. \gfi{However, even during \gfiii{high} memory capacity pressure, the Zswap cache maintains a significantly high hit rate. We conclude that Zswap cache is an effective cache for the swap space.}

\begin{figure}[ht]
\centering
\centering
   \centering
     \includegraphics[width=\linewidth]{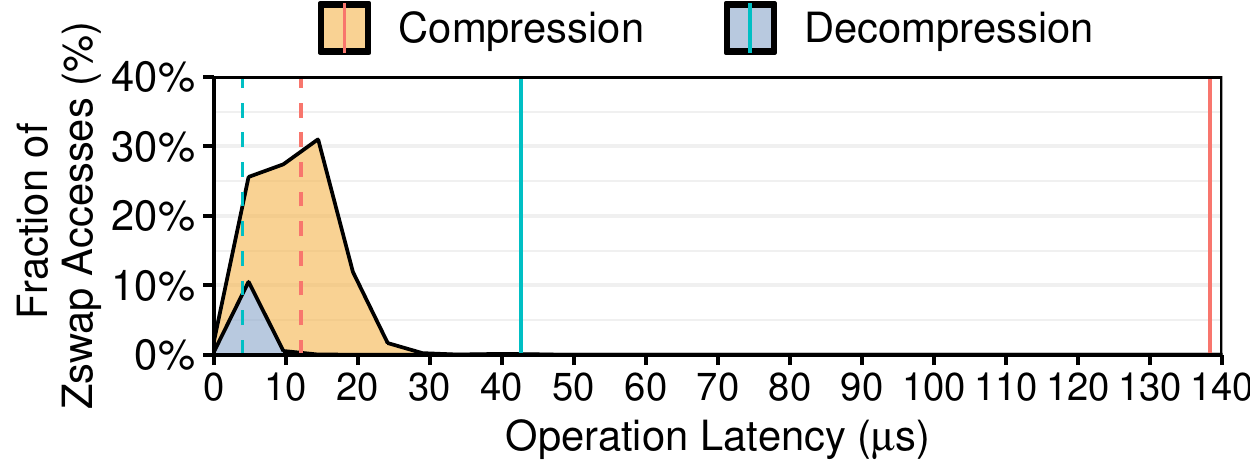}  
   \caption{\gfii{Zswap cache c}ompression/decompression latency \gfii{distribution}. \gfii{Dashed (solid) lines represents average (maximum) compression/decompression latency.}}
   \label{fig_step_1_zswap_compression_decompression}
\end{figure}

\gfi{Figure~\ref{fig_step_1_zswap_compression_decompression} shows} the distribution of the compression and decompression latencies \gfi{in the} Zswap \gfi{cache} during the \gfii{execution of the memory capacity pressure} test. In the figure, dashed lines represent the average compression/decompression latency, and solid lines represent the maximum compression/decompression latency. \gfi{We make two observations. First,} the decompression latency, which is \gfi{on} the critical path of Chrome execution when a page fault happens, is \SI{3.9}{\micro\second} on average \gfii{during the execution of the memory capacity pressure test} (minimum of \SI{1.5}{\micro\second}, and maximum of \SI{42.6}{\micro\second}). \gfii{We observe that} \gfi{98.7\%} of the decompression requests have a latency of less than \SI{10}{\micro\second}. \gfi{Second}, \sgi{the} compression \gfi{latency is larger,} \sgi{with an average latency of \SI{12.1}{\micro\second}} (minimum of \SI{1.5}{\micro\second}, and maximum of \SI{138.2}{\micro\second}). As a comparison, the Intel Optane SSD read latency is \SI{22.4}{\micro\second} on average (minimum of \SI{9.0}{\micro\second}, and maximum of \SI{5380}{\micro\second}). \gfi{We conclude that Zswap is an effective caching \gfiii{mechanism} for the swap space, since it provides significantly lower access latency than directly accessing the swap device. }

\begin{figure}[ht]
  \vspace{-10pt}
\begin{subfigure}{\linewidth}
  \centering
  \includegraphics[width=0.9\linewidth]{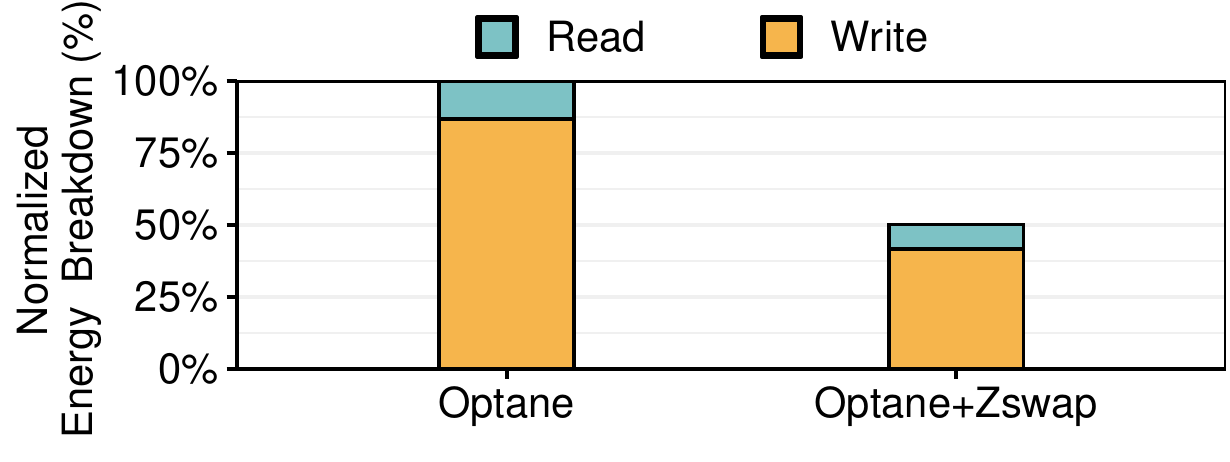}  
  \vspace{-10pt}
  \caption{Effect of Zswap cache on system energy.}
  \label{fig_step_1_zswap_io_traffica}
\end{subfigure}
\par\bigskip
\begin{subfigure}{\linewidth}
  \centering
  \includegraphics[width=0.9\linewidth]{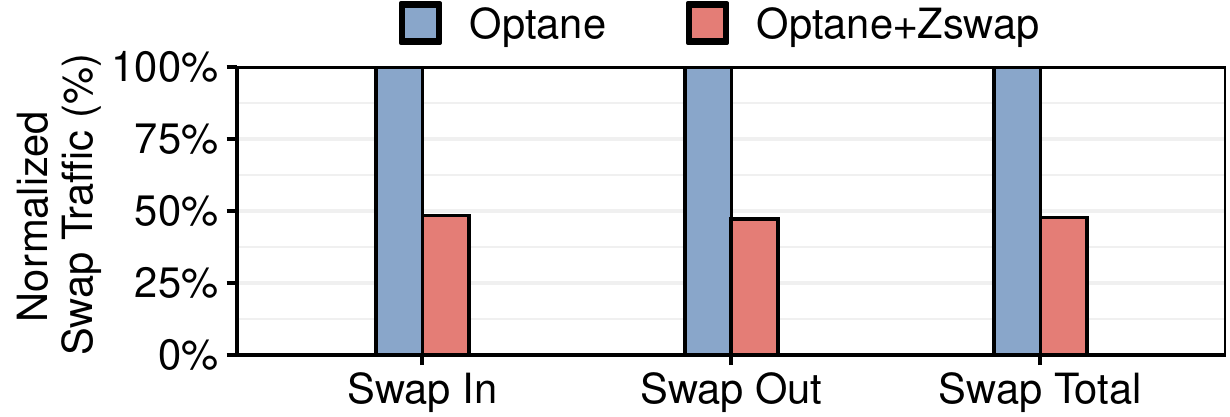}  
  \caption{Effect of Zswap cache on swap traffic.}
  \label{fig_step_1_zswap_io_trafficb}
\end{subfigure}
 \caption{Effect of Zswap cache on system energy and swap traffic.}
 \label{fig_step_1_zswap_io_traffic}
\end{figure}

\gfi{W}e \gfi{also study the system energy savings} \sgi{that the} Zswap \gfi{cache provides}.  Figure~\ref{fig_step_1_zswap_io_traffica} shows the energy savings \sgi{when Zswap is enabled for} the Optane configuration. \sgi{We make \ieeea{three} observations from the figure. 
First, we} observe that enabling Zswap reduce\gfi{s} overall \gfi{system} energy consumption by 2$\times$. 
\sgi{Second, we} observe that the majority of the energy is spent on write requests. To understand \gfi{these} energy results, we analyze \gfi{in Figure~\ref{fig_step_1_zswap_io_trafficb}} the amount of memory \gfi{s}wap-\gfi{i}n and \gfi{s}wap-\gfi{o}ut \gfi{activity} during the execution of the test. \gfi{As shown in Figure~\ref{fig_step_1_zswap_io_trafficb}, with Zswap cache enabled, swap-in and swap-out activity reduces by 2.06$\times$ and 2.11$\times$, respectively.} \gfi{This large} reduction in swap activity directly translates to a reduction in energy consumption. 
\ieeea{Third, even with Zswap enabled, the Optane+Zswap configuration consumes 34.75$\times$ the energy of the baseline configuration. This large increase in energy consumption is due to 
(i) an increase in swap activity since the Optane+Zswap configuration enables significantly more tab switches than the baseline and 
(ii) the high energy cost of write operations to the Optane device when swapping out pages~\cite{chen2012energy}.}
\gfi{We conclude that the Zswap cache greatly reduces energy consumption due to a \gfiii{large} reduction of swap traffic.} \ieeea{However, the Optane+Zswap configuration still significantly increases the energy consumption compared to the baseline. We expect that such energy can be further reduced by employing techniques to reduce the energy cost of write operations on NVM devices~\cite{guo2018latency, choi2017nvm,swami2016secret, chen2012energy}.}

\label{r2.r2.1}\Copy{R2/2/1}{\noindent \ieeearevii{\hl{\textbf{Lifetime Analysis.} One characteristic of NVM devices is their limited write endurance, i.e., a memory cell in the Optane device becomes unreliable beyond a certain number of writes. Therefore, we evaluate how Optane's limited write endurance affects the lifetime of our system when employing the Intel Optane SSD as a swap space. We compare the Intel Optane SSD lifetime (in years) when executing Chrome tab switching and scrolling activities in a system with and without Zswap. To do so, we adopt the lifetime model in~\mbox{\cite{lee2009architecting}}, which estimates the lifetime of a memory module
driven by the access patterns observed in our Chrome workload. 
We assume a conservative Optane cell endurance of $10^{6}$ writes~\mbox{\cite{optaneendurance}} (i.e., the same cell endurance of PCM-based memory cells~\mbox{\cite{chang2016improving,aghaei2014prolonging,ferreira2010increasing}}) and an \emph{optimistic} wear-leveling mechanism that evenly distributes write requests across all cells of the Intel Optane media (which the Intel Optane SSD is reported to implement~\mbox{\cite{IntelOp22:online}}). 
Our model shows that the lifetime of the Optane configuration without (with) Zswap enabled when executing our Chrome workload is \mbox{\ltzswapoff} (\mbox{\ltzswapon}) years. Such an expected lifetime can be hard to obtain in practice because it is unlikely that the wear-\mbox{\gfcr{leveling}} algorithm will be able to distribute \emph{all} writes across the Optane device \emph{equally}. However, prior works propose more realistic wear-\mbox{\gfcr{leveling}} mechanisms that can achieve up to 53\% of the lifetime of an optimistic wear-\mbox{\gfcr{leveling}} mechanism~\mbox{\cite{qureshi2009enhancing}}. Therefore, employing such a wear-\mbox{\gfcr{leveling}} mechanism can still guarantee a high lifetime for our Optane-based system without (with) Zswap enabled of \mbox{\ltzswapoffhf} (\mbox{\ltzswaponhf}) years.}}}

\gfii{\sgii{Summarizing our findings,} the Zswap cache is an effective caching mechanism that reduces the swap traffic and system energy consumption when utilizing the Intel Optane SSD as a swap space. These benefits come at the cost of a \gfiv{small} increase in the number of high-latency tab switches \gfiii{and a \gfiv{small} decrease in open tab count}.}

\subsection{Effect of Using Different NVM Devices}
\label{sec_step3}

In the previous sections, we show that enabling Intel Optane SSD as swap space for consumer devices can provide \gfi{significant} benefits due to \gfi{the extended main} memory space \gfi{it provides}. However, it is essential to understand if we can achieve similar results using cheaper \gfi{state-of-the-art} \gfi{NAND-flash-based SSDs} \sgi{in place of the} Intel Optane SSD. \gfi{\sgi{We} aim to study whether state-of-the-art NAND-flash-based SSDs that are already \sgi{widely used} \gfii{(e.g., Micron~\cite{micronslc} and Transcend~\cite{transcendssd} NAND-flash-based SSDs)} can provide similar benefits for our workloads as we have observed using the Intel Optane \sgi{SSD in Sections~\ref{sec_step1_1} and \ref{sec_step1_2}.}}

\gfi{There are three major} differences between \gfi{the} Intel Optane SSD and \gfi{a} traditional \gfi{NAND-flash-based} SSD\gfi{:} (1)~\gfi{lower} access latency \gfi{in the Intel Optane SSD}, (2)~\gfi{higher} endurance \gfi{in the Intel Optane SSD}, and (3)~\gfi{higher cost of the Intel Optane SSD} device. First, as previous work\gfi{s}~\gfi{\cite{harris2020ultra,lee2019asynchronous,zhang2018performance,chien2018characterizing,yang2020exploring,hady2017platform,wu2019exploiting,imamura2018reducing,wu2021storage}} \sgi{show}, performing a \SI{4}{\kilo\byte} random read using \sgi{the} Intel Optane SSD is \sgi{approximately} \gfi{6$\times$} faster \gfi{than \sgi{using a traditional NAND-flash-based} SSD}. Second, \gfi{the} Intel Optane SSD can provide 10$\times$ \gfiii{the} endurance \gfi{\gfiii{of} a \sgi{traditional NAND-flash-based} SSD}~\cite{optane_pe_cycles}. \gfi{Third}, \gfi{a} traditional \gfi{NAND-flash-based} SSD \gfi{is} \sgi{approximately} 3$\times$ cheaper than \gfi{the} Intel Optane SSD (\$0.50/GB~\gfi{~\cite{optanePrice}} vs. \$1.5/GB~\gfi{~\cite{ssdprice}}, respectively). 

We compare the number of \gfii{open} Chrome tabs and the tab switch latency when using \gfi{the} Intel Optane SSD \gfi{(Optane configuration)} \gfi{versus} an NVMe NAND\gfi{-}flash\gfi{-}based SSD as \gfi{the} swap device \gfi{(NANDFlash configuration)}. We choose a \SI{16}{\giga\byte} M.2 NVMe NAND\gfi{-}flash\gfi{-}based SSD for our experiments. We also evaluate the effect of enabling Zswap when using the \sgi{NAND-flash-based} SSD \gfi{(NANDFlash+Zswap)}.

\begin{figure}[h]
  \centering
  \includegraphics[width=\linewidth]{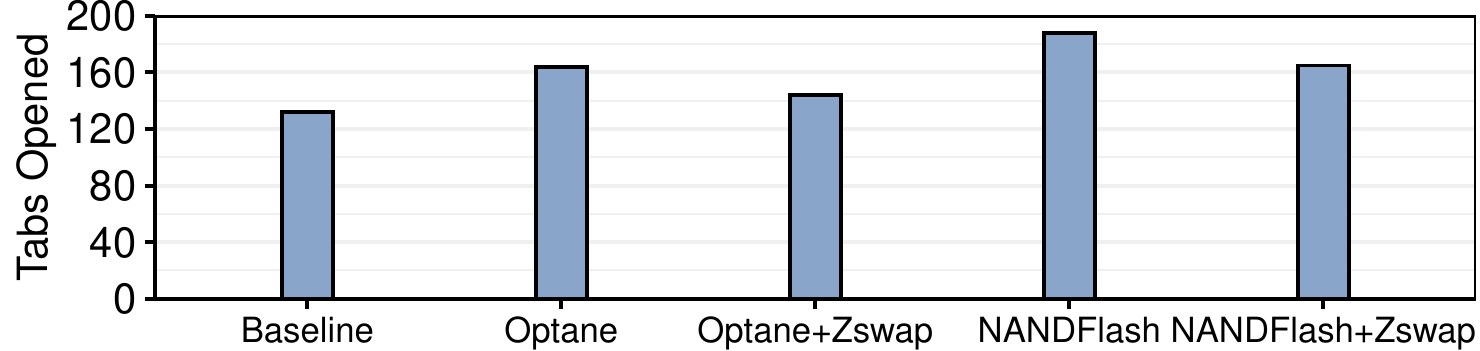}
   \caption{Number of \gfi{open tabs:} Baseline vs.\ Optane vs.\ \gfi{NANDFlash}.}
  \label{fig_slc_tabs_opend}
\vspace{-10pt}   
\end{figure}

\label{r1.6}\Copy{R1/6}{Figure~\ref{fig_slc_tabs_opend} compares the number of \gfii{open} Chrome tabs \gfi{under five} configurations (baseline, Optane, Optane+Zswap, \gfi{NANDFlash}, \gfi{NANDFlash}+ZSwap). We make \gfi{two} observations. First, the \gfi{NANDFlash} configuration \gfi{enables} 14\% more \gfi{open} tabs than \gfi{the} \gfi{Optane configuration}.  \ieeearev{This is due to the \texttt{RAM\_vs\_swap\_weight} parameter. This kernel parameter defines the effort (in terms of swap activity) that the system will \sgfb{demand} to allocate more memory. For our Optane configuration, we empirically choose the \texttt{RAM\_vs\_swap\_weight} value that gives the best trade-off regarding swap activity, number of tabs open, and tab switch latency. However, since we use the NANDFlash configuration only as a reference, we utilize the default \texttt{RAM\_vs\_swap\_weight} value that the kernel suggests. Thus, even though the kernel can allocate more memory in the NANDFlash configuration than in the Optane configuration, more Chrome tabs result in higher swap activity.}} Second, when enabling \gfi{a Zswap cache using 20\% of the DRAM capacity in the NANDFlash configuration (\sgi{the same capacity as} in the Optane+Zswap configuration, we evaluate in Section~\ref{sec_step1_2})}, the number of \gfi{open} tabs reduces by 12\%, due to \gfi{the} decrease in the \gfi{available} \gfi{DRAM} space.  

\begin{figure}[ht]
\begin{subfigure}{\linewidth}
  \centering
  \includegraphics[width=\linewidth]{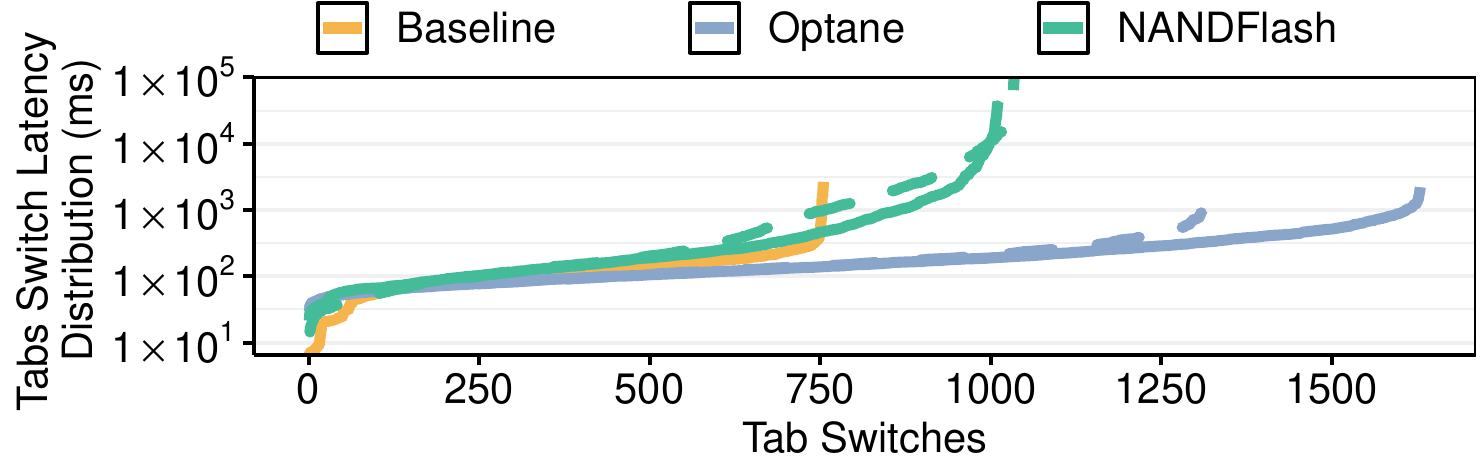}  
  \vspace{-15pt}
  \caption{Tab switch latency distribution. Dashed lines represent Zswap enabled.}
  \label{fig_slc_tabs_switch_latency}
\end{subfigure}
\par\bigskip
\begin{subfigure}{\linewidth}
  \centering
  \includegraphics[width=\linewidth]{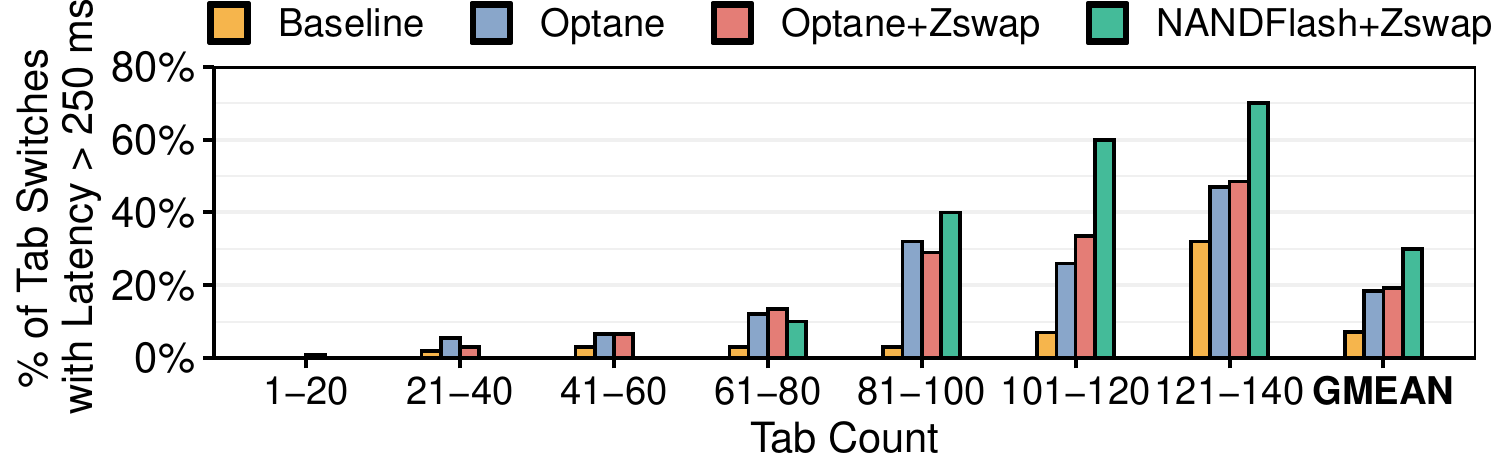} 
  \caption{High-latency tab switch distribution.}
  \label{fig_slc_tabs_switch_latency_distribution}
\end{subfigure}
\caption{Tab switch latency:  Baseline vs.\ Optane vs.\ \gfi{NANDFlash}.}
\label{fig_slc_latency}
\end{figure}

\gfi{Figure~\ref{fig_slc_latency} shows the tab switch latency distribution \gfi{(Figure~\ref{fig_slc_tabs_switch_latency})} and the distribution of high-latency tab switches (Figure~\ref{fig_slc_tabs_switch_latency_distribution}) for all \gfi{five} configurations. We make two observations \sgi{from the figure}. First, we observe (in Figure~\ref{fig_slc_tabs_switch_latency}) that the high access latency of the NAND-flash-based SSD leads to a significant increase in the tab switch latency in the NANDFlash (NANDFlash+Zswap) configuration compared to the Optane (Optane+Zswap) configuration\gfii{. The average tab switch latency of the NANDFlash (NANDFlash+Zswap) \sgii{configuration} is \gfii{3.6$\times$ (10$\times$)} that of the Optane (Optane+Zswap) configuration.} Second, the number of high-latency tab switches in the NANDFlash configuration (Figure~\ref{fig_slc_tabs_switch_latency_distribution}) \sgi{increases} by 35\% compared to the baseline configuration, and by 39\% \sgi{compared} to the Optane configuration. For \sgi{high tab} counts, the fraction of tabs with high latency is as \sgi{much} as 70\% (for 100--120 tabs) in the NANDFlash+Zswap configuration.} \gfii{We conclude that the NANDFlash configuration provides benefits compared to the baseline configuration, but it \sgii{is unable to approach the performance of} the Optane configuration.}

\gfii{\sgii{Summarizing our findings,} using \sgii{a} state-of-the-art NAND-flash-based SSD to \gfiii{extend} the main memory capacity of the system \sgi{improves performance} compared to the baseline\gfii{, considering the number of \gfiii{additional} open tabs the NANDFlash configuration provides}. However, it cannot \sgi{achieve similar performance \gfiii{as using}} the Intel Optane SSD \gfiii{to extend the main memory capacity} \gfii{due to the \sgii{high access latency of the} NAND-flash-based SSD}, \gfii{which leads to a significantly \sgii{larger} number of high-latency tab switches} \gfiii{than the Intel Optane SSD}. Thus, the \gfiii{Intel} Optane SSD can lead to a better user experience than a \sgii{state-of-the-art NAND-flash-based} SSD.}

%% file: sections/4-system-optimization.tex
\section{System Optimization}
\label{sec_step2}

\gfi{Using the} Intel Optane SSD \gfi{as swap space} allows our system to enjoy \gfi{an extended} memory space, which translates to an average latency improvement for our workload at the cost of \gfi{larger} tail \gfi{latencies}. \gfi{However, longer tail latencies are usually not acceptable. Thus, it is essential} to reduce \gfi{tail latency (i.e., the 99th-percentile latency in our tab switch latency distributions)} for interactive workloads since it affects how the user \gfi{experiences} the system. 

\gfi{T}he goal of this section is to analyze the primary sources of latency \gfi{overheads} that impact the tail \gfi{latency} in the system when \gfi{we} \sgi{make use of the} Intel Optane SSD \gfi{as swap space}. We extensively profile the system when executing the Google Chrome web \gfi{browser} to identify performance bottlenecks caused by the added swap device. We observe that the \sgi{Linux} block I/O layer increases both the average and the 99th\gfi{-percentile} latency for Chrome's page faults significantly, mostly due to I/O scheduling \gfi{issues} and queuing \gfi{delays}, and overheads related to the I/O \gfi{request} completion mechanism. \gfi{To solve these issues}, we tune the system parameters related to the \sgi{Linux} block I/O layer, aiming to improve \sgi{the} 99th\gfi{-percentile} latency \gfi{for} tab \gfi{switches}. 

In this section, we limit the number of open tabs in our experiments to 50, as \gfi{50 tabs are} enough to generate moderate memory \gfii{capacity} pressure in our system \gfi{and thus examine tail latencies}.

\subsection{Profiling the Chrome Browser}
\label{sec_profiling_chrome}
\begin{figure}[ht]
  \centering
  \includegraphics[width=\linewidth]{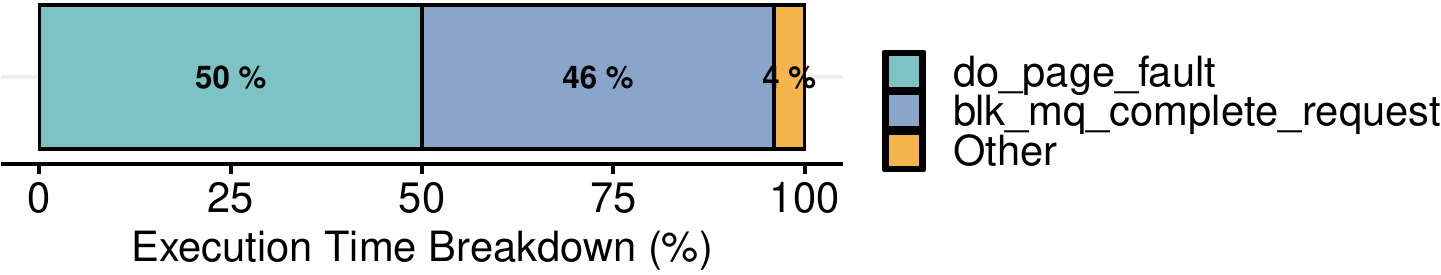}
  \caption{\texttt{perf} results for a Chrome tab.}
 \label{fig_perf_chrome}
 \vspace{-10pt}
\end{figure}

We extensively profile \sgi{the} Google Chrome \gfi{browser's} activity while running the memory \gfi{capacity} pressure test \gfi{when the Intel Optane SSD is \sgi{employed} as swap space}. We use the \texttt{perf} profiling tool~\gfi{\cite{de2010new} to collect the execution time breakdown of each Chrome tab (including kernel activity).}  Figure~\ref{fig_perf_chrome} shows a simplified execution breakdown of one of the \gfi{representative} \gfi{Chrome} tabs that \gfi{demonstrates \sgi{long-latency} switching times}. We observe from the figure that the tab spends more than 96\% of its execution \sgi{time} on kernel modules that manage I/O \gfi{requests (i.e., the \textit{do\_page\_fault} kernel function, which issues I/O requests in case of a page fault operation; and the \textit{blk\_mq\_complete\_request} kernel function, which receives and \gfiii{completes} the processing of I/O requests issued to the swap device)}. We observe that \gfi{(i)}~50\% of the execution \gfiii{time} is spent on issuing a block I/O request due to page fault\sgi{s}\gfi{, and (ii)}~46\% of the \gfi{execution} \sgi{time} is spent on processing the requested page once the data is received from the swap device. The remaining 4\% of the execution time is spent on Chrome's internal processes and other kernel calls. Therefore, \gfi{we conclude that} the block I/O layer is the \emph{primary} source of \gfi{the tail latency} overhead.

\subsection{{Linux} Block I/O Layer}

\begin{figure}[ht]
  \centering
  \includegraphics[width=\linewidth]{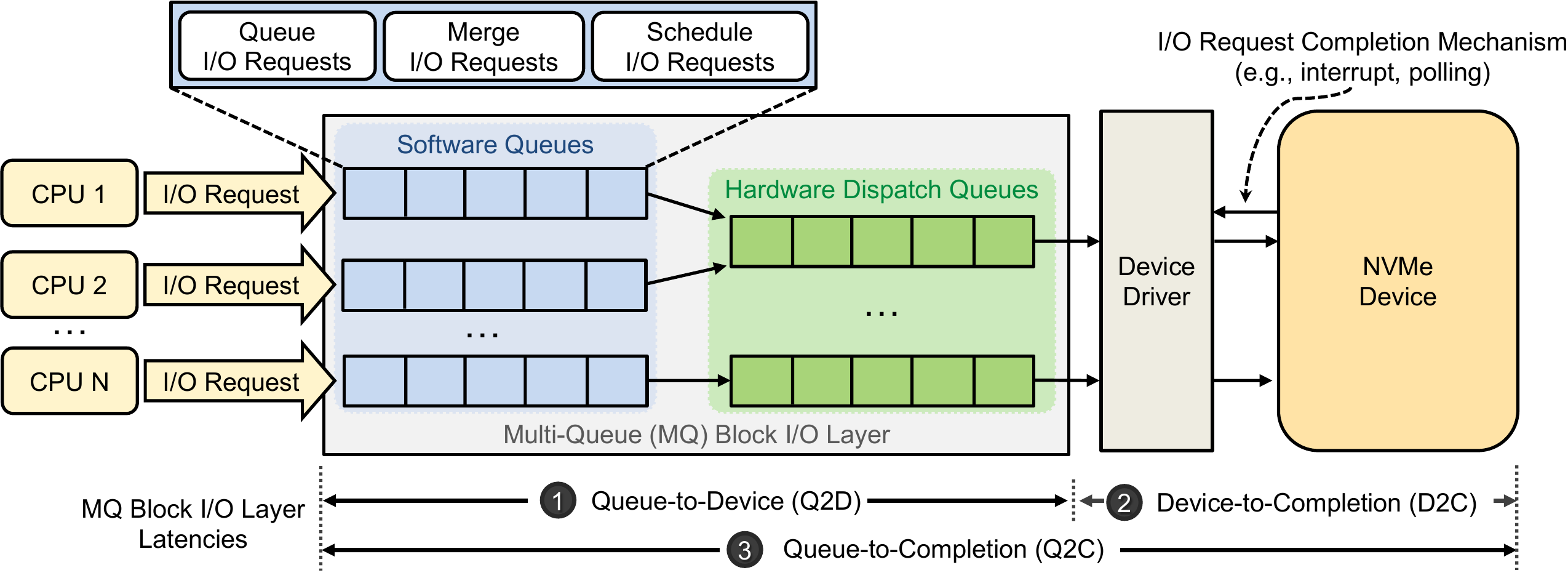}  
  \caption{\gfii{Linux block} I/O layer.}
  \label{fig_block_io_layer}
\end{figure}

The block I/O layer is the \gfi{Linux} kernel layer responsible for managing block I/O devices (e.g., \sgi{magnetic hard drives}, SSDs)~\cite{bovet2005understanding,chen2012energy}. It is a key \gfi{system} component since accessing block \gfi{I/O} devices involves issuing high\gfi{-}latency and power-hungry operations to the block device. Therefore, the block I/O layer is highly optimized to ensure low latency and high throughput from block devices. Figure~\ref{fig_block_io_layer} illustrates the primary operations the block I/O layer \sgi{performs} when an application or the kernel issues an I/O request (e.g., read from a file, page fault). \gfi{The block I/O layer} works in three main steps. 

\begin{figure*}
\Copy{R1/7}{
\begin{minipage}{\linewidth}
    \captionsetup{type=figure} 
    \centering
    \begin{tabular}{c}
       \includegraphics[width=\linewidth]{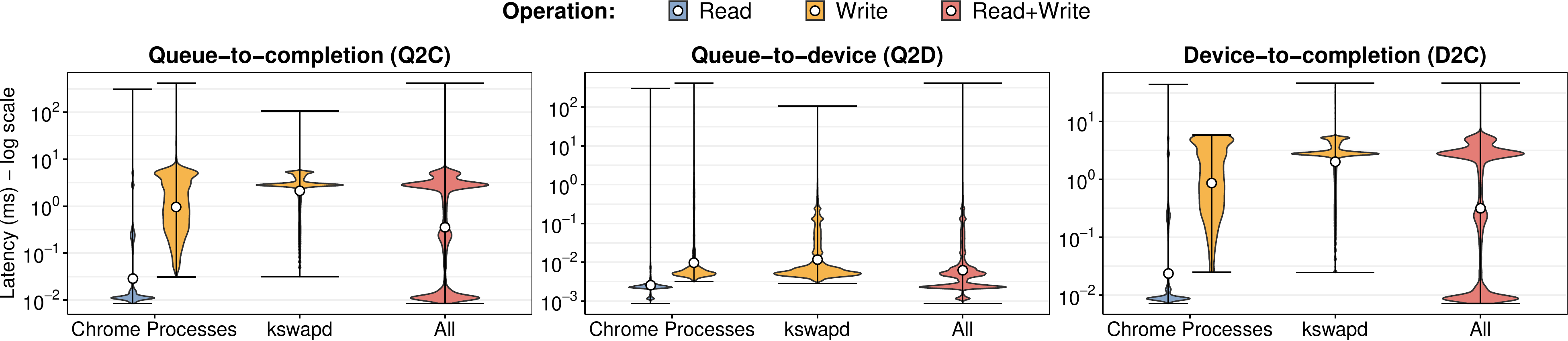}
    \end{tabular}
    \caption{\gfiv{\texttt{Q2C}, \texttt{Q2D}, and \gfiv{\ieeearev{{\texttt{D2C}}}} \gfiv{latency} \gfiii{distribution} for (i) Chrome processes, (ii) \texttt{kswapd}, and (iii) Chrome processes+\texttt{kswapd} (\emph{All}) \gfiv{during the execution of the memory capacity pressure test}.} \gfiii{Error bars depict the minimum and maximum data point values, and a bubble depicts average value of each category.}} 
    \label{fig_blk_trace_chrome}

  \end{minipage}
  }
\end{figure*}



First, when the block I/O layer receives a block I/O request, \sgi{it queues} the request in a request queue \gfii{(called \emph{software queue})}, which is unique per \gfii{CPU}. \gfi{Second}, it attempts to merge and sort requests based on the sector number to avoid \gfi{costly} seek operations \gfi{in the device}. \gfi{Third}, it schedules the requests \gfi{using} \gfi{an} I/O scheduler. The scheduled request is \gfii{stored in a dispatch queue (called \emph{hardware dispatch queue}). Requests stored in the hardware dispatch queue are} issued to the device driver and eventually reach the block device. Once the device completes executing the request, it sends the response back to the block I/O layer, which finalizes the execution of the request by either \sgi{(1)~}waking up the requester, in case an \gfi{interrupt-based} (IRQ) \gfi{I/O request} completion mechanism is employed~\gfi{\cite{bovet2005understanding}}; or \sgi{(2)~forwarding} the response to the requester, in case a polling-based \gfiii{I/O request} completion  mechanism is employed~\gfi{\cite{bovet2005understanding}}. \gfi{Recent} Linux kernels have adopted a \sgi{multi}-queue \sgi{(MQ)} block I/O layer~\gfi{\cite{bjorling2013linux}}, since modern block devices (e.g., NVMe devices~\gfi{\cite{tavakkol2018flin}}) can execute many requests \gfi{concurrently}~\gfi{\cite{bjorling2013linux,tavakkol2018mqsim,tavakkol2018flin}}. \gfi{T}he \sgi{MQ block} I/O layer employs one request queue per CPU core for each block device.

To monitor and analyze the performance implications of the block I/O layer in the system, we make use of the \texttt{blktrace} tool~\cite{brunelle2007blktrace}. \texttt{blktrace} monitors the activity of the block I/O layer, and provides detailed \sgi{timing} information for each \gfi{major} operation. We analyze three important \sgi{timings that the tool reports}: \gfi{\emph{(i) queue-to-device (\texttt{Q2D})}}, the time from when a block I/O request enters the block I/O layer \gfi{to} the time the request is issued to the block device, including queuing, merging, and scheduling \gfii{(\ding{182} in Figure~\ref{fig_block_io_layer})};  \gfi{\emph{(ii) device-to-completion (\texttt{D2C})}}, the time it takes the block device to complete the request \gfii{(\ding{183} in Figure~\ref{fig_block_io_layer})}; and  \gfi{\emph{(iii) queue-to-completion (\texttt{Q2C})}}, the total \gfi{end-to-end} time for a block I/O request to complete\gfi{, i.e., \texttt{Q2C} = \texttt{Q2D} + \texttt{D2C}} \gfii{(\ding{184} in Figure~\ref{fig_block_io_layer})}.

During the execution of the memory \gfii{capacity} pressure test, we profile the block I/O layer using \texttt{blktrace}. We analyze the latencies for I/O requests that Chrome processes and the kernel memory management \gfi{unit} (\texttt{kswapd}) issue. Figure~\ref{fig_blk_trace_chrome} shows the \gfi{\texttt{Q2C},} \texttt{Q2D}, \gfi{and} \texttt{D2C}  latencies during the execution of the test. The \gfi{end-to-end block I/O latency (i.e., }\texttt{Q2C} latency\gfi{)} is \SI{1.80}{\milli\second}, on average \gfiii{across both Chrome processes and \texttt{kswapd} read and write I/O requests} (min. of \SI{0.0085}{\milli\second}, max. of \SI{414.11}{\milli\second}). The \texttt{Q2D} latency is \SI{0.03}{\milli\second}, on average (min. of \SI{0.00087}{\milli\second}, max. of \SI{413.80}{\milli\second}). The \texttt{D2C} latency is \SI{1.78}{\milli\second}, on average (min. of \SI{0.00723}{\milli\second}, max. of \SI{46.10}{\milli\second}). We make \sgi{three observations from the reported latencies}. First, Chrome \gfi{processes} (\texttt{chrome}, \texttt{Chrome\_IOThread}, and \texttt{CompositorTileW}) issue \sgi{high-latency} I/O requests (up to \SI{414.11}{\milli\second}). Most of \sgi{these} requests are read requests caused by page faults. Second, \gfi{\texttt{kswapd} is responsible \sgi{for} issuing \gfiv{a} majority of \gfiv{the} write request\gfiv{s} to the I/O device.}\footnote{\gfiv{The \texttt{kswapd} process issues \emph{only} write I/O requests since its goal is to free memory by reclaiming inactive memory pages. 
\juan{If} an inactive memory page is dirty, the \texttt{kswapd} process writes 
\juan{this dirty page} to swap space.}} Third, \gfiii{for high-latency I/O requests, }most of the I/O latency comes from the block I/O layer \gfi{rather than from the swap device}. While the device latency \gfi{(i.e., \texttt{D2C})} is at most \gfiii{\SI{46.10}{\milli\second}}, the latency of queuing and scheduling requests in the block I/O layer \gfi{(i.e., the \texttt{Q2D} latency)} dominate\gfi{s} the execution time of \gfiii{high-latency I/O} requests (\gfi{as observed by the high maximum \texttt{Q2D} \gfiii{latency} of \SI{413.80}{\milli\second}}).

\gfiii{Based on this analysis, we conclude that the block I/O layer operations are the \sgiii{primary bottleneck} in high-latency block I/O requests.} To alleviate this bottleneck, we investigate how two system optimizations impact block I/O latencies and, consequently, Chrome performance. \gfi{We investigate the effect of (1) different block I/O schedulers \gfi{(Section~\ref{sec_io_schedulers})} and (2) different \gfi{I/O request} completion mechanisms \gfi{(Section~\ref{sec_io_completition})} on system's performance.}

\subsection{Optimization 1: Block I/O Schedulers}
\label{sec_io_schedulers}

\begin{figure*}[t!]
  \centering
  \includegraphics[width=\linewidth]{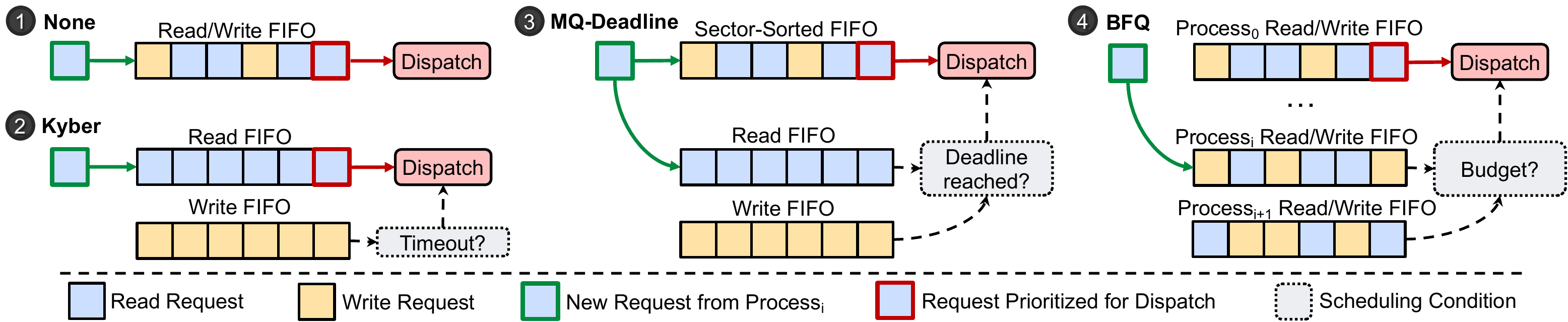}  
  \caption{\gfiv{Four} \gfiii{Linux block} I/O layer schedulers.}
  \label{fig_io_schedulers}
\end{figure*}

The \sgiii{block} I/O layer \gfiii{of the Linux kernel} provides four different \gfiv{multi-queue} I/O schedulers. \gfiii{These} schedulers vary in complexity, and \gfiv{their aim and ability}  to \gfiii{consider and exploit} different properties of \gfiii{different} block device\gfiii{s}. Therefore, it is essential to tailor the system to make use of the I/O scheduler that matches \gfiii{the} requirements \gfiii{and characteristics of the Intel Optane SSD}. 

\gfiii{We first briefly explain each of the four I/O schedulers. The four I/O schedulers are called None~\cite{bjorling2013linux}, Kyber~\cite{blkmqKyb12}, MQ-Deadline~\cite{mqdeadline}, and budget-fair queuing (BFQ)~\cite{bovet2005understanding}. Figure~\ref{fig_io_schedulers} illustrates the operation of the four I/O schedulers.} The None \gfiii{I/O} scheduler (\ding{182} in Figure~\ref{fig_io_schedulers}) is the simplest I/O scheduler. It employs a simple first-\sgiii{in} first-out (FIFO) request queue. Therefore, it does \gfiii{\emph{not}} reorder requests. The request queue includes both \sgii{read and write} \gfiii{I/O requests}. Due to its simplicity, \gfiii{the None I/O scheduler} incurs low overhead, \sgiii{but} does not guarantee any quality-of-service.

The Kyber \gfiii{I/O} scheduler (\ding{183} in Figure~\ref{fig_io_schedulers}) maintains two separate request queues, one for synchronous block \gfiii{I/O} read requests, and another for asynchronous \gfiii{block} \gfiii{I/O} write requests.  It prioritizes requests in the read queue \gfiii{over those in the write queue}\gfiv{, unless a write request has been outstanding for too long, i.e., it times out by reaching the target write access latency} \ieeea{({\gfiv{the default target write access latency for the Kyber I/O scheduler is \SI{10}{\milli\second}~\cite{blkmqKyb12})}}.} \gfiii{As such}, it is also a simple I/O scheduler \gfiv{that aims to provide better service to read requests.} 

The MQ-Deadline \gfiii{I/O} scheduler (\ding{184} in Figure~\ref{fig_io_schedulers}) employs three different request queues: read FIFO, write FIFO, and a sorted FIFO. The sorted FIFO maintains read and write \gfiii{I/O} requests \gfiii{that} are sorted by the sector number \gfiii{they are to access}. The scheduler prioritizes \gfiii{I/O} requests from the \gfiv{sector-}sorted \gfiii{FIFO} unless \sgiii{any} request in either the read or write queue \gfiii{is} about to \gfiii{violate its service} deadline. The default deadline is \SI{500}{\milli\second} \sgiii{for read \gfiii{I/O} requests, and \SI{5}{\second} for write \gfiii{I/O} requests}. 

\gfiii{T}he BFQ \gfiii{I/O} scheduler (\ding{185} in Figure~\ref{fig_io_schedulers}) is the most complex I/O scheduler \gfiii{among} \gfiii{the} four, which \gfiii{leads to} 
high \gfiii{scheduling} overhead. \gfiv{
\juan{The BFQ I/O scheduler guarantees} fairness \juan{across processes} by distributing the throughput of the block I/O device proportionally to each process via an indirectly assigned weight value. The BFQ I/O scheduler} employs a\gfiii{n I/O} request queue per process \gfiv{and assigns} \gfiii{an} I/O budget per \gfiv{I/O request queue}. \gfiv{It assigns I/O budgets, measured in number of sectors, proportionally to a process's I/O activity. 
\juan{This way,} I/O-bound processes with sequential I/O requests are assigned a large I/O budget, while processes with short and sporadic I/O requests are assigned a small I/O budget}. \gfiv{
\juan{The BFQ I/O scheduler} uses a variant of the worst-case fair weighted fair queuing+ (WF2Q+) scheduling algorithm~\cite{bennett1997hierarchical} to select an I/O request queue to be serviced (
\juan{typically} the I/O request queue with the lowest I/O budget). The selected I/O request queue is} \gfiii{prioritized, \gfiv{and its I/O requests are exclusively serviced}} until its I/O budget finishes. \gfiv{
\juan{As a result,} the BFQ I/O scheduler guarantees a fraction of the device throughput to each process.}  The \gfiv{Linux} kernel \gfiv{
uses} the BFQ I/O scheduler \juan{by default,} since it usually provides high system responsiveness and fairness\gfiii{, even though its scheduling decision overhead is higher than the other three I/O schedulers}.

\begin{figure}[ht]
\vspace{-5pt}
\begin{subfigure}{\linewidth}
  \centering
  \includegraphics[width=\linewidth]{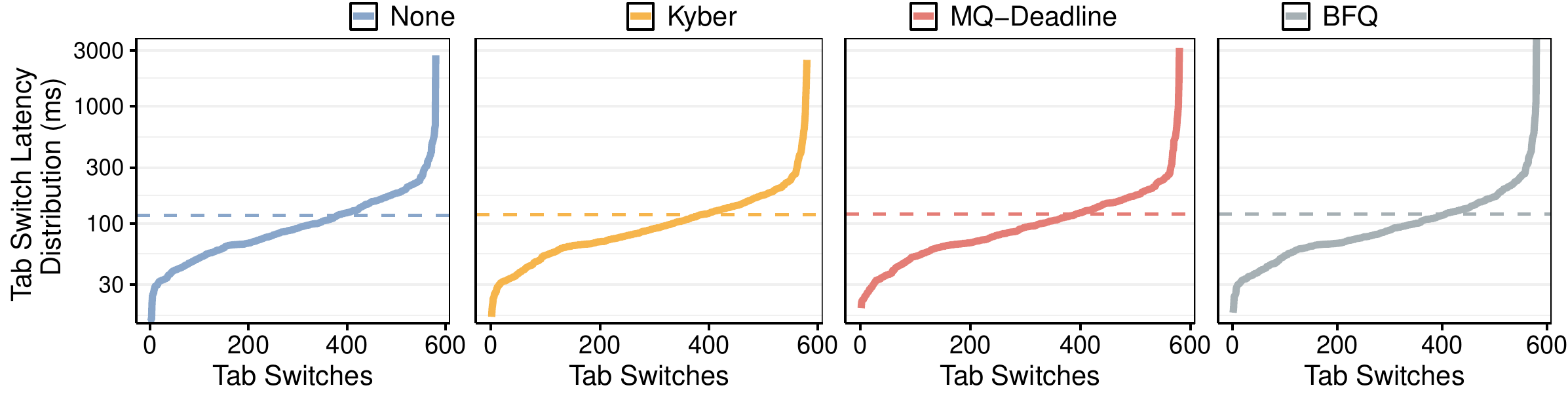}  
  \caption{Tab switch latency distribution. Horizontal dashed lines represent average values. \vspace{10pt}
}
  \label{fig_tab_switch_io_schedulers}
\end{subfigure}
\begin{subfigure}{\linewidth}
  \centering
  \includegraphics[width=\linewidth]{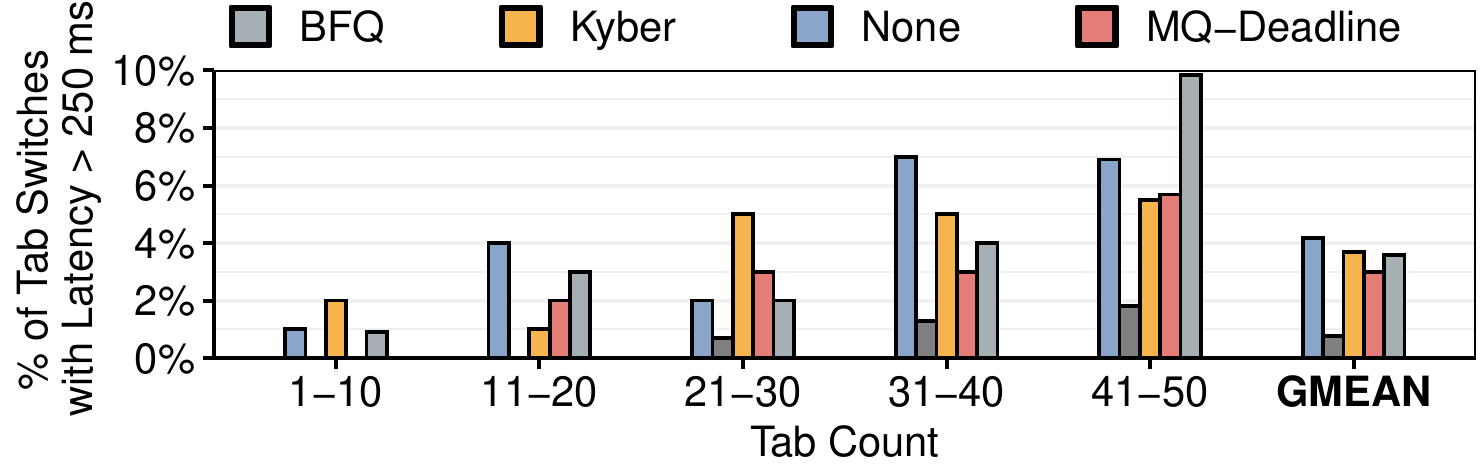}  
  \caption{High-latency tab switch distribution.}
  \label{fig_high_latency_tab_switch_io_schedulers}
\end{subfigure}
\caption{Tab switch latency: Optane with different I/O schedulers.}
\label{fig_latency_io_schedulers}
\vspace{-10pt}
\end{figure}

 Figure~\ref{fig_tab_switch_io_schedulers} shows the tab switch latency for the \gfiii{Optane} configuration \gfiii{under the four different \gfiii{I/O} schedulers}. We observe that, on average, the \gfiii{four} different I/O schedulers provide similar tab switch latencies. The average tab switch latency for the \gfiii{None, Kyber, MQ-Deadline, and BFQ I/O schedulers is \SI{116}{\milli\second}, \SI{119}{\milli\second}, \SI{120}{\milli\second}, and \SI{118}{\milli\second}}, respectively. However, when we examine the \gfiii{fraction} of \gfiii{high-latency} tab switches in Figure~\ref{fig_high_latency_tab_switch_io_schedulers}, we observe that the \gfiii{fraction of high-latency tab switches} increases \gfiii{significantly} with \gfiii{open} tab count for the default BFQ I/O scheduler. \gfiii{The BFQ I/O scheduler leads to} the largest number of high-latency tab switches when the system has 41--50 tabs open, while \gfiii{the Kyber I/O scheduler} provides the lowest number of high-latency tab switches for the \gfiii{same} tab-count range\gfiii{,} reducing the \gfiii{fraction} of high-latency tab switches by \gfiii{44}\% compared to the BFQ I/O scheduler. \gfiii{Therefore, \gfiv{to reduce} \gfiii{Chrome's} tail latency, \gfiii{the Kyber I/O scheduler \sgiii{can potentially} be a better \gfiii{I/O} scheduler than the default BFQ I/O scheduler.}}

\begin{figure}[ht]
\vspace{-5pt}
 \centering
     \includegraphics[width=\linewidth]{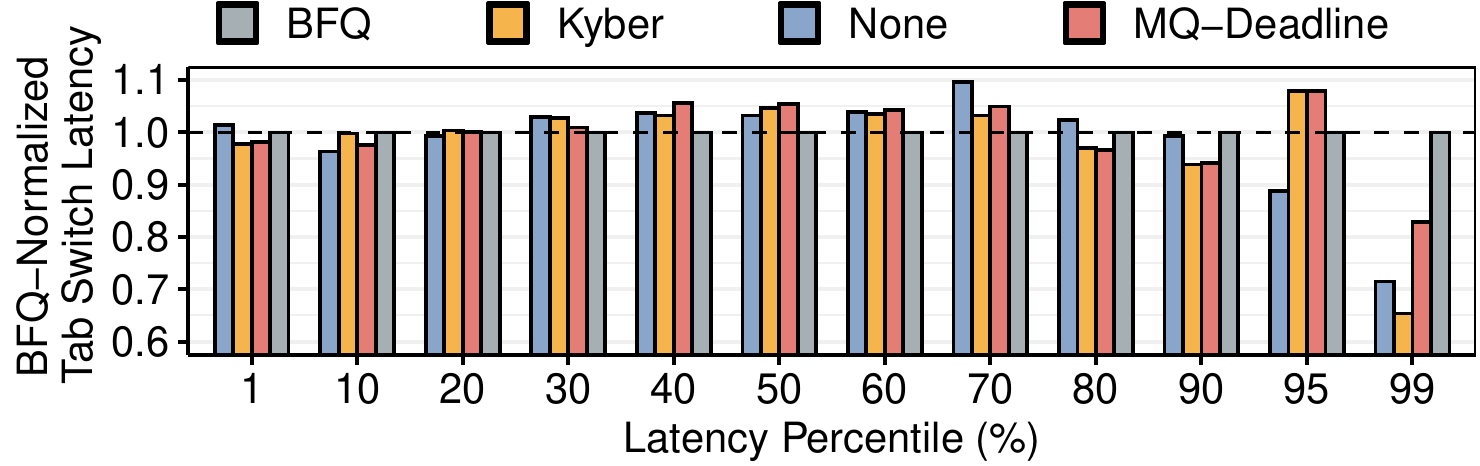}
  \caption{\gfiii{Normalized t}ab switch \gfiv{latency} \gfiii{of four} schedulers\gfiv{, categorized across different latency percentiles}. \gfiii{Y-axis is} normalized to the default BFQ \gfiii{I/O} scheduler.}
  \label{fig_normalized_latency_vs_schedulers}
  \vspace{-5pt}
\end{figure}

We \gfiii{further} analyze the tab switch latency distribution in Figure~\mbox{\ref{fig_normalized_latency_vs_schedulers}} for the \gfiii{four} I/O schedulers. \gfiii{The figure shows the tab switch latency percentiles (along the x-axis) and the corresponding tab switch latency (y-axis), normalized to the values when the system employs the default BFQ I/O scheduler.} We make three observations. First, at the tail \sgiii{(99th-percentile) latency}, the  \gfiii{None, Kyber, and MQ-Deadline I/O schedulers} significantly reduce the tab switch latency compared to the default BFQ I/O scheduler, \gfiii{by 29\%, 35\%, and 18\%,} respectively. Second, the BFQ I/O scheduler provides the lowest tab switch latency for latency percentiles outside the tail. The \gfiii{None/Kyber/MQ-Deadline} I/O scheduler\gfiii{s} slightly increase  BFQ's tab switch latency by \gfiii{3\%/3/\%/10\%}, 4\%/4\%/4\%, and \gfiii{5\%/4\%/5\%} \gfiii{at} the 50th-, 60th-, and 70th-\sgiii{percentile latencies}, respectively. Third, when moving closer to the latency percentiles at the tail (i.e., from the 80th-\sgiii{percentile latency to the 95th-percentile latency}), we observe that the \gfiii{None, Kyber, and MQ-Deadline} I/O schedulers \gfiv{all} provide lower tab switch \gfiii{latencies} than the BFQ I/O scheduler. By employing the Kyber/MQ-Deadline I/O schedulers, the tab switch latency reduces by 3\%/4\% and 7\%/6\% compared to the BFQ I/O scheduler for the 80th- and 90th-\sgiii{percentile latencies}. \sgiii{At the 95th\gfiii{-percentile latency}}, the None I/O scheduler reduces the tab switch latency by 12\% compared to the BFQ I/O scheduler. Therefore, we conclude that even though the default BFQ I/O scheduler provides good performance \gfiii{overall (except at the tail)}, alternative I/O schedulers can \gfiii{greatly} improve \gfcr{tail latency} performance\gfiii{, thereby improving user experience}.

\begin{figure}[ht]
\vspace{-7pt}
 \centering
     \includegraphics[width=\linewidth]{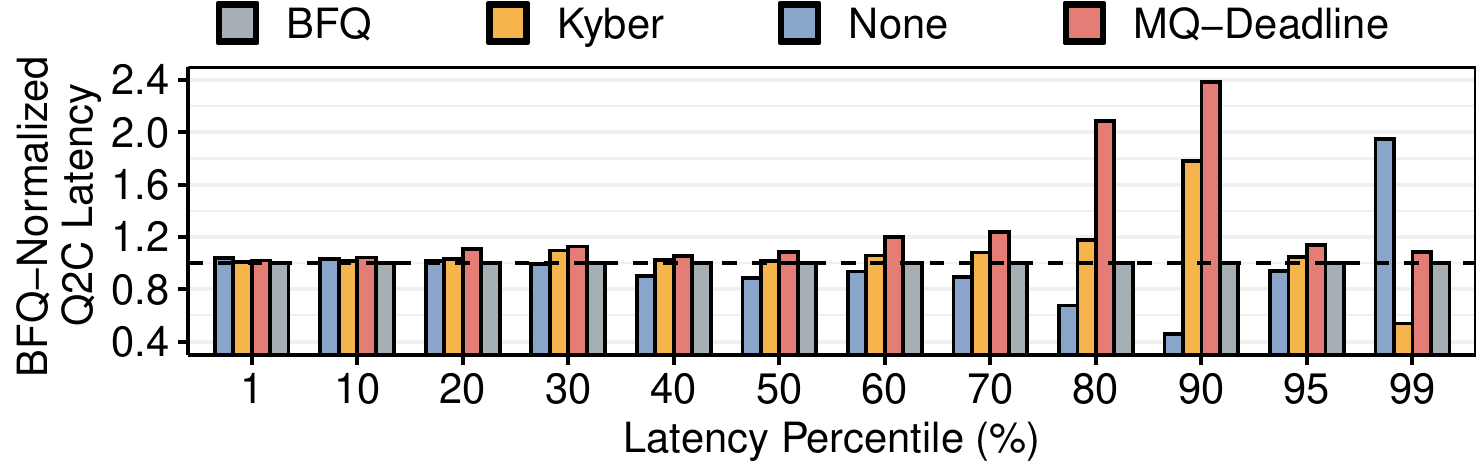}
   \caption{\texttt{Q2C} latency \gfiii{of four} I/O schedulers. \gfiii{Y-axis is} normalized to the default BFQ \gfiii{I/O} scheduler.} 
   \label{fig_q2c_io_schedulers}
   \vspace{-5pt}
\end{figure}

To further understand the I/O schedulers' impact on Chrome \gfiii{browser} performance, we analyze the \texttt{Q2C} \gfiii{latency (i.e., the end-to-end I/O request latency)} for each I/O \sgiii{scheduler}. \sgiii{Figure}~\ref{fig_q2c_io_schedulers} depicts the \gfiii{end-to-end I/O request (i.e.,} \texttt{Q2C}\gfiii{)} latency \sgiii{percentiles} (\sgiii{along} the x-axis) and the corresponding \texttt{Q2C} latency (y-axis), normalized to the \gfiv{\texttt{Q2C} latency} values for the BFQ \gfiii{I/O} scheduler. We make three observations. First, the None \gfiii{I/O} scheduler reduces \texttt{Q2C} \gfiii{latency} in \gfiii{\gfiv{a} majority of the} latency \gfiii{percentiles}. However, such a reduction does \gfiii{\emph{not}} directly translate to tab switch latency improvements \gfiii{(as seen in Figure~\ref{fig_normalized_latency_vs_schedulers})}. \gfiii{This happens because the None I/O scheduler does not enforce any ordering among read and write I/O requests. Since (i)~Chrome mostly issues read I/O requests during the execution of our test (as Figure~\ref{fig_blk_trace_chrome} shows) and (ii)~the Intel Optane SSD \sgiii{internally handles read and write requests} equally (as characterized by prior work~\cite{wu2019towards}), the None I/O scheduler delays the execution of Chrome's read I/O operations \gfiii{by executing write I/O requests}, which hurts Chrome's performance.} Second, the Kyber \gfiii{I/O} scheduler provides the best \texttt{Q2C} reduction for the 99th-percentile latency, which \gfiii{directly} translates to better tab switch \gfiii{latency (Figure~\ref{fig_normalized_latency_vs_schedulers})}. \gfiii{The Kyber I/O scheduler \sgiii{improves Chrome's} performance since it better fits Chrome's I/O request characteristics by: (i)~reducing \gfiii{the average I/O queuing, merging, and scheduling latencies (i.e., \texttt{Q2D} latencies) by \gfiii{up to 3.7$\times$ that of the BFQ I/O scheduler} \gfiv{(not shown)}; and (ii)~\sgiii{prioritizing} read I/O requests over writes. }} Third, even though Kyber's \texttt{Q2C} latency is larger than BFQ's \texttt{Q2C} latency (Figure~\mbox{\ref{fig_q2c_io_schedulers}}) at the 90th\gfiii{-}percentile \gfiii{latency}, Kyber's \texttt{Q2C} latency for read requests is slightly lower (by 15\%) than BFQ's \texttt{Q2C} latency for read requests \gfiv{(not shown)}, which translates to a faster tab switch latency \gfiii{for Kyber at the 90th-percentile latency}. However, at the 95th\gfiii{-}percentile \gfiii{latency},  Kyber's \texttt{Q2C} latencies for both read and write requests are larger than \sgiii{BFQ's} \texttt{Q2C} latencies. At the 99th\gfiii{-}percentile \gfiii{latency}, the system's high memory \gfiii{capacity} pressure highlights the high overhead \gfiii{of} the BFQ \gfiii{I/O} scheduler, leading to a significant increase in BFQ's \texttt{Q2C} latency for both read and write \gfiv{requests.} \gfiii{BFQ's \texttt{Q2C} latency increases by 99\% from the 95th\gfiii{-}percentile \gfiii{latency} to the 99\gfiii{-}percentile \gfiii{latency}}. \gfiii{In contrast}, \gfiii{Kyber's} \texttt{Q2C} latency increases by only 19\% when moving from the 95th\gfiii{-}percentile \gfiii{latency} to the 99\gfiii{-}percentile \gfiii{latency}. \gfiii{W}e conclude that the Kyber I/O scheduler can reduce tail latency for \gfiii{the Chrome browser} since it (i)~provides low overhead I/O scheduling decisions for already critical I/O requests while (ii)~matching the access pattern of our workload by prioritizing read accesses over writes. 

\begin{figure}[ht]
\vspace{-7pt}
 \Copy{R2/4Bfig}{
 \centering

     \includegraphics[width=\linewidth]{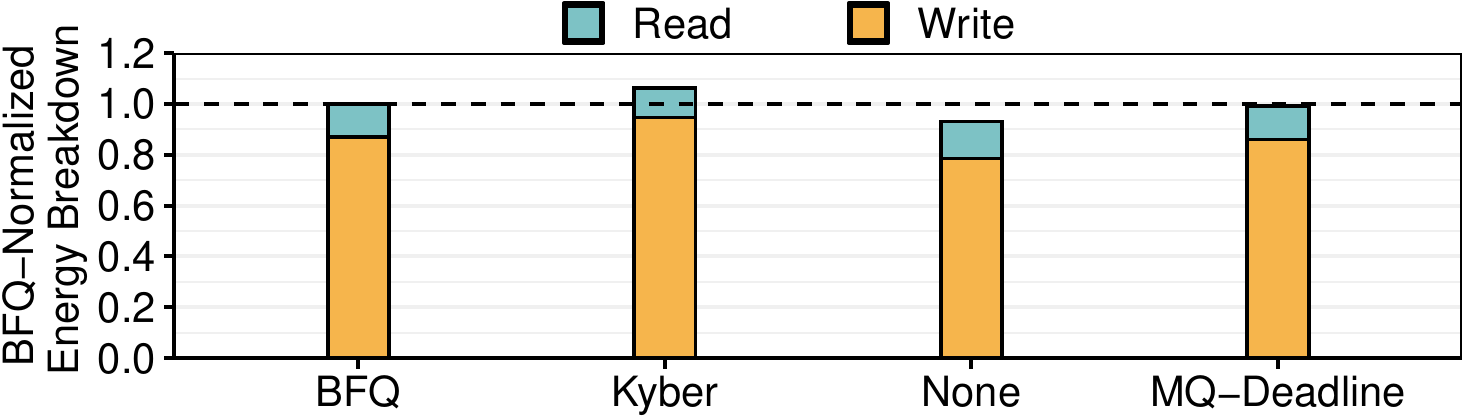}
     
   \caption{\ieeearev{Energy consumption: Optane with different I/O schedulers. Y-axis is normalized to the default BFQ I/O scheduler.}} 
   \label{fig_energy_io_schedulers}
   }
   \vspace{-5pt}
\end{figure}

\label{r2.4b}\Copy{R2/4B}{\noindent \textbf{Energy Analysis.} \ieeearev{Figure~\mbox{\ref{fig_energy_io_schedulers}} compares the impact of the different I/O schedulers (x-axis) on the average memory subsystem energy consumption \sgfb{(which includes the energy consumption of main memory and swap space)} for the Optane configuration (y-axis; normalized to the baseline BFQ I/O scheduler). We use the same energy model described in Section~\mbox{\ref{sec_step1_1}} for our analysis. We make two key observations from the figure. First, we observe from the figure that all four I/O scheduler mechanisms achieve a \emph{similar} memory subsystem energy consumption during the execution of our test. Second, the Kyber I/O scheduler slightly \emph{increases} energy consumption by 6.9\%, while the None I/O scheduler slightly \emph{decreases} energy consumption by 6.3\% compared to the baseline BFQ I/O scheduler. This is due to an increase in the number of write I/O requests the Kyber I/O scheduler produces compared to the None I/O scheduler. Recall that the Kyber I/O scheduler uses dedicated queues for read and write quest and dispatch write requests using a pre-defined threshold, while the None I/O scheduler uses a single queue for read and write requests and dispatch both read/write requests using a first-in-first-out approach. This leads to the Kyber I/O scheduler prioritizing more write requests from background kernel processes than the None I/O scheduler and the baseline BFQ mechanism. In contrast, the None I/O scheduler serves the processes that generate an I/O request first (in our case, Chrome processes issuing read I/O requests due to moderate-to-high swap activity). We conclude that since the available I/O schedulers mostly target improving the throughput of block I/O devices, they achieve a similar memory subsystem energy consumption.}  }

\gfiii{\gfiv{We conclude that,} we can reduce \gfiv{especially} the block I/O \gfiv{tail} latency by employing different I/O schedulers \gfiii{(e.g., Kyber)} \gfiv{from the default Linux block I/O scheduler}. However,} the high latency overhead related to managing I/O requests \gfiii{is} still large in comparison to the actual device time \gfiii{(as Figure~\ref{fig_blk_trace_chrome} shows)}. \gfiv{Therefore, we evaluate a second optimization in Section~\ref{sec_io_completition}.} 

\subsection{Optimization 2: {Interrupts} vs.\ Polling {Based I/O Request Completion}}
\label{sec_io_completition}

Another key component of the \sgiii{Linux} block I/O layer that directly impacts I/O performance is the I/O \gfiii{request} completion \gfiii{mechanism}. There are two main I/O \gfiii{request} completion \gfiii{mechanisms} in current Linux systems: 
\sgiii{(1)~}\gfiii{interrupt-based (\gfiii{i.e.,} IRQ-based)} \gfiii{I/O request} completion~\gfiii{\cite{bovet2005understanding}} and 
\sgiii{(2)~}\gfiii{p}olling-based \gfiii{I/O request} completion~\cite{yang2012poll}. \gfiii{Interrupt}-based \gfiii{I/O request} completion employs an asynchronous operation model. When a process issues a block I/O request, the OS puts the process to sleep and context switches to another process. When the \gfiii{I/O} response arrives, the device driver receives an interrupt and wakes up the sender process. \gfiv{In contrast, p}olling-based \gfiii{I/O request} completion employs a synchronous operation model. When a process issues a block I/O request, the process \gfiii{continuously polls in} the CPU waiting for the \gfiii{I/O} request to complete \sgiii{(i.e., instead of going to sleep, the process continually executes CPU instructions to \gfiv{check} the current status of the I/O request)}. 

While \gfiii{interrupt}-based \gfiii{I/O request} completion may incur large system overheads due to context switching, polling-based \gfiii{I/O request} completion imposes a high CPU load to the system. Previous work~\cite{le2017latency} proposes a hybrid \gfiii{I/O request completion} mechanism, which targets fast NVM devices. In this hybrid \gfiii{I/O request completion mechanism}, when a process issues a\gfiii{n I/O} request, the OS puts the process to sleep, similar to the IRQ mode. However, \gfiii{to remove} the context switch latency \gfiii{from \sgiii{the} critical path of I/O request completion}, the OS wakes up the process after some predefined sleep delay \gfiii{time} \texttt{t}. Then, the process polls \gfiii{the I/O request queue for completion of its request} until the response arrives from the block device. \gfiii{The hybrid I/O request completion mechanism can improve performance compared to polling since it reduces the number \gfcr{of} CPU cycles spent on polling.}

\gfiii{The hybrid I/O request completion mechanism works in two modes: (1)~\emph{fixed latency}, where the user \gfiii{sets} the sleep delay \gfiii{time} \texttt{t} to a specific latency; and (2)~\emph{adaptive latency}, where the OS dynamically sets the sleep delay \gfiii{time} \texttt{t} by attempting to estimate when the I/O request will complete~\cite{Queuesys14:online}. In the adaptive \gfiii{latency} mode, the OS monitors the completion time of the different types of I/O requests, and then utilizes half of the average of the I/O request completion time for a particular I/O request type as the sleep delay \gfiii{time} \texttt{t} \sgiii{for} future I/O requests \sgiii{of that type}. Based on this estimation, the OS puts the process that issues I/O requests to sleep before entering a polling loop. The adaptive \gfiii{latency} mode is enabled \sgiii{by setting} \texttt{t} to 0.}

We evaluate how different I/O \gfiii{request} completion \sgiii{mechanisms} impact \gfiii{the} Chrome \gfiii{browser} performance. The OS employs the \gfiii{interrupt}-based \gfiii{I/O request completion mechanism} by default. However, Intel \gfiii{recommends} enabling the Hybrid \gfiii{technique} when using \gfiii{the} Intel Optane SSD for enterprise computing~\cite{hybridoptane}. \gfiii{P}revious work \cite{le2017latency} advocates that fast NVM-based devices can benefit from the \gfiii{p}olling\gfiii{-based I/O request completion mechanism,} since the context switch latency \gfiii{\sgiii{incurred during} interrupt-based I/O request completion} can be larger than the device access latency. We evaluate \sgiii{all three I/O request completion mechanisms to identify the best-performing one to use for Chrome with an Intel Optane SSD} \gfiv{used as swap space}. \gfiii{In the Hybrid I/O request competition mechanism, we evaluate the adaptive latency mode (by setting \texttt{t} = 0) and fixed latency mode, where we \sgiii{evaluate two values of \texttt{t} (\SI{2}{\micro\second} and \SI{4}{\micro\second})}}. 

\begin{figure}[ht]
\begin{subfigure}{\linewidth}
    \vspace{-8pt}
  \centering
  \includegraphics[width=\linewidth]{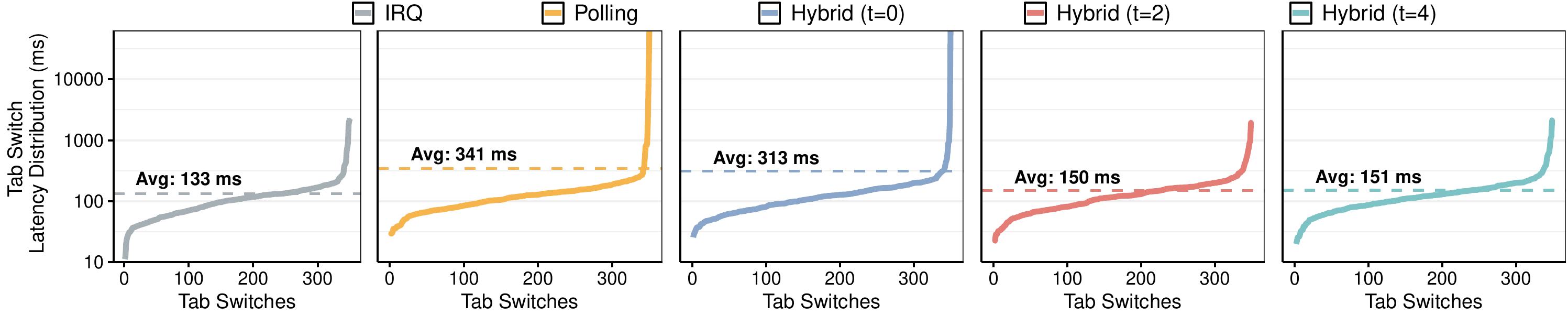}  
  \caption{Tab switch latency distribution. \gfiii{Horizontal dashed lines represent average values.} \vspace{10pt}
}
  \label{fig_tab_switch_latency_io_models_all}
\end{subfigure}
\begin{subfigure}{\linewidth}
  \centering
    \includegraphics[width=\textwidth]{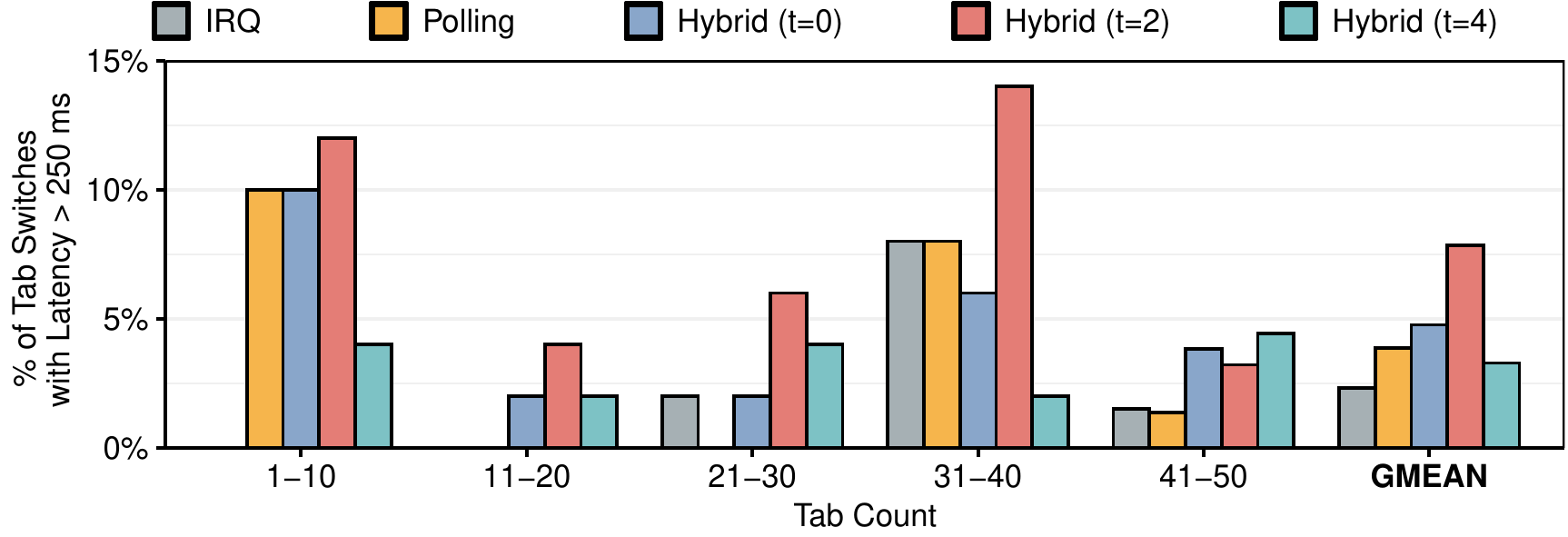}
 \caption{High-latency tab switch latency distribution.}
    \label{fig_irq_high_latency}
\end{subfigure}
\caption{Tab switch latency: Optane with different I/O completion mechanisms.}
\label{fig_tab_switch_latency_io_models}
\vspace{-4pt}
\end{figure}

Figure~\ref{fig_tab_switch_latency_io_models} shows the tab switch latency \gfiii{distribution \gfiii{(Figure~\ref{fig_tab_switch_latency_io_models_all}) and the distribution of high-latency tab switches for the three I/O completion mechanisms (Figure~\ref{fig_irq_high_latency})}} for the \gfiii{Optane} configuration with Zswap enabled, when \gfiii{we employ three} different \gfiii{I/O request} completion \sgiii{mechanisms}\gfiii{: (1)~interrupt-based (IRQ); \sgiii{(2)~}polling-based (Polling); and \sgiii{(3)~}Hybrid with \texttt{t} = 0 \ieeea{(i.e., adaptive latency mode)}, \texttt{t} = \SI{2}{\micro\second}, and \texttt{t} = \SI{4}{\micro\second}}. We make \gfiii{three} observations. \gfiii{First,} the \gfiv{interrupt-based mechanism} provides the \gfiii{lowest average} tab switch latency, \gfiii{with an average tab switch latency reduction of 60\% compared to the Polling mechanism, \gfiii{which provides the highest average tab switch latency;} and \gfiii{11}\% compared to the Hybrid (t = \gfiii{2}) \gfiii{mechanism, which provides the second\gfcr{-}best average tab switch latency} \gfiii{(Figure~\ref{fig_tab_switch_latency_io_models_all})}.} 
Second, on average \gfiii{across tab counts}, \gfiii{the interrupt-based I/O request completion \sgiii{mechanism} leads to the lowest number of high-latency tab switches, \sgii{with only \gfiii{2.3\%} of tab switches being high latency,} versus \gfiii{3.9\%/4.8\%/7.8\%/3.2\%} from the \sgiii{Polling, Hybrid (t=0), Hybrid (t=2), and Hybrid (t=4) mechanisms}, respectively \gfiii{(Figure~\ref{fig_irq_high_latency})}.} 
\gfiii{Third, we observe that \sgiii{the Polling mechanism} eliminates high-latency tab switches when 11--30 tabs are opened. \gfiv{With only a small number of open Chrome tabs (1--10 tabs), Chrome issues few I/O requests 
\juan{due to} the system's low swap traffic (as Figure~\ref{fig_memory_consumption} shows). In such case, the Polling mechanism increases the number of high-latency tab switches compared to the interrupt-based I/O completion mechanism since, in the event of an I/O request, the system cannot context switch to Chrome tab, which will \juan{likely} not issue an I/O request, thus increasing tab switch latency. On the other hand, when the number of \juan{open} Chrome tabs increases, and consequently the swap traffic (Figure~\ref{fig_memory_consumption})}, the system can \gfiv{sometimes} leverage idle CPU time to wait for the completion of Chrome's I/O \gfiv{requests}\gfiii{, reducing} I/O request latency. We conclude that, on average, maintaining the interrupt-based I/O request completion mechanism is still the best approach for consumer devices. \gfiii{However, enabling polling can \gfiv{sometimes} be a good alternative.}}

\begin{figure}[ht]
    \centering
    \includegraphics[width=\linewidth]{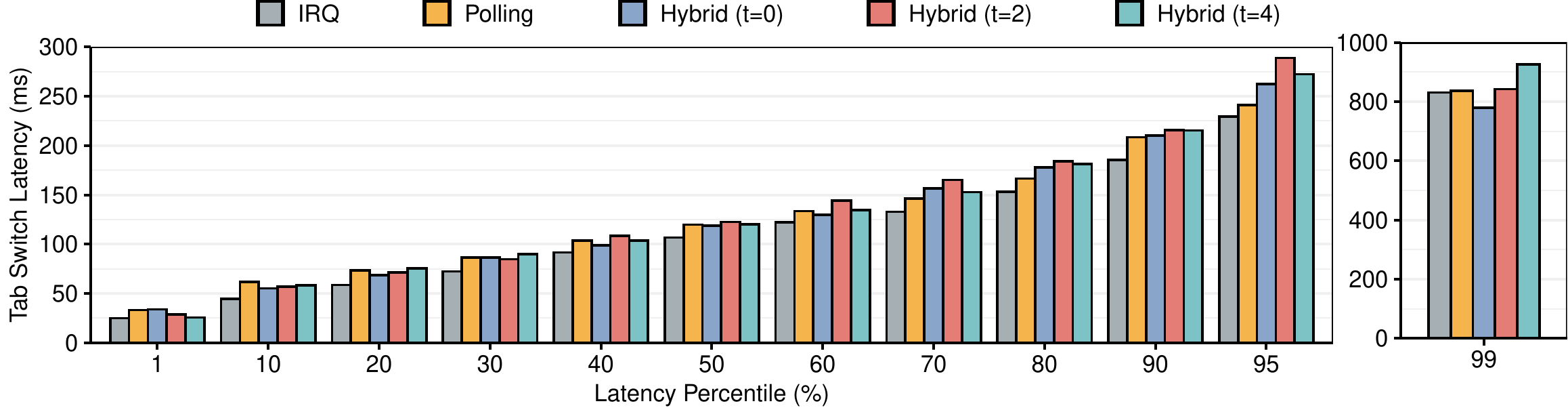}
 \caption{Tab switch latency distribution: different I/O completion \gfiii{mechanisms}.}
    \label{fig_99_irq_models}
\end{figure}

Figure~\mbox{\ref{fig_99_irq_models}} shows the tab switch latency distribution for the \gfiii{three} different I/O completion \gfiii{mechanisms}. We make two observations. First, the \gfiii{interrupt-based \sgiii{(IRQ)} I/O completion} \sgiii{mechanism} provides the \sgiii{lowest tab switch latency of the evaluated mechanisms} up to the 95th-\gfiii{percentile latency}. On average, it reduces tab switch latency by 12\%, 14\%, 20\%, and 17\% compared to the Polling, \gfiii{Hybrid (t=0), Hybrid (t=2), \sgiii{and Hybrid (t=4) mechanisms},} respectively. Second, when examining the 99th-\gfiii{percentile latency}, the Hybrid (t=0) \gfiii{mechanism} reduces the tab switch latency by 7\% compared to the IRQ \gfiii{mechanism}. This happens because at high memory \gfiii{capacity} pressure, most of the processes are waiting for I/O. 
\sgiii{As a result, the OS has few opportunities (if any) to switch in a process that can make forward progress, and there is therefore little to no performance cost in keeping the waiting processes awake to poll the CPU and eliminating the context switch overhead that they would incur with IRQ.} \gfiii{We conclude that the Hybrid I/O request completion mechanism is a good solution to reduce tail latency for Chrome \sgiii{when using an Intel Optane SSD}.}

\begin{figure}[ht]
 \centering
 \Copy{R2/4Cfig}{
    \includegraphics[width=\linewidth]{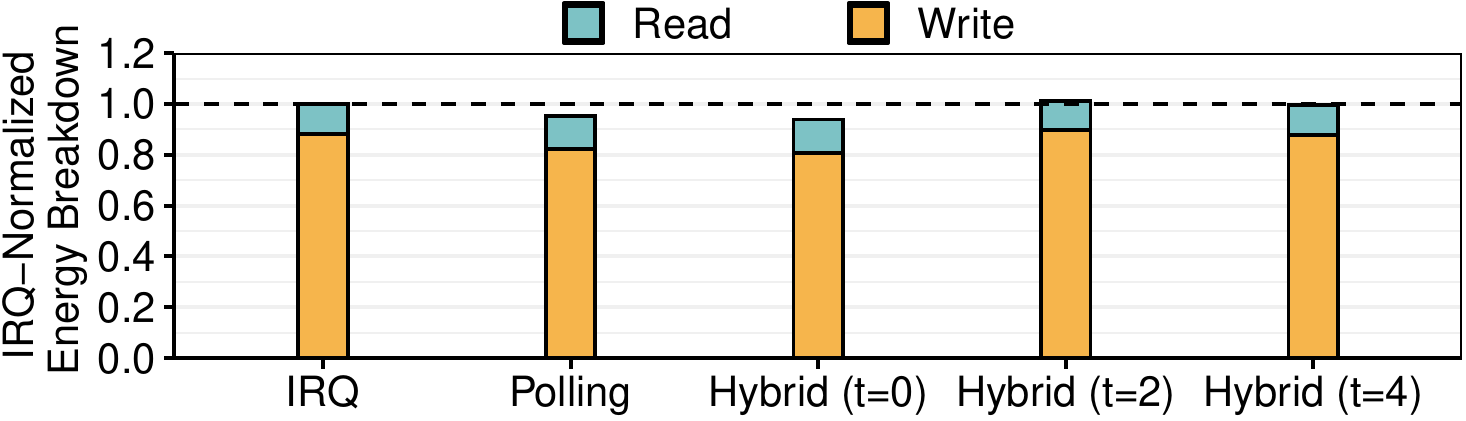}
   \caption{\ieeearev{{Energy consumption: Optane with different I/O completion mechanisms. Y-axis is normalized to the default IRQ I/O completion mechanism.}}} 
   \label{fig_energy_irq_models}
   }
\end{figure}

\label{r2.4c}\Copy{R2/4C}{\noindent \textbf{Energy Analysis.} \ieeearev{Figure~\mbox{\ref{fig_energy_irq_models}} compares the impact of the different I/O completion mechanisms (x-axis) on the average memory subsystem energy consumption for the Optane configuration (y-axis; normalized to the baseline IRQ completion mechanism). We use the same energy model described in Section~\mbox{\ref{sec_step1_1}} for our analysis. The figure shows that the different I/O completion \sgfb{mechanisms have}{ little to no impact on the average memory subsystem energy consumption. This is because such mechanisms do \emph{not} employ any optimization targeting the reduction of I/O traffic or prioritization of I/O requests.}}}

\gfiii{\gfiv{We conclude that} \gfiv{(1)} the interrupt-based I/O request completion mechanism provides the best average performance for the Chrome web browser\gfiv{, but (2) a}t the tail \gfiv{latency} (i.e., 99th-percentile latency), the hybrid I/O request completion mechanism can further reduce I/O request latency.} \gfiv{We, therefore, believe that there needs to be further research into new I/O request completion mechanisms that provide both the best average and tail performance. } 

\section{Key Takeaways}
\label{sec_takeaways}

\gfiii{To summarize, our experimental analysis reveals that extending the main memory space by using the Intel Optane SSD as NVM-based swap space for DRAM provides a cost-effective way to alleviate DRAM scalability issues. However, naively integrating the Intel Optane SSD into the system leads to several system-level overheads that can negatively impact overall performance and energy efficiency. We mitigate such overheads by \gfiv{examining and evaluating} system optimizations driven by our analyses. }

\gfiii{We provide the following \sgii{six} key takeaways from our empirical analyses: }

\begin{enumerate}[noitemsep, leftmargin=*, topsep=0pt]
\item \gfiii{\textit{Effect of Intel Optane SSD as swap space~(Section~\ref{sec_step1_1}).} \gfiii{R}educing DRAM size and \gfiii{extending the main memory space with the} Intel Optane SSD \gfiii{as swap space provides benefits for the Chrome browser, since it can (a)~increase the number of open tabs, and (b)~reduce \gfiv{system} cost. However, it also} leads to an increase in the number of tab \gfiii{switches} with \gfiii{high} latency \gfiii{compared to the baseline.}}

\item \gfiii{\textit{Reducing tail latency by enabling Zswap~(Section~\ref{sec_step1_2}).} Zswap is a good mechanism to reduce I/O traffic \gfiii{introduced by the Intel Optane SSD,} at the cost of a small increase in tab \gfiv{switch} latency at large tab counts. \gfiii{The Zswap cache} reduce\gfiii{s system} energy by 2$\times$ (\sgiii{compared to the} Intel Optane \gfiii{SSD} without Zswap enabled), at the cost of increasing the high\gfiii{-}latency \gfiii{tab switches} by 4\% and reducing the number of open tabs by 12\%.}

\item \gfiii{\textit{Effect of using different NVM devices~(Section~\ref{sec_step3}).} \gfiii{A state-of-the-art NAND-flash-based SSD provides benefits over both the baseline and the Intel Optane SSD. Importantly, it enables more \gfiii{Chrome} tabs to be open. These benefits come due to the larger \sgiii{effective main memory} capacity provided by the state-of-the-art NAND-flash-based SSD over \sgiii{the baseline configuration}.} \gfiii{Unfortunately, these benefits come at the cost of higher tab switch latencies, compared to both the baseline and Optane configurations, due to the much longer device latencies of NAND flash memory. These large tab switch latencies degrade user experience.} \sgiii{Taking both performance and user experience into account, emerging NVM-based SSDs such as the Intel Optane SSD are quite promising to employ in consumer devices, providing performance benefits without the \gfiv{undesirable} user experience trade-offs incurred by NAND-flash-based SSDs.} }

\item \gfiii{\textit{System bottlenecks caused by NVMs~(Section~\ref{sec_profiling_chrome}).} The \gfiii{Linux} block I/O layer is \gfiv{a} key \sgiii{system bottleneck when the Intel Optane SSD is used as swap space}. \gfiii{We can mitigate some of the overheads caused by the block I/O layer} by (a)~employing an I/O scheduler that meets the requirements of the application's access pattern and (b)~using \gfiii{different I/O request} completion mechanisms. }
    
\item \gfiii{\textit{Optimization 1: block I/O schedulers (Section~\ref{sec_io_schedulers}).} \sgiii{We} can reduce tab switch latency by \gfiii{changing the default BFQ I/O scheduler \gfiii{in the system that uses the Intel Optane SSD as swap space}. We reduce 95th- and 99th-percentile latencies by} employing the None and the Kyber \gfiii{I/O} scheduler\gfiii{s, respectively}, as those I/O schedulers reduce \gfiii{I/O scheduling overheads and fit the I/O access pattern of the Chrome web browser.}}
    
\item \gfiii{\textit{Optimization 2: interrupt- vs.\ polling-based I/O request completion (Section~\ref{sec_io_completition}).} \gfiii{On average, the interrupt-based I/O \gfiv{request} completion mechanism provides the best performance for the \gfiii{system with the Intel Optane SSD device. However\gfcr{,} t}}he \gfiii{Hybrid} I/O \gfiv{request} completion mechanism can help \sgiii{reduce} \gfiii{99th-percentile latency} for block I/O requests.}

\end{enumerate}

Based on our analysis, we conclude that \gfiii{there is a large optimization space to be explore in order to \emph{efficiently} adopt emerging NVMs in consumer devices. For example, we believe that one of} the main issue\gfiv{s} the system suffers from when executing interactive workloads is that scheduling decisions made by the OS do not consider the response time expected \sgiii{by} the workload. \gfiii{E}xposing such information to the OS \gfiii{could} reduce tail latency and allow the scheduler to take action according to the needs of a particular workload (e.g., by prioritizing the workload with the shorter or more urgent response deadline\gfiv{s}).  We leave the \gfiv{design,} implementation\gfiv{,} and evaluation of such \gfiv{ideas} for future work.

\subsection{{Overall Limitations of the Technology}}
\label{sec:limitations}

\Copy{R2/4D}{\ieeearev{{Even though employing the Intel Optane SSD as a swap space can lead to several benefits in terms of cost and performance, it can also impact overall system energy consumption and }\ieeearevi{{lifetime}}{. We provide the following two key takeaways from our empirical analyses that highlight the limitations of NVM-based swap space in consumer devices:}}

\begin{enumerate}[noitemsep, leftmargin=*, topsep=0pt]
    \item \ieeearev{{\textit{Effect of Intel Optane SSD as swap space on energy consumption.} Integrating Intel Optane SSD as a swap space increases average memory subsystem energy consumption }\sgfb{{to}}{ 69.5$\times$ that of the baseline ZRAM-based system configuration }\ieeearevi{{(Section~\mbox{\ref{sec_step1_1}}; Figure~\mbox{\ref{fig_step_1_zram_optane_energy}})}}\ieeearev{{. This happens }}\sgfb{{due to the higher swap activity enabled by the Optane-based swap space}}{ (Section~\mbox{\ref{sec_step1_1}}). Such an increase in energy consumption can be mitigated by employing a Zswap cache, which reduces the increase caused by the Optane-based swap space to 34.75$\times$ that of the baseline }\ieeearevi{{(Section~\mbox{\ref{sec_step1_2}; Figure~\mbox{\ref{fig_step_1_zswap_io_traffica}}})}}{. Unfortunately, tuning the block I/O scheduler (Section~\mbox{\ref{sec_io_schedulers}}) and I/O completion mechanism (Section~\mbox{\ref{sec_io_completition}}) do \emph{not} lead to significant energy savings for the Optane configuration, since such optimizations primarily target improving the throughput and latency of I/O operations, rather than reducing energy consumption.}} 
    
    \item \ieeearev{{\textit{Effect of Intel Optane SSD as swap space on system }}\ieeearevi{{\textit{lifetime}}}{. The Intel Optane SSD, as an NVM-based device, suffers from limited write endurance, which can impact }\ieeearevi{{the lifetime of the}}{ system. Based on our analysis (Section~\mbox{\ref{sec_step1_2}}), we observe that it would take an Optane-based system (without Zswap) running our Chrome web browser \ieeearevii{\hl{\mbox{\ltzswapoff}}} years to experience a write-endurance failure. Enabling Zswap increases the }\ieeearevi{{lifetime}}{ of the Optane-based system to \ieeearevii{\hl{\mbox{\ltzswapon}}} years.}}  
\end{enumerate}

{Many prior works~\cite{lee2009architecting,song2020improving,yoon2013techniques,choi2017nvm,guo2018latency,yavits2020wolfram,aghaei2014prolonging,chang2016improving,chen2012age,cheng2016efficient,fan2014wl,han2015enhanced,im2014differentiated,joo2010energy,liu2014application,qureshi2009enhancing,qureshi2011practical,zhou2009durable} aim to reduce the impact of emerging NVMs on overall system energy consumption and }\ieeearevi{{lifetime}}{. The great majority of such works aim to 
(i) reduce the number of write operations the system issue to the NVM device using techniques such as caching~\mbox{\cite{lee2009architecting,yoon2013techniques}}, write-aware data mapping and data allocation algorithms~\mbox{\cite{song2020improving,yoon2013techniques,choi2017nvm}}, and data compression~\mbox{\cite{guo2018latency}}; and
(ii) distribute write operations across NVM cells using diverse wear-\gfcr{leveling} techniques~\mbox{\cite{yavits2020wolfram,aghaei2014prolonging,chang2016improving,chen2012age,cheng2016efficient,fan2014wl,han2015enhanced,im2014differentiated,joo2010energy,liu2014application,qureshi2009enhancing,qureshi2011practical,zhou2009durable}}. We believe such approaches can be employed to mitigate the limitations of NVMs in consumer devices. We leave such analyses for future work.} }

%% file: sections/6-relatedwork.tex
\section{Related Work}
\label{sec_related_work}

\gfiii{To our knowledge, this is the first work that (i)~comprehensively analyzes the impact of extending the main memory space of consumer devices using \emph{real} off-the-shelf emerging \sgiii{NVM-based SSDs}, and (ii)~proposes practical system-level optimizations that can mitigate the tail latency of interactive workloads when employing emerging \sgiii{NVM-based SSDs} in the system.}
\sgiii{We discuss the large body of related work on NVM using four broad categories.}

\gfiii{\textbf{Enabling NVM-Based Swap Space for Mobile Devices.}} 
Several past works~\cite{zhong2014building, kim2015cause, liu2017non, zhong2017building, kim2019analysis, zhu2017smartswap, kim2018comparison} \gfiii{investigate} how to \gfiii{efficiently} enable swap-based NVMs for mobile devices. \gfiii{\sgiii{Unlike} our work, these past works do not utilize \emph{real} NVM devices to evaluate their mechanisms \gfiii{\sgiii{or their system-level implications} on a \emph{real} mobile system}. Thus, it is not \gfiii{fully} clear if their \gfiii{results and} insights can be easily translated to a real system employing a real NVM device. We briefly describe the key mechanisms \sgiii{proposed by each of these works}.}

\sgiii{\gfiv{Two p}rior works~\cite{zhong2014building, liu2017non}} propose to improve swap performance and the lifetime of \sgiii{byte-addressable NVM devices being used as swap space in smartphones}. \sgiii{The\gfiv{se} works} emulate swapping behavior by creating a swap area inside DRAM (similar to our ZRAM configuration). \sgiii{CAUSE~\cite{kim2015cause} is} a hybrid memory architecture for mobile devices that leverages application access patterns to allocate memory either in DRAM or in NVM \gfiii{within a hybrid DRAM-NVM} memory architecture. Similarly, \sgiii{Kim et al.~\cite{kim2019analysis} employ} an NVM-based swap space for Android devices, which leverage\gfiii{s} hot/cold data to manage swap activity \gfiii{between DRAM and NVM}. \sgiii{Zhong et al.~\cite{zhong2017building} aim} to reduce write endurance issues related to NVM devices by identifying and swapping \gfiii{cold} pages \gfiii{from DRAM to} NVM-based swap space in smartphones. \sgiii{SmartSwap~\cite{zhu2017smartswap} predicts the most-rarely-used} applications to be dynamically swapped to a flash\gfiii{-memory}-based swap space \sgiii{ahead of time}. \sgiii{Kim et al.~\cite{kim2018comparison} compare} two swap \gfiii{space} organizations for mobile devices\gfiii{: (1)}~a hierarchical swap architecture, where NVM-based swap \gfiii{space} is used as a cache for a larger flash\gfiii{-memory}-base\gfiv{d} swap \gfiv{space}; and \gfiii{(2)}~a \sgiii{hybrid} swap architecture, where both NVM and flash devices are used \gfiii{as} a single\gfiii{-}level swap space. \sgiii{As part of this work, the authors propose SPP-CLOCK~\cite{kim2018comparison}}, a mechanism to \gfiii{identify} hot/cold data to manage swap activity.

\gfiii{We believe that many of the mechanisms \sgiii{proposed} by \gfiii{these} prior works can be adapted \gfiii{to be employed} in our system \gfiii{to further improve performance and lifetime}. \gfiii{We} leave such \gfiii{studies} \gfiii{to} future works.}

\textbf{Improving Block I/O Latency for Fast NVMe Devices.} 
Previous works~\mbox{\cite{lee2019asynchronous, lee2020case, tavakkol2018flin, oh2020h, shin2014path, vuvcinic2014dc, zhang2018flashshare,liu2022towards}} propose several techniques to mitigate block I/O latencies for fast NVMe devices. \gfiii{These techniques \sgiii{include} software~\cite{lee2019asynchronous,oh2020h,shin2014path,vuvcinic2014dc,zhang2018flashshare,liu2022towards} and hardware solutions~\cite{tavakkol2018flin,lee2020case} to provide lower I/O access latency~\cite{lee2019asynchronous,vuvcinic2014dc,zhang2018flashshare}, page fault handling~\cite{lee2020case}, and I/O scheduling~\cite{tavakkol2018flin,oh2020h,liu2022towards}. Even though \gfiii{these techniques are} promising \gfiii{solutions} to reduce the high block I/O latencies, \gfiii{they require} substantial changes in the hardware and the software stack, which \gfiv{are} outside the scope of this work\gfiv{, but can also be used in our proposed system}.}

Another body of work~\mbox{\cite{caulfield2012providing,kim2016nvmedirect, scargall2020introducing, meza2013case, yang2017spdk, OpenMPDK, peter2015arrakis, kim2017user, kwon2017strata}} aims to completely remove the block I/O layer from the system by providing programming models that enable the user to directly access data from fast NVMe devices. Even though this is a promising solution, it involves several challenges such as code refactoring and security\gfiii{, which can be a promising direction for future work.}

\textbf{Real NVM Devices in Real Systems.} 
Since the release of the Intel Optane SSD, \gfiv{various} works~\cite{wu2019exploiting,zhang2018performance,wu2017early,jia2020flash,han2020splitkv,ke2018lirs,chien2018characterizing,liu2020nvm,lu2021case} \gfiii{have experimentally shown} that the Intel Optane SSD can improve performance, energy\gfiii{, and} cost for different workloads \gfiii{(\sgiii{e.g.}, databases~\cite{wu2019exploiting,zhang2018performance,lu2021case}, high-performance computing~\cite{wu2017early}, key-value stores~\cite{jia2020flash,han2020splitkv}, machine learning~\cite{ke2018lirs,chien2018characterizing}, query processing~\cite{liu2020nvm}).} \gfiii{Our work differs from these works since we (i)~target a different \sgiii{family of workloads} (i.e., interactive \gfiv{consumer} workloads, \gfiv{and in} particular the Google Chrome web browser) and (ii)~employ the Intel Optane SSD as an extension of main memory, instead of as a \gfiv{separate} storage device.}

\textbf{Hybrid DRAM-NVM Memory Systems.} 
\gfiii{A large body of works~\cite{lee2009architecting,qureshi2009scalable,lee2010phase,lee2010phasecacm,kultursay2013evaluating,zhou2009durable,DAC-2009-DhimanAR, wang2019panthera,salkhordeh2019analytical,li2017utility,yoon2012row,meza2012enabling,singh2022sibyl,oh2015sqlite,li2022multi,raybuck2021hemem} \sgiii{propose} to use NVMs as an alternative technology to DRAM, where the NVM completely \sgiii{replaces} DRAM as the main memory device~\cite{lee2009architecting,qureshi2009scalable,lee2010phase,lee2010phasecacm,kultursay2013evaluating,zhou2009durable}, or is incorporated \sgiii{into} the memory hierarchy alongside DRAM \sgiii{to create} a hybrid DRAM--NVM memory system~\cite{DAC-2009-DhimanAR,wang2019panthera,salkhordeh2019analytical,li2017utility,yoon2012row,meza2012enabling,oh2015sqlite,raybuck2021hemem}. \sgiii{Unlike} our work, these works \ieeea{either} \sgiii{(1)~}use NVMs as part of main memory, and not as swap space, which increases the complexity of the memory architecture; \sgiii{\ieeea{or} (2)~}\gfiv{mainly} leverage simulation infrastructures to evaluate their proposals, and thus, do not examine \emph{real} NVM devices \gfiv{and their implications on real systems with real measurement data}.}

%% file: sections/7-conclusion.tex
\section{Conclusion}
\label{sec_conclusion}

\Copy{R2/4E}{\gfiii{In this paper, we comprehensively evaluate the performance implications of \sgiii{leveraging real} emerging \gfiv{NVMs} as an extension of main memory space in \sgiii{real} consumer devices, while targeting interactive workloads. We \sgiii{employ} a state-of-the-art NVM-based SSD device (i.e., the Intel Optane SSD) as swap space \sgiii{for DRAM, which increases the effective main memory capacity} in our system. We observe that \sgiii{using the Intel Optane SSD can improve the average and \gfcr{tail latency} performance of the Chrome web browser,} compared to a baseline system with double the amount of DRAM,
and \sgiii{to} a system \sgiii{where a state-of-the-art NAND-flash-based SSD is used for the swap space}. \gfiv{We identify that the Linux block I/O layer becomes a major source of performance overhead when \juan{the} main memory space is extended using NVM, primarily due to (i) I/O scheduling bottlenecks; and (ii) overheads related to the asynchronous operation of the I/O request completion mechanism. We mitigate some of these overheads by proposing two system optimizations that can better leverage the characteristics of our workloads and the NVM.} 
\ieeearev{{We also evaluate the limitations of real emerging NVMs in consumer devices and conclude that real systems need to employ solutions to mitigate the issues associated with energy increase and }\ieeearevi{{lifetime}}{ degradation NVM devices introduce.}} We conclude that emerging NVMs are a cost-effective solution to alleviate the DRAM capacity bottleneck in consumer devices. We hope that the results of our study can inspire and drive novel hardware and software optimizations in future NVM-based computing systems.}
}

%% file: main.bbl
\begin{thebibliography}{100}
\providecommand{\url}[1]{#1}
\csname url@samestyle\endcsname
\providecommand{\newblock}{\relax}
\providecommand{\bibinfo}[2]{#2}
\providecommand{\BIBentrySTDinterwordspacing}{\spaceskip=0pt\relax}
\providecommand{\BIBentryALTinterwordstretchfactor}{4}
\providecommand{\BIBentryALTinterwordspacing}{\spaceskip=\fontdimen2\font plus
\BIBentryALTinterwordstretchfactor\fontdimen3\font minus \fontdimen4\font\relax}
\providecommand{\BIBforeignlanguage}[2]{{%
\expandafter\ifx\csname l@#1\endcsname\relax
\typeout{** WARNING: IEEEtran.bst: No hyphenation pattern has been}%
\typeout{** loaded for the language `#1'. Using the pattern for}%
\typeout{** the default language instead.}%
\else
\language=\csname l@#1\endcsname
\fi
#2}}
\providecommand{\BIBdecl}{\relax}
\BIBdecl

\bibitem{chromebook}
{Google LLC}, ``{Chromebook},'' \url{https://www.google.com/chromebook/}.

\bibitem{emarketer2016slowing}
eMarketer, ``{Slowing Growth Ahead for Worldwide Internet Audience},'' 2016.

\bibitem{reddi2018two}
V.~J. Reddi, H.~Yoon, and A.~Knies, ``{Two Billion Devices and Counting},'' \emph{IEEE Micro}, 2018.

\bibitem{armqualcomm2014}
{ARM and Qualcomm}, ``{Enabling the Next Mobile Computing Revolution with Highly Integrated ARMv8-A Based SoCs},'' {White Paper}, 2014.

\bibitem{halpern2016mobile}
M.~Halpern, Y.~Zhu, and V.~J. Reddi, ``{Mobile CPU's Rise to Power: Quantifying the Impact of Generational Mobile CPU Design Trends on Performance, Energy, and User Satisfaction},'' in \emph{HPCA}, 2016.

\bibitem{CanalysN72}
Canalys, ``{Chromebooks Lead PC Revival in Q1 2021 With 275\% Growth},'' \url{https://rb.gy/jm7xu}, 2021.

\bibitem{heater2017chromebook}
B.~Heater, ``{As Chromebook Sales Soar in Schools, Apple and Microsoft Fight Back},'' \emph{TechCrunch}, 2017.

\bibitem{dennard1968dram}
R.~H. Dennard, ``{Field-Effect Transistor Memory},'' U.S. Patent 3,387,286, 1968.

\bibitem{mutlu2014memorybook}
Y.~Kim and O.~Mutlu, ``{Memory Systems},'' in \emph{Computing Handbook, Third Edition: Computer Science and Software Engineering}.\hskip 1em plus 0.5em minus 0.4em\relax Taylor \& Francis, 2014.

\bibitem{bovet2005understanding}
D.~P. Bovet and M.~Cesati, \emph{{Understanding the Linux Kernel: From I/O Ports to Process Management}}, 3rd~ed.\hskip 1em plus 0.5em minus 0.4em\relax O'Reilly Media, Inc., 2005.

\bibitem{tanenbaum1997operating}
A.~S. Tanenbaum and A.~S. Woodhull, \emph{{Operating Systems: Design and Implementation}}.\hskip 1em plus 0.5em minus 0.4em\relax Prentice Hall Englewood Cliffs, 1997.

\bibitem{lecturevirtualmemory}
O.~Mutlu, ``{Lecture Notes for Digital Design and Computer Architecture -- Lecture 23b: Virtual Memory},'' \url{https://rb.gy/qzj7r}, 2020.

\bibitem{badr2020mocktails}
M.~Badr, C.~Delconte, I.~Edo, R.~Jagtap, M.~Andreozzi, and N.~E. Jerger, ``{Mocktails: Capturing the Memory Behaviour of Proprietary Mobile Architectures},'' in \emph{ISCA}, 2020.

\bibitem{boroumand2018google}
A.~Boroumand, S.~Ghose, Y.~Kim, R.~Ausavarungnirun, E.~Shiu, R.~Thakur, D.~Kim, A.~Kuusela, A.~Knies, P.~Ranganathan \emph{et~al.}, ``{Google Workloads for Consumer Devices: Mitigating Data Movement Bottlenecks},'' in \emph{ASPLOS}, 2018.

\bibitem{mohan2017storage}
J.~Mohan, D.~Purohith, M.~Halpern, V.~Chidambaram, and V.~J. Reddi, ``{Storage on Your Smartphone Uses More Energy Than You Think},'' in \emph{HotStorage}, 2017.

\bibitem{amiraliphd}
A.~Boroumand, ``{Practical Mechanisms for Reducing Processor-Memory Data Movement in Modern Workloads},'' Ph.D. dissertation, Carnegie Mellon University, 2020.

\bibitem{nelsonsize}
R.~Nelson, ``{The Size of iPhone’s Top Apps Has Increased by 1,000\% in Four Years},'' \url{https://sensortower.com/blog/ios-app-size-growth}, 2017.

\bibitem{lebeck2020end}
N.~Lebeck, A.~Krishnamurthy, H.~M. Levy, and I.~Zhang, ``{End the Senseless Killing: Improving Memory Management for Mobile Operating Systems},'' in \emph{USENIX ATC}, 2020.

\bibitem{mutlu2013memory}
O.~Mutlu, ``{Memory Scaling: A Systems Architecture Perspective},'' in \emph{IMW}, 2013.

\bibitem{mutlu2015research}
O.~Mutlu and L.~Subramanian, ``{Research Problems and Opportunities in Memory Systems},'' \emph{SUPERFRI}, 2015.

\bibitem{kim2014flipping}
Y.~Kim, R.~Daly, J.~Kim, C.~Fallin, J.~H. Lee, D.~Lee, C.~Wilkerson, K.~Lai, and O.~Mutlu, ``{Flipping Bits in Memory Without Accessing Them: An Experimental Study of DRAM Disturbance Errors},'' in \emph{ISCA}, 2014.

\bibitem{mutlu2019rowhammer}
O.~Mutlu and J.~S. Kim, ``{RowHammer: A Retrospective},'' \emph{TCAD}, 2019.

\bibitem{kim2020revisiting}
J.~S. Kim, M.~Patel, A.~G. Ya{\u{g}}l{\i}k{\c{c}}{\i}, H.~Hassan, R.~Azizi, L.~Orosa, and O.~Mutlu, ``{Revisiting RowHammer: An Experimental Analysis of Modern DRAM Devices and Mitigation Techniques},'' in \emph{ISCA}, 2020.

\bibitem{mutlu2015main}
O.~Mutlu, ``{Main Memory Scaling: Challenges and Solution Directions},'' in \emph{More than Moore Technologies for Next Generation Computer Design}, 2015.

\bibitem{kang2014co}
U.~Kang, H.-S. Yu, C.~Park, H.~Zheng, J.~Halbert, K.~Bains, S.~Jang, and J.~S. Choi, ``{Co-Architecting Controllers and DRAM to Enhance DRAM Process Scaling},'' in \emph{The Memory Forum}, 2014.

\bibitem{hong2010memory}
S.~Hong, ``{Memory Technology Trend and Future Challenges},'' in \emph{IEDM}, 2010.

\bibitem{kanev_isca2015}
S.~Kanev, J.~P. Darago, K.~Hazelwood, P.~Ranganathan, T.~Moseley, G.-Y. Wei, and D.~Brooks, ``{Profiling a Warehouse-Scale Computer},'' in \emph{ISCA}, 2015.

\bibitem{mutlu2017rowhammer}
O.~Mutlu, ``{The RowHammer Problem and Other Issues We may Face as Memory Becomes Denser},'' in \emph{DATE}, 2017.

\bibitem{ghose2018your}
S.~Ghose, A.~G. Yaglik{\c{c}}i, R.~Gupta, D.~Lee, K.~Kudrolli, W.~X. Liu, H.~Hassan, K.~K. Chang, N.~Chatterjee, A.~Agrawal \emph{et~al.}, ``{What Your DRAM Power Models Are Not Telling You: Lessons from a Detailed Experimental Study},'' in \emph{SIGMETRICS}, 2018.

\bibitem{liu2013experimental}
J.~Liu, B.~Jaiyen, Y.~Kim, C.~Wilkerson, and O.~Mutlu, ``{An Experimental Study of Data Retention Behavior in Modern DRAM Devices: Implications for Retention Time Profiling Mechanisms},'' in \emph{ISCA}, 2013.

\bibitem{frigo2020trrespass}
P.~Frigo, E.~Vannacc, H.~Hassan, V.~Van Der~Veen, O.~Mutlu, C.~Giuffrida, H.~Bos, and K.~Razavi, ``{TRRespass: Exploiting the Many Sides of Target Row Refresh},'' in \emph{SP}, 2020.

\bibitem{liu2012raidr}
J.~Liu, B.~Jaiyen, R.~Veras, and O.~Mutlu, ``{RAIDR: Retention-Aware Intelligent DRAM Refresh},'' in \emph{ISCA}, 2012.

\bibitem{patel2017reach}
M.~Patel, J.~S. Kim, and O.~Mutlu, ``{The Reach Profiler (REAPER): Enabling the Mitigation of DRAM Retention Failures via Profiling at Aggressive Conditions},'' in \emph{ISCA}, 2017.

\bibitem{qureshi2015avatar}
M.~K. Qureshi, D.~Kim, S.~Khan, P.~J. Nair, and O.~Mutlu, ``{AVATAR: A Variable-Retention-Time (VRT) Aware Refresh for DRAM Systems},'' in \emph{DSN}, 2015.

\bibitem{mandelman2002challenges}
J.~A. Mandelman, R.~H. Dennard, G.~B. Bronner, J.~K. DeBrosse, R.~Divakaruni, Y.~Li, and C.~J. Radens, ``{Challenges and Future Directions for the Scaling of Dynamic Random-Access Memory (DRAM)},'' \emph{IBM JRD}, 2002.

\bibitem{khan2014efficacy}
S.~Khan, D.~Lee, Y.~Kim, A.~R. Alameldeen, C.~Wilkerson, and O.~Mutlu, ``{The Efficacy of Error Mitigation Techniques for DRAM Retention Failures: A Comparative Experimental Study},'' in \emph{SIGMETRICS}, 2014.

\bibitem{khan2016parbor}
S.~Khan, D.~Lee, and O.~Mutlu, ``{PARBOR: An Efficient System-Level Technique to Detect Data-Dependent Failures in DRAM},'' in \emph{DSN}, 2016.

\bibitem{khan2017detecting}
S.~Khan, C.~Wilkerson, Z.~Wang, A.~R. Alameldeen, D.~Lee, and O.~Mutlu, ``{Detecting and Mitigating Data-Dependent DRAM Failures by Exploiting Current Memory Content},'' in \emph{MICRO}, 2017.

\bibitem{lee2015adaptive}
D.~Lee, Y.~Kim, G.~Pekhimenko, S.~Khan, V.~Seshadri, K.~Chang, and O.~Mutlu, ``{Adaptive-Latency DRAM: Optimizing DRAM Timing for the Common-Case},'' in \emph{HPCA}, 2015.

\bibitem{lee2017design}
D.~Lee, S.~Khan, L.~Subramanian, S.~Ghose, R.~Ausavarungnirun, G.~Pekhimenko, V.~Seshadri, and O.~Mutlu, ``{Design-Induced Latency Variation in Modern DRAM Chips: Characterization, Analysis, and Latency Reduction Mechanisms},'' in \emph{SIGMETRICS}, 2017.

\bibitem{chang2017understandingphd}
K.~K. Chang, ``{Understanding and Improving the Latency of DRAM-Based Memory Systems},'' Ph.D. dissertation, Carnegie Mellon University, 2017.

\bibitem{chang2017understandingsigmetrics}
K.~K. Chang, A.~G. Ya{\u{g}}l{\i}k{\c{c}}{\i}, S.~Ghose, A.~Agrawal, N.~Chatterjee, A.~Kashyap, D.~Lee, M.~O'Connor, H.~Hassan, and O.~Mutlu, ``{Understanding Reduced-Voltage Operation in Modern DRAM Devices: Experimental Characterization, Analysis, and Mechanisms},'' in \emph{SIGMETRICS}, 2017.

\bibitem{chang2016understanding}
K.~K. Chang, A.~Kashyap, H.~Hassan, S.~Ghose, K.~Hsieh, D.~Lee, T.~Li, G.~Pekhimenko, S.~Khan, and O.~Mutlu, ``{Understanding Latency Variation in Modern DRAM Chips: Experimental Characterization, Analysis, and Optimization},'' in \emph{SIGMETRICS}, 2016.

\bibitem{chang2014improving}
K.~K.-W. Chang, D.~Lee, Z.~Chishti, A.~R. Alameldeen, C.~Wilkerson, Y.~Kim, and O.~Mutlu, ``{Improving DRAM Performance by Parallelizing Refreshes with Accesses},'' in \emph{HPCA}, 2014.

\bibitem{meza2015revisiting}
J.~Meza, Q.~Wu, S.~Kumar, and O.~Mutlu, ``{Revisiting Memory Errors in Large-Scale Production Data Centers: Analysis and Modeling of New Trends from the Field},'' in \emph{DSN}, 2015.

\bibitem{david2011memory}
H.~David, C.~Fallin, E.~Gorbatov, U.~R. Hanebutte, and O.~Mutlu, ``{Memory Power Management via Dynamic Voltage/Frequency Scaling},'' in \emph{ICAC}, 2011.

\bibitem{deng2011memscale}
Q.~Deng, D.~Meisner, L.~Ramos, T.~F. Wenisch, and R.~Bianchini, ``{MemScale: Active Low-Power Modes for Main Memory},'' in \emph{ASPLOS}, 2011.

\bibitem{yauglikcci2022understanding}
A.~G. Ya{\u{g}}l{\i}k{\c{c}}{\i}, H.~Luo, G.~F. de~Oliviera, A.~Olgun, M.~Patel, J.~Park, H.~Hassan, J.~S. Kim, L.~Orosa, and O.~Mutlu, ``{Understanding RowHammer Under Reduced Wordline Voltage: An Experimental Study Using Real DRAM Devices},'' in \emph{DSN}, 2022.

\bibitem{orosa2021deeper}
L.~Orosa, A.~G. Yaglikci, H.~Luo, A.~Olgun, J.~Park, H.~Hassan, M.~Patel, J.~S. Kim, and O.~Mutlu, ``{A Deeper Look into RowHammer's Sensitivities: Experimental Analysis of Real DRAM Chips and Implications on Future Attacks and Defenses},'' in \emph{MICRO}, 2021.

\bibitem{hassan2021uncovering}
H.~Hassan, Y.~C. Tugrul, J.~S. Kim, V.~Van~der Veen, K.~Razavi, and O.~Mutlu, ``{Uncovering In-DRAM RowHammer Protection Mechanisms: A New Methodology, Custom RowHammer Patterns, and Implications},'' in \emph{MICRO}, 2021.

\bibitem{lee2009architecting}
B.~C. Lee, E.~Ipek, O.~Mutlu, and D.~Burger, ``{Architecting Phase Change Memory as a Scalable DRAM Alternative},'' in \emph{ISCA}, 2009.

\bibitem{qureshi2009scalable}
M.~K. Qureshi, V.~Srinivasan, and J.~A. Rivers, ``{Scalable High Performance Main Memory System Using Phase-Change Memory Technology},'' in \emph{ISCA}, 2009.

\bibitem{lee2010phase}
B.~C. Lee, P.~Zhou, J.~Yang, Y.~Zhang, B.~Zhao, E.~Ipek, O.~Mutlu, and D.~Burger, ``{Phase-Change Technology and the Future of Main Memory},'' \emph{IEEE Micro}, 2010.

\bibitem{lee2010phasecacm}
B.~C. Lee, E.~Ipek, O.~Mutlu, and D.~Burger, ``{Phase Change Memory Architecture and the Quest for Scalability},'' \emph{CACM}, 2010.

\bibitem{kultursay2013evaluating}
E.~K{\"u}lt{\"u}rsay, M.~Kandemir, A.~Sivasubramaniam, and O.~Mutlu, ``{Evaluating STT-RAM as an Energy-Efficient Main Memory Alternative},'' in \emph{ISPASS}, 2013.

\bibitem{zhou2009durable}
P.~Zhou, B.~Zhao, J.~Yang, and Y.~Zhang, ``{A Durable and Energy Efficient Main Memory Using Phase Change Memory Technology},'' in \emph{ISCA}, 2009.

\bibitem{wong2010phase}
H.-S.~P. Wong, S.~Raoux, S.~Kim, J.~Liang, J.~P. Reifenberg, B.~Rajendran, M.~Asheghi, and K.~E. Goodson, ``{Phase Change Memory},'' \emph{Proc. IEEE}, 2010.

\bibitem{meza2012case}
J.~Meza, J.~Li, and O.~Mutlu, ``{A Case for Small Row Buffers in Non-Volatile Main Memories},'' in \emph{ICCD}, 2012.

\bibitem{meza2013case}
J.~Meza, Y.~Luo, S.~Khan, J.~Zhao, Y.~Xie, and O.~Mutlu, ``{A Case for Efficient Hardware/Software Cooperative Management of Storage and Memory},'' in \emph{WEED}, 2013.

\bibitem{song2020improving}
S.~Song, A.~Das, O.~Mutlu, and N.~Kandasamy, ``{Improving Phase Change Memory Performance with Data Content Aware Access},'' in \emph{ISMM}, 2020.

\bibitem{song2021aging}
S.~Song, A.~Das, O.~Mutlu, and N.~Kandasamy, ``{Aging-Aware Request Scheduling for Non-Volatile Main Memory},'' in \emph{ASP-DAC}, 2021.

\bibitem{song2019enabling}
S.~Song, A.~Das, O.~Mutlu, and N.~Kandasamy, ``{Enabling and Exploiting Partition-Level Parallelism (PALP) in Phase Change Memories},'' \emph{TECS}, 2019.

\bibitem{atwood2018pcm}
G.~Atwood, ``{PCM Applications and an Outlook to the Future},'' in \emph{Phase Change Memory: Device Physics, Reliability and Applications}.\hskip 1em plus 0.5em minus 0.4em\relax Springer International Publishing, 2017.

\bibitem{bock2011analyzing}
S.~Bock, B.~Childers, R.~Melhem, D.~Moss{\'e}, and Y.~Zhang, ``{Analyzing the Impact of Useless Write-Backs on the Endurance and Energy Consumption of PCM Main Memory},'' in \emph{ISPASS}, 2011.

\bibitem{burr2008overview}
G.~W. Burr, B.~N. Kurdi, J.~C. Scott, C.~H. Lam, K.~Gopalakrishnan, and R.~S. Shenoy, ``{Overview of Candidate Device Technologies for Storage-Class Memory},'' \emph{IBM JRD}, 2008.

\bibitem{du2013bit}
Y.~Du, M.~Zhou, B.~R. Childers, D.~Moss{\'e}, and R.~Melhem, ``{Bit Mapping for Balanced PCM Cell Programming},'' in \emph{ISCA}, 2013.

\bibitem{ferreira2010increasing}
A.~P. Ferreira, M.~Zhou, S.~Bock, B.~Childers, R.~Melhem, and D.~Moss{\'e}, ``{Increasing PCM Main Memory Lifetime},'' in \emph{DATE}, 2010.

\bibitem{jiang2012fpb}
L.~Jiang, Y.~Zhang, B.~R. Childers, and J.~Yang, ``{FPB: Fine-Grained Power Budgeting to Improve Write Throughput of Multi-Level Cell Phase Change Memory},'' in \emph{MICRO}, 2012.

\bibitem{jiang2013hardware}
L.~Jiang, Y.~Du, B.~Zhao, Y.~Zhang, B.~R. Childers, and J.~Yang, ``{Hardware-Assisted Cooperative Integration of Wear-Leveling and Salvaging for Phase Change Memory},'' \emph{TACO}, 2013.

\bibitem{kannan2016energy}
S.~Kannan, M.~Qureshi, A.~Gavrilovska, and K.~Schwan, ``{Energy Aware Persistence: Reducing Energy Overheads of Memory-Based Persistence in NVMs},'' in \emph{PACT}, 2016.

\bibitem{qureshi2011pay}
M.~K. Qureshi, ``{Pay-As-You-Go: Low-Overhead Hard-Error Correction for Phase Change Memories},'' in \emph{MICRO}, 2011.

\bibitem{qureshi2010improving}
M.~K. Qureshi, M.~M. Franceschini, and L.~A. Lastras-Montano, ``{Improving Read Performance of Phase Change Memories via Write Cancellation and Write Pausing},'' in \emph{HPCA}, 2010.

\bibitem{qureshi2010morphable}
M.~K. Qureshi, M.~M. Franceschini, L.~A. Lastras-Monta{\~n}o, and J.~P. Karidis, ``{Morphable Memory System: A Robust Architecture for Exploiting Multi-Level Phase Change Memories},'' in \emph{ISCA}, 2010.

\bibitem{sebastian2017temporal}
A.~Sebastian, T.~Tuma, N.~Papandreou, M.~Le~Gallo, L.~Kull, T.~Parnell, and E.~Eleftheriou, ``{Temporal Correlation Detection Using Computational Phase-Change Memory},'' \emph{Nature Commun.}, 2017.

\bibitem{wang2015exploit}
R.~Wang, L.~Jiang, Y.~Zhang, L.~Wang, and J.~Yang, ``{Exploit Imbalanced Cell Writes to Mitigate Write Disturbance in Dense Phase Change Memory},'' in \emph{DAC}, 2015.

\bibitem{yue2013accelerating}
J.~Yue and Y.~Zhu, ``{Accelerating Write by Exploiting PCM Asymmetries},'' in \emph{HPCA}, 2013.

\bibitem{zhou2012writeback}
M.~Zhou, Y.~Du, B.~Childers, R.~Melhem, and D.~Moss{\'e}, ``{Writeback-Aware Partitioning and Replacement for Last-Level Caches in Phase Change Main Memory Systems},'' \emph{TACO}, 2012.

\bibitem{zhou2013writeback}
M.~Zhou, Y.~Du, B.~R. Childers, R.~Melhem, and D.~Moss{\'e}, ``{Writeback-Aware Bandwidth Partitioning for Multi-Core Systems with PCM},'' in \emph{PACT}, 2013.

\bibitem{yoon2013techniques}
H.~Yoon, N.~Muralimanohar, J.~Meza, O.~Mutlu, and N.~P. Jouppi, ``{Techniques for Data Mapping and Buffering to Exploit Asymmetry in Multi-Level Cell (Phase Change) Memory},'' SAFARI Research Group, Tech. Rep. TR-SAFARI-2013-002, 2013.

\bibitem{DAC-2009-DhimanAR}
G.~Dhiman, R.~Z. Ayoub, and T.~Rosing, ``{PDRAM: A Hybrid PRAM and DRAM Main Memory System},'' in \emph{DAC}, 2009.

\bibitem{wang2013low}
K.~Wang, J.~Alzate, and P.~K. Amiri, ``{Low-Power Non-Volatile Spintronic Memory: STT-RAM and Beyond},'' \emph{J. Phys. D: Appl. Phys}, 2013.

\bibitem{chen2010advances}
E.~Chen, D.~Apalkov, Z.~Diao, A.~Driskill-Smith, D.~Druist, D.~Lottis, V.~Nikitin, X.~Tang, S.~Watts, S.~Wang \emph{et~al.}, ``{Advances and Future Prospects of Spin-Transfer Torque Random Access Memory},'' \emph{TMAG}, 2010.

\bibitem{diao2007spin}
Z.~Diao, Z.~Li, S.~Wang, Y.~Ding, A.~Panchula, E.~Chen, L.-C. Wang, and Y.~Huai, ``{Spin-Transfer Torque Switching in Magnetic Tunnel Junctions and Spin-Transfer Torque Random Access Memory},'' \emph{Journal of Physics: Condensed Matter}, 2007.

\bibitem{hosomi2005novel}
M.~Hosomi, H.~Yamagishi, T.~Yamamoto, K.~Bessho, Y.~Higo, K.~Yamane, H.~Yamada, M.~Shoji, H.~Hachino, C.~Fukumoto \emph{et~al.}, ``{A Novel Nonvolatile Memory with Spin Torque Transfer Magnetization Switching: Spin-RAM},'' in \emph{IEDM}, 2005.

\bibitem{raychowdhury2009design}
A.~Raychowdhury, D.~Somasekhar, T.~Karnik, and V.~De, ``{Design Space and Scalability Exploration of 1T-1STT MTJ Memory Arrays in the Presence of Variability and Disturbances},'' in \emph{IEDM}, 2009.

\bibitem{akinaga2010resistive}
H.~Akinaga and H.~Shima, ``{Resistive Random Access Memory (ReRAM) Based on Metal Oxides},'' \emph{Proc. IEEE}, 2010.

\bibitem{wong2012metal}
H.-S.~P. Wong, H.-Y. Lee, S.~Yu, Y.-S. Chen, Y.~Wu, P.-S. Chen, B.~Lee, F.~T. Chen, and M.-J. Tsai, ``{Metal--Oxide RRAM},'' \emph{Proc. IEEE}, 2012.

\bibitem{yang2013memristive}
J.~J. Yang, D.~B. Strukov, and D.~R. Stewart, ``{Memristive Devices for Computing},'' \emph{Nature Nanotechnology}, 2013.

\bibitem{kund2005conductive}
M.~Kund, G.~Beitel, C.-U. Pinnow, T.~Rohr, J.~Schumann, R.~Symanczyk, K.~Ufert, and G.~Muller, ``{Conductive Bridging RAM (CBRAM): An Emerging Non-Volatile Memory Technology Scalable to Sub 20nm},'' in \emph{IEDM}, 2005.

\bibitem{bondurant1990ferroelectronic}
D.~Bondurant, ``{Ferroelectronic RAM Memory Family for Critical Data Storage},'' \emph{Ferroelectrics}, 1990.

\bibitem{harris2020ultra}
B.~Harris and N.~Altiparmak, ``{Ultra-Low Latency SSDs' Impact on Overall Energy Efficiency},'' in \emph{HotStorage}, 2020.

\bibitem{wu2021storage}
K.~Wu, Z.~Guo, G.~Hu, K.~Tu, R.~Alagappan, R.~Sen, K.~Park, A.~C. Arpaci-Dusseau, and R.~H. Arpaci-Dusseau, ``{The Storage Hierarchy is Not a Hierarchy: Optimizing Caching on Modern Storage Devices with Orthus},'' in \emph{FAST}, 2021.

\bibitem{wang2019panthera}
C.~Wang, H.~Cui, T.~Cao, J.~Zigman, H.~Volos, O.~Mutlu, F.~Lv, X.~Feng, and G.~H. Xu, ``{Panthera: Holistic Memory Management for Big Data Processing Over Hybrid Memories},'' in \emph{PLDI}, 2019.

\bibitem{salkhordeh2019analytical}
R.~Salkhordeh, O.~Mutlu, and H.~Asad, ``{An Analytical Model for Performance and Lifetime Estimation of Hybrid DRAM-NVM Main Memories},'' \emph{TC}, 2019.

\bibitem{yoon2012row}
H.~Yoon, J.~Meza, R.~Ausavarungnirun, R.~A. Harding, and O.~Mutlu, ``{Row Buffer Locality Aware Caching Policies for Hybrid Memories},'' in \emph{ICCD}, 2012.

\bibitem{meza2012enabling}
J.~Meza, J.~Chang, H.~Yoon, O.~Mutlu, and P.~Ranganathan, ``{Enabling Efficient and Scalable Hybrid Memories Using Fine-Granularity DRAM Cache Management},'' \emph{CAL}, 2012.

\bibitem{h10}
{Intel Corp.}, ``{Intel{\textsuperscript{\textregistered}} Optane{\texttrademark} Memory H10 with Solid State Storage},'' \url{https://rb.gy/f682j}.

\bibitem{izraelevitz2019basic}
J.~Izraelevitz, J.~Yang, L.~Zhang, J.~Kim, X.~Liu, A.~Memaripour, Y.~J. Soh, Z.~Wang, Y.~Xu, S.~R. Dulloor \emph{et~al.}, ``{Basic Performance Measurements of the Intel Optane DC Persistent Memory Module},'' arXiv:1903.05714 [cs.AR], 2019.

\bibitem{psaropoulos2019bridging}
G.~Psaropoulos, I.~Oukid, T.~Legler, N.~May, and A.~Ailamaki, ``{Bridging the Latency Gap Between NVM and DRAM for Latency-Bound Operations},'' in \emph{DaMoN}, 2019.

\bibitem{lee2019asynchronous}
G.~Lee, S.~Shin, W.~Song, T.~J. Ham, J.~W. Lee, and J.~Jeong, ``{Asynchronous I/O Stack: A Low-Latency Kernel I/O Stack for Ultra-Low Latency SSDs},'' in \emph{USENIX ATC}, 2019.

\bibitem{zhang2018performance}
J.~Zhang, P.~Li, B.~Liu, T.~G. Marbach, X.~Liu, and G.~Wang, ``{Performance Analysis of 3D XPoint SSDs in Virtualized and Non-Virtualized Environments},'' in \emph{ICPADS}, 2018.

\bibitem{chien2018characterizing}
S.~W. Chien, S.~Markidis, C.~P. Sishtla, L.~Santos, P.~Herman, S.~Narasimhamurthy, and E.~Laure, ``{Characterizing Deep-Learning I/O Workloads in TensorFlow},'' in \emph{PDSW-DISCS}, 2018.

\bibitem{yang2020exploring}
J.~Yang, B.~Li, and D.~J. Lilja, ``{Exploring Performance Characteristics of the Optane 3D XPoint Storage Technology},'' \emph{TOMPECS}, 2020.

\bibitem{hady2017platform}
F.~T. Hady, A.~Foong, B.~Veal, and D.~Williams, ``{Platform Storage Performance with 3D XPoint Technology},'' \emph{Proc. IEEE}, 2017.

\bibitem{wu2019exploiting}
K.~Wu, A.~Arpaci-Dusseau, R.~Arpaci-Dusseau, R.~Sen, and K.~Park, ``{Exploiting Intel Optane SSD for Microsoft SQL Server},'' in \emph{DaMoN}, 2019.

\bibitem{imamura2018reducing}
S.~Imamura and E.~Yoshida, ``{Reducing CPU Power Consumption for Low-Latency SSDs},'' in \emph{NVMSA}, 2018.

\bibitem{optanePrice}
{Amazon.com, Inc.}, ``{Intel Optane Memory Module 16GB M.2 80mm PCIe 3.0 20nm 3D XPoint MEMPEK1W016GA},'' \url{https://amzn.to/33Z6bws}.

\bibitem{dramcost}
{DRAMeXchange}, ``{World Leading DRAM and NAND Flash Market Research Firm, with More than a Decade of Most Authoritative Database},'' \url{https://www.dramexchange.com/}.

\bibitem{peng2019system}
I.~B. Peng, M.~B. Gokhale, and E.~W. Green, ``{System Evaluation of the Intel Optane Byte-Addressable NVM},'' in \emph{MemSys}, 2019.

\bibitem{metzler2015prototyping}
B.~Metzler and A.~Trivedi, ``{Prototyping Byte-Addressable NVM Access},'' in \emph{OpenFabrics Developers Workshop}, 2015.

\bibitem{hassan2015energy}
A.~Hassan, H.~Vandierendonck, and D.~S. Nikolopoulos, ``{Energy-Efficient Hybrid DRAM/NVM Main Memory},'' in \emph{PACT}, 2015.

\bibitem{chauhan2016nvmove}
H.~Chauhan, I.~Calciu, V.~Chidambaram, E.~Schkufza, O.~Mutlu, and P.~Subrahmanyam, ``{NVMOVE: Helping Programmers Move to Byte-Based Persistence},'' in \emph{INFLOW}, 2016.

\bibitem{yoon2014efficient}
H.~Yoon, J.~Meza, N.~Muralimanohar, N.~P. Jouppi, and O.~Mutlu, ``{Efficient Data Mapping and Buffering Techniques for Multilevel Cell Phase-Change Memories},'' \emph{TACO}, 2014.

\bibitem{li2017utility}
Y.~Li, S.~Ghose, J.~Choi, J.~Sun, H.~Wang, and O.~Mutlu, ``{Utility-Based Hybrid Memory Management},'' in \emph{CLUSTER}, 2017.

\bibitem{zhang2021chameleondb}
W.~Zhang, X.~Zhao, S.~Jiang, and H.~Jiang, ``{ChameleonDB: A Key-Value Store for Optane Persistent Memory},'' in \emph{EuroSys}, 2021.

\bibitem{bae20182b}
D.-H. Bae, I.~Jo, Y.~A. Choi, J.-Y. Hwang, S.~Cho, D.-G. Lee, and J.~Jeong, ``{2B-SSD: The Case for Dual, Byte- and Block-Addressable Solid-State Drives},'' in \emph{ISCA}, 2018.

\bibitem{kim2018optimized}
S.~Kim and J.-S. Yang, ``{Optimized I/O Determinism for Emerging NVM-Based NVMe SSD in an Enterprise System},'' in \emph{DAC}, 2018.

\bibitem{cai2017error}
Y.~Cai, S.~Ghose, E.~F. Haratsch, Y.~Luo, and O.~Mutlu, ``{Error Characterization, Mitigation, and Recovery in Flash-Memory-Based Solid-State Drives},'' \emph{Proc. IEEE}, 2017.

\bibitem{luo2018improving}
Y.~Luo, S.~Ghose, Y.~Cai, E.~F. Haratsch, and O.~Mutlu, ``{Improving 3D NAND Flash Memory Lifetime by Tolerating Early Retention Loss and Process Variation},'' in \emph{SIGMETRICS}, 2018.

\bibitem{luo2018heatwatch}
Y.~Luo, S.~Ghose, Y.~Cai, E.~F. Haratsch, and O.~Mutlu, ``{HeatWatch: Improving 3D NAND Flash Memory Device Reliability by Exploiting Self-Recovery and Temperature Awareness},'' in \emph{HPCA}, 2018.

\bibitem{cai2017vulnerabilities}
Y.~Cai, S.~Ghose, Y.~Luo, K.~Mai, O.~Mutlu, and E.~F. Haratsch, ``{Vulnerabilities in MLC NAND Flash Memory Programming: Experimental Analysis, Exploits, and Mitigation Techniques},'' in \emph{HPCA}, 2017.

\bibitem{luo2016enabling}
Y.~Luo, S.~Ghose, Y.~Cai, E.~F. Haratsch, and O.~Mutlu, ``{Enabling Accurate and Practical Online Flash Channel Modeling for Modern MLC NAND Flash Memory},'' \emph{JSAC}, 2016.

\bibitem{cai2015read}
Y.~Cai, Y.~Luo, S.~Ghose, and O.~Mutlu, ``{Read Disturb Errors in MLC NAND Flash Memory: Characterization, Mitigation, and Recovery},'' in \emph{DSN}, 2015.

\bibitem{lecturenand}
O.~Mutlu, ``{Lecture Notes for Computer Architecture -- Lecture 26: Flash Memory and Solid-State Drives},'' \url{https://rb.gy/xqis8}, 2020.

\bibitem{ke2018lirs}
Z.-L. Ke, H.-Y. Cheng, and C.-L. Yang, ``{LIRS: Enabling Efficient Machine Learning on NVM-Based Storage via a Lightweight Implementation of Random Shuffling},'' arXiv:1810.04509 [cs.AR], 2018.

\bibitem{liu2020nvm}
X.~Liu, Y.~Pan, Y.~Li, G.~Wang, and X.~Liu, ``{An NVM SSD-Optimized Query Processing Framework},'' in \emph{CIKM}, 2020.

\bibitem{han2020splitkv}
S.~Han, D.~Jiang, and J.~Xiong, ``{SplitKV: Splitting IO Paths for Different Sized Key-Value Items with Advanced Storage Devices},'' in \emph{HotStorage}, 2020.

\bibitem{papagiannis2020optimizing}
A.~Papagiannis, G.~Xanthakis, G.~Saloustros, M.~Marazakis, and A.~Bilas, ``{Optimizing Memory-Mapped I/O for Fast Storage Devices},'' in \emph{USENIX ATC}, 2020.

\bibitem{jia2020flash}
Y.~Jia and F.~Chen, ``{From Flash to 3D XPoint: Performance Bottlenecks and Potentials in RocksDB with Storage Evolution},'' in \emph{ISPASS}, 2020.

\bibitem{wu2019towards}
K.~Wu, A.~Arpaci-Dusseau, and R.~Arpaci-Dusseau, ``{Towards an Unwritten Contract of Intel Optane SSD},'' in \emph{HotStorage}, 2019.

\bibitem{zhong2014building}
K.~Zhong, T.~Wang, X.~Zhu, L.~Long, D.~Liu, W.~Liu, Z.~Shao, and E.~H.-M. Sha, ``{Building High-Performance Smartphones via Non-Volatile Memory: The Swap Approach},'' in \emph{EMSOFT}, 2014.

\bibitem{kim2015cause}
Y.~Kim, M.~Imani, S.~Patil, and T.~S. Rosing, ``{CAUSE: Critical Application Usage-Aware Memory System Using Non-Volatile Memory for Mobile Devices},'' in \emph{ICCAD}, 2015.

\bibitem{liu2017non}
D.~Liu, K.~Zhong, X.~Zhu, Y.~Li, L.~Long, and Z.~Shao, ``{Non-Volatile Memory Based Page Swapping for Building High-Performance Mobile Devices},'' \emph{TC}, 2017.

\bibitem{zhong2017building}
K.~Zhong, D.~Liu, L.~Long, J.~Ren, Y.~Li, and E.~H.-M. Sha, ``{Building NVRAM-Aware Swapping Through Code Migration in Mobile Devices},'' \emph{TPDS}, 2017.

\bibitem{kim2019analysis}
J.~Kim and H.~Bahn, ``{Analysis of Smartphone I/O Characteristics --- Toward Efficient Swap in a Smartphone},'' \emph{IEEE Access}, 2019.

\bibitem{zhu2017smartswap}
X.~Zhu, D.~Liu, K.~Zhong, J.~Ren, and T.~Li, ``{SmartSwap: High-Performance and User Experience Friendly Swapping in Mobile Systems},'' in \emph{DAC}, 2017.

\bibitem{kim2018comparison}
J.~Kim and H.~Bahn, ``{Comparison of Hybrid and Hierarchical Swap Architectures in Android by Using NVM},'' \emph{JSTS}, 2018.

\bibitem{zhong2014dr}
K.~Zhong, X.~Zhu, T.~Wang, D.~Zhang, X.~Luo, D.~Liu, W.~Liu, and E.~H.-M. Sha, ``{DR. Swap: Energy-Efficient Paging for Smartphones},'' in \emph{ISLPED}, 2014.

\bibitem{kim2017application}
S.-H. Kim, J.~Jeong, and J.-S. Kim, ``{Application-Aware Swapping for Mobile Systems},'' \emph{TECS}, 2017.

\bibitem{kim2019ezswap}
J.~Kim, C.~Kim, and E.~Seo, ``{$ ezswap $: Enhanced Compressed Swap Scheme for Mobile Devices},'' \emph{IEEE Access}, 2019.

\bibitem{kim2020maintaining}
J.~Kim and H.~Bahn, ``{Maintaining Application Context of Smartphones by Selectively Supporting Swap and Kill},'' \emph{IEEE Access}, 2020.

\bibitem{guo2015mars}
W.~Guo, K.~Chen, H.~Feng, Y.~Wu, R.~Zhang, and W.~Zheng, ``{$ MARS $: Mobile Application Relaunching Speed-Up through Flash-Aware Page Swapping},'' \emph{IEEE Trans. Comput.}, 2015.

\bibitem{liang2020acclaim}
Y.~Liang, J.~Li, R.~Ausavarungnirun, R.~Pan, L.~Shi, T.-W. Kuo, and C.~J. Xue, ``{Acclaim: Adaptive Memory Reclaim to Improve User Experience in Android Systems},'' in \emph{USENIX ATC}, 2020.

\bibitem{chrome}
{Google LLC}, ``{Chrome Browser},'' \url{https://www.google.com/chrome/}.

\bibitem{memorypressure}
{Chromium Project}, ``{MemoryPressure Tast Test},'' \url{https://rb.gy/j1ft7}.

\bibitem{intel2018900p}
{Intel Corp.}, ``{Intel Optane SSD 900P Series},'' 2018.

\bibitem{gutierrez2011full}
A.~Gutierrez, R.~G. Dreslinski, T.~F. Wenisch, T.~Mudge, A.~Saidi, C.~Emmons, and N.~Paver, ``{Full-System Analysis and Characterization of Interactive Smartphone Applications},'' in \emph{IISWC}, 2011.

\bibitem{huang2014moby}
Y.~Huang, Z.~Zha, M.~Chen, and L.~Zhang, ``{Moby: A Mobile Benchmark Suite for Architectural Simulators},'' in \emph{ISPASS}, 2014.

\bibitem{pandiyan2013performance}
D.~Pandiyan, S.-Y. Lee, and C.-J. Wu, ``{Performance, Energy Characterizations and Architectural Implications of an Emerging Mobile Platform Benchmark Suite - MobileBench},'' in \emph{IISWC}, 2013.

\bibitem{popper2017google}
B.~Popper, ``{Google Announces Over 2 Billion Monthly Active Devices on Android},'' \url{https://rb.gy/yyk1b}, 2017.

\bibitem{googleshare}
{Net Applications}, ``{Market Share Statistics for Internet Technologies},'' \url{https://www.netmarketshare.com/}.

\bibitem{blink}
{Chromium Project}, ``{Blink Rendering Engine},'' \url{https://rb.gy/j32v9}.

\bibitem{skia}
{Google LLC}, ``{Skia Graphics Library},'' \url{https://skia.org/}.

\bibitem{reis2009isolating}
C.~Reis and S.~D. Gribble, ``{Isolating Web Programs in Modern Browser Architectures},'' in \emph{EuroSys}, 2009.

\bibitem{barth2008security}
A.~Barth, C.~Jackson, C.~Reis, T.~Team \emph{et~al.}, ``{The Security Architecture of the Chromium Browser},'' in \emph{Technical Report}.\hskip 1em plus 0.5em minus 0.4em\relax Stanford University, 2008.

\bibitem{httparchieve}
{HTTP Archive}, \url{http://httparchive.org/}.

\bibitem{rientjes2010oom}
D.~Rientjes, ``{OOM Killer Rewrite; When the Kernel Runs Out of Memory},'' \emph{LinuxCon Boston}, 2010.

\bibitem{collins2011android}
C.~Collins, M.~Galpin, and M.~K{\"a}ppler, \emph{{Android in Practice}}.\hskip 1em plus 0.5em minus 0.4em\relax Manning Publications Co., 2011.

\bibitem{pixel}
{Google LLC}, ``{Pixel Smartphones},'' \url{https://www.google.com/pixel/}.

\bibitem{jennings2013transparent}
S.~Jennings, ``{Transparent Memory Compression in Linux},'' \emph{LinuxCon}, 2013.

\bibitem{shiu2015system}
E.~Shiu and S.~Prakash, ``{System Challenges and Hardware Requirements for Future Consumer Devices: From Wearable to ChromeBooks and Devices In-Between},'' in \emph{VLSI Technology}, 2015.

\bibitem{shiu2017driving}
E.~Shiu and S.~Lim, ``{Driving Innovation in Memory Architecture of Consumer Hardware with Digital Photography and Machine Intelligence Use Cases},'' in \emph{IMW}, 2017.

\bibitem{pekhimenko2013linearly}
G.~Pekhimenko, V.~Seshadri, Y.~Kim, H.~Xin, O.~Mutlu, P.~B. Gibbons, M.~A. Kozuch, and T.~C. Mowry, ``{Linearly Compressed Pages: A Low-Complexity, Low-Latency Main Memory Compression Framework},'' in \emph{MICRO}, 2013.

\bibitem{MemoryCo67}
{Chromium Project}, ``{Memory Coordinator},'' \url{https://bit.ly/3lOW7w7}, 2016.

\bibitem{grigorikPerf}
I.~Grigorik, ``{High Performance Networking in Chrome},'' The Performance of Open Source Applications: Speed, Precision, and a Bit of Serendipity, 2013.

\bibitem{lohrImpatient}
S.~Lohr, ``{For Impatient Web Users, an Eye Blink Is Just Too Long to Wait},'' \url{https://rb.gy/50zon}, 2012.

\bibitem{Chi2016}
P.~Chi, S.~Li, C.~Xu, T.~Zhang, J.~Zhao, Y.~Liu, Y.~Wang, and Y.~Xie, ``{PRIME: A Novel Processing-in-Memory Architecture for Neural Network Computation in ReRAM-Based Main Memory},'' in \emph{ISCA}, 2016.

\bibitem{song2018graphr}
L.~Song, Y.~Zhuo, X.~Qian, H.~Li, and Y.~Chen, ``{GraphR: Accelerating Graph Processing Using ReRAM},'' in \emph{HPCA}, 2018.

\bibitem{song2017pipelayer}
L.~Song, X.~Qian, H.~Li, and Y.~Chen, ``{PipeLayer: A Pipelined ReRAM-Based Accelerator for Deep Learning},'' in \emph{HPCA}, 2017.

\bibitem{yao2017face}
P.~Yao, H.~Wu, B.~Gao, S.~B. Eryilmaz, X.~Huang, W.~Zhang, Q.~Zhang, N.~Deng, L.~Shi, H.-S.~P. Wong \emph{et~al.}, ``{Face Classification Using Electronic Synapses},'' \emph{Nature Communications}, 2017.

\bibitem{hu2016dot}
M.~Hu, J.~P. Strachan, Z.~Li, E.~M. Grafals, N.~Davila, C.~Graves, S.~Lam, N.~Ge, J.~J. Yang, and R.~S. Williams, ``{Dot-Product Engine for Neuromorphic Computing: Programming 1T1M Crossbar to Accelerate Matrix-Vector Multiplication},'' in \emph{DAC}, 2016.

\bibitem{gopalan2011demonstration}
C.~Gopalan, Y.~Ma, T.~Gallo, J.~Wang, E.~Runnion, J.~Saenz, F.~Koushan, P.~Blanchard, and S.~Hollmer, ``{Demonstration of Conductive Bridging Random Access Memory (CBRAM) in Logic CMOS Process},'' \emph{Solid-State Electronics}, 2011.

\bibitem{jana2015conductive}
D.~Jana, S.~Roy, R.~Panja, M.~Dutta, S.~Z. Rahaman, R.~Mahapatra, and S.~Maikap, ``{Conductive-Bridging Random Access Memory: Challenges and Opportunity for 3D Architecture},'' \emph{NRL}, 2015.

\bibitem{cha2020conductive}
J.-H. Cha, S.~Y. Yang, J.~Oh, S.~Choi, S.~Park, B.~C. Jang, W.~Ahn, and S.-Y. Choi, ``{Conductive-Bridging Random-Access Memories for Emerging Neuromorphic Computing},'' \emph{Nanoscale}, 2020.

\bibitem{scott1989ferroelectric}
J.~F. Scott and C.~A.~P. De~Araujo, ``{Ferroelectric Memories},'' \emph{Science}, 1989.

\bibitem{scott2007applications}
J.~Scott, ``{Applications of Modern Ferroelectrics},'' \emph{Science}, 2007.

\bibitem{mikolajick2001feram}
T.~Mikolajick, C.~Dehm, W.~Hartner, I.~Kasko, M.~Kastner, N.~Nagel, M.~Moert, and C.~Mazure, ``{FeRAM Technology for High Density Applications},'' \emph{Microelectronics Reliability}, 2001.

\bibitem{webb20163d}
M.~Webb, ``{3D XPoint Status and Forecast},'' in \emph{Flash Memory Summit}, 2016.

\bibitem{cutress2016intel}
I.~Cutress and B.~Tallis, ``{Intel Launches Optane DIMMs Up to 512GB: Apache Pass Is Here!}'' AnandTech, 2016.

\bibitem{specification2002pci}
PCI-SIG, ``{PCI Express Base Specification Revision 5.0, Version 1.0},'' 2019.

\bibitem{patil2019performance}
O.~Patil, L.~Ionkov, J.~Lee, F.~Mueller, and M.~Lang, ``{Performance Characterization of a DRAM-NVM Hybrid Memory Architecture for HPC Applications Using Intel Optane DC Persistent Memory Modules},'' in \emph{MEMSYS}, 2019.

\bibitem{gill2019single}
G.~Gill, R.~Dathathri, L.~Hoang, R.~Peri, and K.~Pingali, ``{Single Machine Graph Analytics on Massive Datasets Using Intel Optane DC Persistent Memory},'' arXiv:1904.07162 [cs.AR], 2019.

\bibitem{wu2020lessons}
Y.~Wu, K.~Park, R.~Sen, B.~Kroth, and J.~Do, ``{Lessons Learned from the Early Performance Evaluation of Intel Optane DC Persistent Memory in DBMS},'' in \emph{DaMoN}, 2020.

\bibitem{weiland2019early}
M.~Weiland, H.~Brunst, T.~Quintino, N.~Johnson, O.~Iffrig, S.~Smart, C.~Herold, A.~Bonanni, A.~Jackson, and M.~Parsons, ``{An Early Evaluation of Intel's Optane DC Persistent Memory Module and Its Impact on High-Performance Scientific Applications},'' in \emph{SC}, 2019.

\bibitem{shanbhag2020large}
A.~Shanbhag, N.~Tatbul, D.~Cohen, and S.~Madden, ``{Large-Scale In-Memory Analytics on Intel{\textregistered} Optane™ DC Persistent Memory},'' in \emph{DaMoN}, 2020.

\bibitem{mironov2019performance}
V.~Mironov, I.~Chernykh, I.~Kulikov, A.~Moskovsky, E.~Epifanovsky, and A.~Kudryavtsev, ``{Performance Evaluation of the Intel Optane DC Memory With Scientific Benchmarks},'' in \emph{MCHPC}, 2019.

\bibitem{yang2020empirical}
J.~Yang, J.~Kim, M.~Hoseinzadeh, J.~Izraelevitz, and S.~Swanson, ``{An Empirical Guide to the Behavior and Use of Scalable Persistent Memory},'' in \emph{FAST}, 2020.

\bibitem{benson_perma_2022}
L.~Benson, L.~Papke, and T.~Rabl, ``{PerMA-Bench: Benchmarking Persistent Memory Access},'' \emph{VLDB Endow.}, 2022.

\bibitem{xiang2022characterizing}
L.~Xiang, X.~Zhao, J.~Rao, S.~Jiang, and H.~Jiang, ``{Characterizing the Performance of Intel Optane Persistent Memory: A Close Look at Its On-DIMM Buffering},'' in \emph{EuroSys}, 2022.

\bibitem{IntelOpt84}
{Tom's Hardware}, ``{Intel Optane DIMM Pricing},'' \url{https://rb.gy/873zd}, 2019.

\bibitem{bittman2020twizzler}
D.~Bittman, P.~Alvaro, P.~Mehra, D.~D. Long, and E.~L. Miller, ``{Twizzler: A Data-Centric OS for Non-Volatile Memory},'' in \emph{USENIX ATC}, 2020.

\bibitem{chromebox}
{Asus, Inc.}, ``{Asus Chromebox 3},'' \url{https://rb.gy/e9cnq}.

\bibitem{wright2009ready}
A.~Wright, ``{Ready for a Web OS?}'' \emph{CACM}, 2009.

\bibitem{i3}
{Intel Corp.}, ``{Intel® Core i3-7100U Processor},'' \url{https://rb.gy/2ifwc}, 2016.

\bibitem{skhynixddr4}
{SK Hynix Inc.}, ``{SK Hynix 4GB DDR4 HMA851S6AFR6N-UH}.''

\bibitem{transcendssd}
{Transcend Information, Inc.}, ``{SATA III M.2 Solid State Drive M.2 SSD 400S},'' 2020.

\bibitem{m10}
{Intel Corp.}, ``{Intel{\textsuperscript{\textregistered}} Optane{\texttrademark} Memory M10 Series},'' \url{https://rb.gy/atuol}.

\bibitem{facebook}
{Facebook, Inc.}, ``{Facebook},'' \url{https://www.facebook.com/}.

\bibitem{instagram}
{Facebook, Inc.}, ``Instagram,'' \url{https://about.instagram.com/about-us/}.

\bibitem{whatsapp}
{Facebook, Inc.}, ``{WhatsApp Messenger},'' \url{https://www.whatsapp.com/}.

\bibitem{telegram}
{Telegram FZ-LLC}, ``{Telegram Messenger},'' \url{https://telegram.org/}.

\bibitem{adobe}
{Adobe Inc.}, ``{Adobe Acrobat Reader},'' \url{https://get.adobe.com/reader/}.

\bibitem{minecraft}
{Mojang Studios}, ``{Minecraft},'' \url{https://www.minecraft.net/}.

\bibitem{guthaus2001mibench}
M.~R. Guthaus, J.~S. Ringenberg, D.~Ernst, T.~M. Austin, T.~Mudge, and R.~B. Brown, ``{MiBench: A Free, Commercially Representative Embedded Benchmark Suite},'' in \emph{WWC}, 2001.

\bibitem{lee1997mediabench}
C.~Lee, M.~Potkonjak, and W.~H. Mangione-Smith, ``{MediaBench: A Tool for Evaluating and Synthesizing Multimedia and Communications Systems},'' in \emph{MICRO}, 1997.

\bibitem{chromium}
{Chromium Project}, ``{The Chromium Project},'' \url{https://rb.gy/osxys}.

\bibitem{perf}
{Linux Kernel Organization, Inc.}, ``{perf: Linux Profiling with Performance Counters},'' https://perf.wiki.kernel.org/index.php/Main\_Page.

\bibitem{youtube}
{Google LLC}, ``{YouTube},'' \url{https://www.youtube.com}.

\bibitem{maps}
{Google LLC}, ``{Google Maps},'' \url{http://maps.google.com/}.

\bibitem{sheets}
{Google LLC}, ``{Google Sheets},'' \url{https://www.google.com/sheets/about/}.

\bibitem{docs}
{Google LLC}, ``{Google Docs},'' \url{https://www.google.com/docs/about/}.

\bibitem{twitter}
{Twitter, Inc.}, ``Twitter,'' \url{https://www.twitter.com/}.

\bibitem{chromiummedia}
{Chromium Project}, ``{Chromium Media},'' \url{https://rb.gy/n4bnw}.

\bibitem{ffmpegDo29}
{FFmpeg Team}, ``{FFmpeg Documentation},'' \url{https://rb.gy/tr393}.

\bibitem{TheWebMP86}
{WebM Project}, ``{WebM},'' \url{https://www.webmproject.org/code/}.

\bibitem{grange2016vp9}
A.~Grange, P.~De~Rivaz, and J.~Hunt, ``{VP9 Bitstream \& Decoding Process Specification},'' \url{https://rb.gy/1cfjh}.

\bibitem{HTMLStan14}
{Web Hypertext Application Technology Working Group}, ``{HTML Living Standard},'' \url{https://html.spec.whatwg.org/multipage/}, 2021.

\bibitem{Mojodocs82}
{Chromium Project}, ``{Mojo},'' \url{https://rb.gy/ynbj8}.

\bibitem{VaAPI61}
{Chromium Project}, ``{VaAPI},'' \url{https://rb.gy/5ri9b}.

\bibitem{srccodec76}
{Chromium Project}, ``{SkGifCodec},'' \url{https://rb.gy/gzenm}.

\bibitem{V8JavaSc93}
{Chromium Project}, ``{V8 JavaScript Engine},'' \url{https://v8.dev/}.

\bibitem{TastTast15}
{Chromium Project}, ``{Tast},'' \url{https://rb.gy/073gv}.

\bibitem{chromeau55}
{Google LLC}, ``{chrome.automation},'' \url{https://rb.gy/t6kxp}.

\bibitem{choe2017intel}
J.~Choe, ``{Intel 3D XPoint Memory Die Removed from Intel Optane PCM (Phase Change Memory)},'' TechInsights, 2017.

\bibitem{chen2012energy}
J.~Chen, R.~C. Chiang, H.~H. Huang, and G.~Venkataramani, ``{Energy-Aware Writes to Non-Volatile Main Memory},'' \emph{OSR}, 2012.

\bibitem{zolnierkiewicz2013efficient}
B.~Zolnierkiewicz, ``{Efficient Memory Management on Mobile Devices},'' \emph{LinuxCon}, 2013.

\bibitem{tanaka2005monitoring}
B.~K. Tanaka, ``{Monitoring Virtual Memory With vmstat},'' \emph{Linux Journal}, 2005.

\bibitem{guo2018latency}
Y.~Guo, Y.~Hua, and P.~Zuo, ``{A Latency-Optimized and Energy-Efficient Write Scheme in NVM-Based Main Memory},'' \emph{TCADICS}, 2018.

\bibitem{choi2017nvm}
J.-H. Choi and G.-H. Park, ``{NVM Way Allocation Scheme to Reduce NVM Writes for Hybrid Cache Architecture in Chip-Multiprocessors},'' \emph{TPDS}, 2017.

\bibitem{swami2016secret}
S.~Swami, J.~Rakshit, and K.~Mohanram, ``{SECRET: Smartly Encrypted Energy Efficient Non-Volatile Memories},'' in \emph{DAC}, 2016.

\bibitem{optaneendurance}
{Intel Corp.}, ``{Intel{\textsuperscript{\textregistered}} Optane{\texttrademark} Memory},'' \url{https://rb.gy/v31hy}.

\bibitem{chang2016improving}
Y.-M. Chang, P.-C. Hsiu, Y.-H. Chang, C.-H. Chen, T.-W. Kuo, and C.-Y.~M. Wang, ``{Improving PCM Endurance with a Constant-Cost Wear Leveling Design},'' \emph{TODAES}, 2016.

\bibitem{aghaei2014prolonging}
H.~Aghaei~Khouzani, Y.~Xue, C.~Yang, and A.~Pandurangi, ``{Prolonging PCM Lifetime Through Energy-Efficient, Segment-Aware, and Wear-Resistant Page Allocation},'' in \emph{ISLPED}, 2014.

\bibitem{IntelOp22:online}
F.~T. Hady, ``{Intel Optane Technology Delivers New Levels of Endurance},'' \url{https://rb.gy/c83ee}, (Accessed on 06/21/2023).

\bibitem{qureshi2009enhancing}
M.~K. Qureshi, J.~Karidis, M.~Franceschini, V.~Srinivasan, L.~Lastras, and B.~Abali, ``{Enhancing Lifetime and Security of PCM-Based Main Memory with Start-Gap Wear Leveling},'' in \emph{MICRO}, 2009.

\bibitem{micronslc}
{Micron Technology, Inc.}, ``{SLC NAND},'' \url{https://rb.gy/855sx}.

\bibitem{optane_pe_cycles}
J.~Handy, ``{Examining 3D XPoint’s 1,000 Times Endurance Benefit – The Memory Guy},'' \url{https://rb.gy/mvd5a}.

\bibitem{ssdprice}
diskprices.com, ``{Disk Prices (US)},'' \url{https://bit.ly/2STO9We}, accessed on April 2, 2020.

\bibitem{de2010new}
A.~C. de~Melo, ``{The New Linux 'perf' Tools},'' in \emph{Linux Kongress}, 2010.

\bibitem{bjorling2013linux}
M.~Bj{\o}rling, J.~Axboe, D.~Nellans, and P.~Bonnet, ``{Linux Block IO: Introducing Multi-Queue SSD Access on Multi-Core Systems},'' in \emph{SYSTOR}, 2013.

\bibitem{tavakkol2018flin}
A.~Tavakkol, M.~Sadrosadati, S.~Ghose, J.~Kim, Y.~Luo, Y.~Wang, N.~M. Ghiasi, L.~Orosa, J.~G{\'o}mez-Luna, and O.~Mutlu, ``{FLIN: Enabling Fairness and Enhancing Performance in Modern NVMe Solid State Drives},'' in \emph{ISCA}, 2018.

\bibitem{tavakkol2018mqsim}
A.~Tavakkol, J.~G{\'o}mez-Luna, M.~Sadrosadati, S.~Ghose, and O.~Mutlu, ``{MQSim: A Framework for Enabling Realistic Studies of Modern Multi-Queue SSD Devices},'' in \emph{FAST}, 2018.

\bibitem{brunelle2007blktrace}
A.~D. Brunelle, ``{Blktrace User Guide},'' 2007.

\bibitem{blkmqKyb12}
O.~Sandoval, ``{Kyber MQ I/O Scheduler},'' \url{https://rb.gy/6azue}, 2017.

\bibitem{mqdeadline}
J.~Axboe, ``{MQ Deadline I/O Scheduler},'' \url{https://rb.gy/xxdro}, 2016.

\bibitem{bennett1997hierarchical}
J.~C. Bennett and H.~Zhang, ``{Hierarchical Packet Fair Queueing Algorithms},'' \emph{TON}, 1997.

\bibitem{yang2012poll}
J.~Yang, D.~B. Minturn, and F.~Hady, ``{When Poll is Better Than Interrupt},'' in \emph{FAST}, 2012.

\bibitem{le2017latency}
D.~Le~Moal, ``{I/O Latency Optimization with Polling},'' in \emph{Vault}, 2017.

\bibitem{Queuesys14:online}
{Linux Kernel Organization, Inc.}, ``{Linux Kernel Documentation: Queue sysfs Files},'' \url{https://rb.gy/9thuf}, 2009.

\bibitem{hybridoptane}
{Intel Corp.}, ``{Tuning the Performance of Intel Optane SSDs on Linux Operating Systems},'' \url{https://rb.gy/uunhr}.

\bibitem{yavits2020wolfram}
L.~Yavits, L.~Orosa, S.~Mahar, J.~D. Ferreira, M.~Erez, R.~Ginosar, and O.~Mutlu, ``{WoLFRaM: Enhancing Wear-Leveling and Fault Tolerance in Resistive Memories using Programmable Address Decoders},'' in \emph{ICCD}, 2020.

\bibitem{chen2012age}
C.-H. Chen, P.-C. Hsiu, T.-W. Kuo, C.-L. Yang, and C.-Y.~M. Wang, ``{Age-Based PCM Wear Leveling with Nearly Zero Search Cost},'' in \emph{DAC}, 2012.

\bibitem{cheng2016efficient}
S.-W. Cheng, Y.-H. Chang, T.-Y. Chen, Y.-F. Chang, H.-W. Wei, and W.-K. Shih, ``{Efficient Warranty-Aware Wear Leveling for Embedded Systems with PCM Main Memory},'' \emph{VLSI}, 2016.

\bibitem{fan2014wl}
J.~Fan, S.~Jiang, J.~Shu, L.~Sun, and Q.~Hu, ``{WL-Reviver: A Framework for Reviving any Wear-Leveling Techniques in the Face of Failures on Phase Change Memory},'' in \emph{DSN}, 2014.

\bibitem{han2015enhanced}
Y.~Han, J.~Dong, K.~Weng, Y.~Wang, and X.~Li, ``{Enhanced Wear-Rate Leveling for PRAM Lifetime Improvement Considering Process Variation},'' \emph{VLSI}, 2015.

\bibitem{im2014differentiated}
S.~Im and D.~Shin, ``{Differentiated Space Allocation for Wear Leveling on Phase-Change Memory-Based Storage Device},'' \emph{TCE}, 2014.

\bibitem{joo2010energy}
Y.~Joo, D.~Niu, X.~Dong, G.~Sun, N.~Chang, and Y.~Xie, ``{Energy-and Endurance-Aware Design of Phase Change Memory Caches},'' in \emph{DATE}, 2010.

\bibitem{liu2014application}
D.~Liu, T.~Wang, Y.~Wang, Z.~Shao, Q.~Zhuge, and E.~H.-M. Sha, ``{Application-Specific Wear Leveling for Extending Lifetime of Phase Change Memory in Embedded Systems},'' \emph{TCAD}, 2014.

\bibitem{qureshi2011practical}
M.~K. Qureshi, A.~Seznec, L.~A. Lastras, and M.~M. Franceschini, ``{Practical and Secure PCM Systems by Online Detection of Malicious Write Streams},'' in \emph{HPCA}, 2011.

\bibitem{lee2020case}
G.~Lee, W.~Jin, W.~Song, J.~Gong, J.~Bae, T.~J. Ham, J.~W. Lee, and J.~Jeong, ``{A Case for Hardware-Based Demand Paging},'' in \emph{ISCA}, 2020.

\bibitem{oh2020h}
K.~Oh, J.~Park, and Y.~I. Eom, ``{H-BFQ: Supporting Multi-Level Hierarchical Cgroup in BFQ Scheduler},'' in \emph{BigCom}, 2020.

\bibitem{shin2014path}
W.~Shin, Q.~Chen, M.~Oh, H.~Eom, and H.~Y. Yeom, ``{OS I/O Path Optimizations for Flash Solid-State Drives},'' in \emph{USENIX ATC}, 2014.

\bibitem{vuvcinic2014dc}
D.~Vu{\v{c}}ini{\'c}, Q.~Wang, C.~Guyot, R.~Mateescu, F.~Blagojevi{\'c}, L.~Franca-Neto, D.~Le~Moal, T.~Bunker, J.~Xu, S.~Swanson \emph{et~al.}, ``{DC Express: Shortest Latency Protocol for Reading Phase Change Memory Over PCI Express},'' in \emph{FAST}, 2014.

\bibitem{zhang2018flashshare}
J.~Zhang, M.~Kwon, D.~Gouk, S.~Koh, C.~Lee, M.~Alian, M.~Chun, M.~T. Kandemir, N.~S. Kim, J.~Kim \emph{et~al.}, ``{FlashShare: Punching Through Server Storage Stack from Kernel to Firmware for Ultra-Low Latency SSDs},'' in \emph{OSDI}, 2018.

\bibitem{liu2022towards}
M.~Liu, H.~Liu, C.~Ye, X.~Liao, H.~Jin, Y.~Zhang, R.~Zheng, and L.~Hu, ``{Towards Low-Latency I/O Services for Mixed Workloads Using Ultra-Low Latency SSDs},'' in \emph{ICS}, 2022.

\bibitem{caulfield2012providing}
A.~M. Caulfield, T.~I. Mollov, L.~A. Eisner, A.~De, J.~Coburn, and S.~Swanson, ``{Providing Safe, User Space Access to Fast, Solid State Disks},'' in \emph{ASPLOS}, 2012.

\bibitem{kim2016nvmedirect}
H.-J. Kim, Y.-S. Lee, and J.-S. Kim, ``{NVMeDirect: A User-Space I/O Framework for Application-Specific Optimization on NVMe SSDs},'' in \emph{HotStorage}, 2016.

\bibitem{scargall2020introducing}
S.~Scargall, ``{Introducing the Persistent Memory Development Kit},'' in \emph{Programming Persistent Memory}.\hskip 1em plus 0.5em minus 0.4em\relax Springer, 2020.

\bibitem{yang2017spdk}
Z.~Yang, J.~R. Harris, B.~Walker, D.~Verkamp, C.~Liu, C.~Chang, G.~Cao, J.~Stern, V.~Verma, and L.~E. Paul, ``{SPDK: A Development Kit to Build High Performance Storage Applications},'' in \emph{CloudCom}, 2017.

\bibitem{OpenMPDK}
{Samsung Electronics Co., Ltd.}, ``{Open Memory Platform Development Kit: User Level NVMe Driver},'' \url{https://github.com/OpenMPDK/uNVMe}.

\bibitem{peter2015arrakis}
S.~Peter, J.~Li, I.~Zhang, D.~R. Ports, D.~Woos, A.~Krishnamurthy, T.~Anderson, and T.~Roscoe, ``{Arrakis: The Operating System Is the Control Plane},'' \emph{TOCS}, 2015.

\bibitem{kim2017user}
H.-J. Kim and J.-S. Kim, ``{A User-Space Storage I/O Framework for NVMe SSDs in Mobile Smart Devices},'' \emph{TCE}, 2017.

\bibitem{kwon2017strata}
Y.~Kwon, H.~Fingler, T.~Hunt, S.~Peter, E.~Witchel, and T.~Anderson, ``{Strata: A Cross Media File System},'' in \emph{{SOSP}}, 2017.

\bibitem{wu2017early}
K.~Wu, F.~Ober, S.~Hamlin, and D.~Li, ``{Early Evaluation of Intel Optane Non-Volatile Memory with HPC I/O Workloads},'' arXiv:1708.02199 [cs.AR], 2017.

\bibitem{lu2021case}
Z.~Lu and Q.~Cao, ``{A Case Study of Migrating RocksDB on Intel Optane Persistent Memory},'' in \emph{NAS}, 2021.

\bibitem{singh2022sibyl}
G.~Singh, R.~Nadig, J.~Park, R.~Bera, N.~Hajinazar, D.~Novo, J.~G{\'o}mez-Luna, S.~Stuijk, H.~Corporaal, and O.~Mutlu, ``{Sibyl: Adaptive and Extensible Data Placement in Hybrid Storage Systems Using Online Reinforcement Learning},'' in \emph{ISCA}, 2022.

\bibitem{oh2015sqlite}
G.~Oh, S.~Kim, S.-W. Lee, and B.~Moon, ``{SQLite Optimization with Phase Change Memory for Mobile Applications},'' \emph{VLDB Endow.}, 2015.

\bibitem{li2022multi}
Y.~Li, L.~Zeng, G.~Chen, C.~Gu, F.~Luo, W.~Ding, Z.~Shi, and J.~Fuentes, ``{A Multi-Hashing Index for Hybrid DRAM-NVM Memory Systems},'' \emph{JSA}, 2022.

\bibitem{raybuck2021hemem}
A.~Raybuck, T.~Stamler, W.~Zhang, M.~Erez, and S.~Peter, ``{HeMem: Scalable Tiered Memory Management for Big Data Applications and Real NVM},'' in \emph{SOSP}, 2021.

\end{thebibliography}
